%% file: LASr_1.tex
\documentclass[fleqn,usenatbib]{mnras}

\usepackage{graphicx}	
\usepackage{amsmath}	
\usepackage{amssymb}	
\usepackage{multicol}        
\usepackage{bm}		
\usepackage{pdflscape}	

\usepackage[T1]{fontenc}
\usepackage{ae,aecompl}

\usepackage{newtxtext,newtxmath}

\usepackage{longtable,lscape}
\include{macros}

 \title[LASr I. Galaxy sample and MIR colour selection]{Local AGN Survey (LASr): I. Galaxy sample, infrared colour selection and predictions for AGN within 100\,Mpc}

   \author[D. Asmus et al.]{D.~Asmus,$^{1,2}$\thanks{E-mail: d.asmus@soton.ac.uk}
           C.~L.~Greenwell,$^{1}$
   		   P.~Gandhi,$^{1}$
   		   P.~G.~Boorman,$^{1,3}$ 
   		   J.~Aird,$^{4}$
   		   D.~M.~Alexander,$^{5}$
   		   \newauthor 
   		   R.~J.~Assef,$^{6}$
   		   R.~D.~Baldi,$^{1,7,8}$
   		   R.~I.~Davies,$^9$
   		   S.~F.~H\"onig,$^{1}$
   		   C.~Ricci,$^{4}$
   		   D.~J.~Rosario,$^4$
   		   \newauthor 
   		   M.~Salvato,$^{8}$
   		   F.~Shankar,$^{1}$
   		   and D.~Stern$^{10}$\\
   		   $^1$Department of Physics \& Astronomy, University of Southampton, Hampshire SO17 1BJ, Southampton, United Kingdom\\
             $^2$European  Southern Observatory, Casilla 19001, Santiago 19, Chile\\
             $^3$Czech Academy of Sciences, N\'arodn\'i 3, 117 20 Star\'e Mesto, Czechia\\
             $^4$Department of Physics \& Astronomy, University of Leicester, University Road, Leicester LE1 7RJ, UK\\
             $^5$Centre for Extragalactic Astronomy, Department of Physics, Durham University, South Road, Durham, DH1 3LE, UK\\           
             $^6$N\'ucleo de Astronom\'ia de la Facultad de Ingeniería y Ciencias, Universidad Diego Portales, Av. Ej\'ercito Libertador 441, Santiago\\
             $^7$Dipartimento di Fisica, Universit\'a degli Studi di Torino, via Pietro Giuria 1, 10125 Torino, Italy\\
             $^8$INAF - Istituto di Astrofisica e Planetologia Spaziali, via Fosso del Cavaliere 100, I-00133 Roma, Italy\\
             $^9$Max Planck Institute for Extraterrestrial Physics(MPE), Giessenbachstr. 1, 85748 Garching, Germany\\
             $^{10}$Jet Propulsion Laboratory, California Institute of Technology, 4800 Oak Grove Drive, Pasadena, CA 91109, USA
             }

\date{Accepted 2020 Mar 12. Received 2019 Dec 19}

\pubyear{2020}

\begin{document}
\label{firstpage}
\pagerange{\pageref{firstpage}--\pageref{lastpage}}
\maketitle
 
\begin{abstract}
In order to answer some of the major open questions in the fields of supermassive black hole (SMBH) and galaxy evolution, a complete census of SMBH growth, i.e., active galactic nuclei (AGN), is required.
Thanks to deep all-sky surveys, such as those by the {\it Wide-field Infrared Survey Explorer} (\wise) and the {\it Spectrum-Roentgen-Gamma} ({\it SRG}) missions, this task is now becoming feasible in the nearby Universe.
Here, we present a new survey, the Local AGN Survey (LASr), with the goal of identifying AGN unbiased against obscuration and determining the intrinsic Compton-thick (CT) fraction.
First, we construct the most complete all-sky sample of galaxies within 100\,Mpc from astronomical databases ($90\%$ completeness for $\log (M_*/M_\odot) \sim 9.4$), four times deeper than the current local galaxy reference, the Two Micron All-Sky Survey Redshift Survey (2MRS), which turns out to miss $\sim 20\%$ of known luminous AGN.
These 49k galaxies serve as parent sample for LASr, called LASr-GPS.
It contains 4.3k already known AGN, $\ge82\%$ of these are estimated to have $\ltw < 10^{42.3}\,$erg\,s$^{-1}$, i.e., are low-luminosity AGN.
As a first method for identifying Seyfert-like AGN, we use \wise-based infrared colours, finding 221 galaxies at $\ltw \ge 10^{42.3}\,$erg\,s$^{-1}$ to host an AGN at $90\%$ reliability, This includes 61 new AGN candidates and implies and optical type~2 fraction of 50 to $71\%$. 
We quantify the efficiency of this technique and estimate the total number of AGN with $\lxi \ge 10^{42}$\,erg\,s$^{-1}$ in the volume to be $362^{+145}_{-116}$ ($8.6^{+3.5}_{-2.8}\,\times$\,10$^{-5}$\,Mpc$^{-3}$).
X-ray brightness estimates indicate the CT fraction to be 40--55\% to explain the {\it Swift} non-detections of the infrared selected objects.
One third of the AGN within 100\,Mpc remain to be identified and we discuss the prospects for the \erositaa all-sky survey. 
\end{abstract}


%

\section{Introduction}\label{sec:intro}
Today it is commonly accepted that all massive galaxies host a supermassive black hole (SMBH) at their centres. 
Furthermore, there is increasing evidence that the SMBHs somehow co-evolve with their host galaxies as, for example, indicated by empirical scaling relations between the SMBH mass and galaxy properties, such as the stellar velocity dispersion or stellar mass of the spheroidal component (e.g., \citealt{kormendy_coevolution_2013, shankar_selection_2016}).
The existence of such relations is somewhat surprising given the many orders of magnitude difference in size between the black hole sphere of influence and the bulk of the galaxy.
This raises the questions of how the feeding of the SMBH exactly works (e.g., \citealt{alexander_what_2012}), and if there is significant feedback from the SMBH onto the host galaxies.
The latter process is postulated by current cosmological simulations to suppress star formation and explain the galaxy population as observed today in the nearby Universe (e.g., \citealt{granato_physical_2004, shankar_new_2006, lapi_quasar_2006}). 

The last decades of research have significantly increased our understanding of SMBH growth (see \citealt{netzer_revisiting_2015} for a recent review). 
We know that SMBHs grow through several phases over cosmic time, during which large amounts of matter are accreted.
During its journey towards the event horizon, the material forms an accretion disk which, due to the release of gravitational energy, emits large amounts of radiation, mostly in the ultraviolet (UV) which then is partly reprocessed by surrounding material and secondary processes.
As a result, the galaxy nuclei appear as bright compact sources, often outshining the rest of the galaxy.
They are called active galactic nuclei (AGN).
AGN are bright emitters across most of the electromagnetic range and, thus, detectable throughout the entire visible Universe which allows us to directly trace SMBH growth over cosmic history.
In addition, AGN can produce strong outflows which are prime candidates for the feedback onto the host galaxy postulated above.
However, to robustly answer which processes are dominating the SMBH growth and the feedback, we require a complete census of the AGN phenomenon.
For example, precise knowledge of the AGN number counts in the local Universe would provide tight constraints on the duty cycle of AGN, radiative efficiencies and the luminosity and accretion rate distributions (e.g., \citealt{martini_quasar_2001, goulding_towards_2010, shankar_self-consistent_2009,  shankar_constraining_2019}).
Such a census is very challenging to carry out.
First of all, the accretion rates of SMBHs span a wide range from essentially zero up to values in excess of the Eddington limit.
Therefore, AGN span a huge range in luminosities from the nearly quiescent Galactic Centre, Sgr\,A$^*$, to the most powerful quasars roughly twelve orders of magnitude more luminous.
Faint AGN are difficult to detect, in particular if they do not outshine their host galaxy at some wavelengths.
Moreover, the majority of SMBH growth seems to be highly obscured from our lines of sight (e.g., \citealt{fabian_obscured_1999, ueda_toward_2014, buchner_obscuration-dependent_2015, ricci_compton-thick_2015}).
So what is the best, i.e., most efficient and least biased, way to find all the AGN?
Our best chance to achieve this is certainly in the nearby Universe, where the sensitivity and angular resolution of our instruments can be used to their largest effect for finding and characterising even highly obscured AGN.
This is the ultimate goal of the new survey presented here, the Local AGN Survey (LASr).
Its design is motivated by the following insights.

\subsection{Selecting AGN in the X-ray regime}
So far, one of the most successful ways to identify AGN has proven to be in the hard X-ray regime ($\gtrsim 10$\,keV).
Here, most AGN are luminous owing to UV photons from the accretion disk being Compton-up-scattered to higher energies by hot electrons.
These electrons are most likely part of a coronal region surrounding the innermost accretion disk. 
As a result, AGN are easily more luminous in X-ray than any other non-transient astronomical objects.
Another advantage is that X-ray emission are less affected by extinction than longer wavelength emission.
Both reasons together make AGN selection at these energies very reliable.
Specifically, the ongoing all-sky scan at 14-195\,keV with the Burst Alert Telescope (BAT; \citealt{barthelmy_burst_2005}) on the \swiftt satellite \citep{gehrels_swift_2004} provided us with the so far least biased local AGN samples  \citep{markwardt_swift/bat_2005,tueller_swift_2008,baumgartner_70_2013}.
Prominent examples are the Luminous Local AGN with Matches Analogues sample (LLAMA; \citealt{davies_insights_2015}; see \citealt{riffel_gemini_2018} for the Northern analogue) and the BAT AGN Spectroscopic Survey (BASS) samples, e.g., after 70\,month scanning time (hereafter B70 AGN sample; \citealt{koss_bat_2017, ricci_bat_2017}).

However, even the BAT AGN samples are restricted in two ways.
First, the sensitivity of this selection method is relatively low because of the low photon counts.
This caveat results in relatively high flux limits, so that even relatively powerful AGN remain undetected by BAT.
Second and more importantly, even at such high energies, Compton-thick (CT) obscuration ($\nh > 1.5 \cdot 10^{24}\cm^{-2}$) extinguishes the intrinsic flux by factors of ten and larger, resulting in a detection bias against CT obscured AGN.
This last point is a severe problem because the intrinsic fraction of CT-obscured AGN is probably around $\sim30\%$ (e.g., \citealt{ricci_compton-thick_2015, lansbury_nustar_2017,georgantopoulos_nustar_2019}, Boorman et al., in prep.), and possibly even up $50\%$ (\citealt{ananna_accretion_2019}; but see \citealt{gandhi_constraints_2007}). 
Both caveats will be somewhat mitigated in the future with deeper \swift/BAT maps although only slowly as the mission has already reached more than eight years of total integration time.
The newest X-ray satellite, the Russian-German ``{\it Spectrum-Roentgen-Gamma}'' (\srg) mission could allow for advance in this matter.
It hosts two telescopes which will perform a four-year all-sky survey at complementary X-ray energies, namely the extended ROentgen Survey with an Imaging Telescope Array (\erosita; \citealt{predehl_erosita_2010,merloni_erosita_2012}) operating at 0.2-10\,keV and the Astronomical Roentgen Telescope - X-ray Concentrator (ART-XC; \citealt{pavlinsky_art-xc_2011, pavlinsky_art-xc_2018}) operating at 4-30\,keV.
In terms of detecting AGN with their X-ray emission described by a typical power-law, these surveys are expected to be approximately ten times deeper than the current \swift/BAT survey.
Thus, these surveys are our best chance to probe the local AGN population at sufficient depth, in particular to detect (or place stringent constraints on) many of the still missing CT AGN.

\subsection{Selecting AGN in the mid-infrared regime}
Complementary to X-ray selection of AGN is selection in the mid-infrared (MIR).
About half of the primary emission from the accretion disk is absorbed by dust, surrounding the AGN probably on parsec scales in a more or less coherent structure (see \citealt{almeida_nuclear_2017} and \citealt{honig_redefining_2019} for recent reviews).
As a result, this dust is heated to temperatures of several hundred Kelvins and radiates thermally with the emission peaking in the MIR ($\sim3$ to $30\um$).
Owing to the more extended and probably clumpy structure of the dust, obscuration becomes a secondary effect at this wavelength regime and usually does not exceed a factor of a few, even in the worst cases \citep{stalevski_dust_2016}. 
This makes MIR emission a formidable tracer of the primary power of the AGN and allows a highly complete selection.
The recent all-sky survey of the {\it Wide-field Infrared Survey Explorer} (\wise; \citealt{wright_wide-field_2010}) allowed for the most progress here in the last years, thanks to its high sensitivity and spectral coverage.
However, AGN selection in the MIR has some major caveats as well, namely, severe contamination by emission of stellar origin.
At shorter wavelengths, $\lesssim 6\um$, this includes radiation of old stars, while at longer wavelengths, $\gtrsim 6\um$, dust heated by young stars in star forming regions can dominate the total MIR emission of galaxy.
Moreover, AGN and intense star formation events often occur together in time and space, e.g., triggered through galaxy interaction and mergers.
Therefore, any AGN selection in the MIR is prone to host contamination.
Finally, both X-ray and MIR selection are biased against low luminosity and low accretion rate objects, in particular if the SMBH accretes radiatively inefficiently (e.g., \citealt{ho_radiatively_2009}).
Such systems can be much more efficiently selected at radio wavelengths (e.g., \citealt{best_sample_2005,padovani_faint_2016, tadhunter_radio_2016,baldi_lemmings_2018}).

\subsection{The new local AGN survey}
The discussion above shows that no single selection technique can lead by itself to a complete, unbiased AGN sample (see \citealt{hickox_obscured_2018} for a comprehensive review on AGN selection).
Instead a combination of techniques is required. 
This is the approach of LASr. 
Specifically, we want to combine the advantages of the high completeness achievable in the MIR and the high level of reliability in the X-rays to identify all efficiently accreting SMBHs.
Applied to the all-sky surveys of \wisee, \erositaa and \artxc, combined with our nearly complete knowledge of the local galaxy population, LASr should allow us to significantly improve our understanding of the local AGN population and construct the most complete AGN census yet in the nearby Universe with particular focus on the highly obscured objects.

LASr will be performed throughout a series of papers, combining different AGN identification techniques to construct a highly complete AGN sample as final result.
In this first paper, we start LASr by selecting the survey volume, assembling the parent sample of galaxies, and employing the first AGN identification technique.
Specifically, we create a list of all known galaxies within the volume  (Sect.~\ref{sec:sam}) called the LASr galaxy parent sample (LASr-GPS).
It will serve as a base sample for the application of different AGN identification techniques.
In this paper, we focus on the MIR and use the \wisee catalogs to first characterise LASr-GPS in terms of completeness and bulge MIR properties (Sect.~\ref{sec:gal_cha}) before starting the AGN census (Sect.~\ref{sec:AGN}).
This first includes the characterisation of the already known AGN in the volume, followed by the application of the first AGN identification technique, namely by \wisee colours.
This is the most easily-available technique, allowing us to find most of the more luminous AGN in the sample, i.e., those that are more luminous than their host galaxy in at least one \wisee band.
Usually, this is the case for AGN with bolometric luminosities $\gtrsim 10^{43}\ergs$ (e.g., \citealt{alexander_x-ray_2005}) and corresponds to AGN classified as ``Seyferts'' based on their optical emission line ratios.
Such AGN probe significant SMBH growth, which seem to be the most relevant for our main science questions, i.e., cases that contribute significantly to the total mass budget of the SMBH and/or cases where sufficient energy is released to have an impact on the host galaxy. 
The big advantage of MIR colour selection is that it is little affected by obscuration bias, allowing us to identify highly obscured AGN with particular focus on new CT candidates. 
We discuss the newly found AGN and CT AGN candidates in Sect.~\ref{sec:candi} and Sect.~\ref{sec:CT}, respectively, including the prospects to detect them in X-rays.
Throughout this work, we will use the so far least biased AGN sample, the B70 AGN sample, in order to characterise the selection steps of LASr AGN.
Specifically, the characterisation of the MIR colour-based AGN identification technique employed here allows us to estimate the total number counts of AGN in our volume (Sect.~\ref{sec:num}).
This paper is then concluded by a comparison of these numbers to luminosity functions from the literature (Sect.~\ref{sec:lfunc}).

In future papers, we will employ additional MIR-based AGN identification techniques, e.g., variability and SED decomposition, as well as present follow-up observations of AGN candidates. The highly complementary X-ray-based AGN identification can then be provided by the \erositaa and \artxcc all-sky surveys once available.

\section{Creation of the galaxy parent sample}\label{sec:sam}
In this section, we first describe the motivation for the selection of the volume for LASr.
Next, we require a galaxy parent sample highly complete in terms of galaxies sufficiently massive to host an AGN, which can then be used to select AGN from.
We will see that current local galaxy samples do not fulfill this criterion so that we have to assemble our own galaxy parent sample.
Finally, we describe the assembly of the galaxy properties relevant for this work, namely the coordinates, redshifts, and distances, allowing us to find the MIR counterparts of the galactic nuclei and compute their luminosities. 

\subsection{Selection of the volume}\label{sec:vol}
We wish to construct a highly complete census of SMBH growth in the local Universe.
The choice of volume to be used for this purpose is motivated by several factors.
\begin{itemize}
    \item In order to obtain a census that is representative for the whole AGN population, the volume needs to be representative of the larger scale, low redshift Universe. 
    It is estimated that cosmological isotropy is reached for length scales of $\sim200 h^{-1}$\,Mpc with $h = H_0/100$\,km\,s$^{-1}$Mpc$^{-1}$ and $H_0$ the Hubble constant \citep{sarkar_testing_2019}.
    
    \item The volume should also be large enough to sample rarer AGN sub-populations in sufficient numbers to yield statistically robust conclusions on their relevance. 
    Here particular emphasis should be on the high luminosity AGN regime because these may dominate the integrated black hole growth and AGN feedback (e.g., \citealt{aird_evolution_2010,fabian_observational_2012}). 
    However, high-luminosity AGN have a low space density. 
    For example, current estimates of the AGN luminosity function in X-rays, e.g., \cite{aird_evolution_2015}, let us expect
    a space density of $\sim 5\,\times\,10^{-7}$\,Mpc$^{-3}$ for AGN with an intrinsic X-ray luminosity of $\lxi \ge 10^{44}\ergs$, e.g., $\sim 20$ objects within a sphere of $200$\,Mpc radius.
    
    \item On the other hand, the volume should be small enough so that the depth of the all-sky surveys, used to identify and characterise the AGN, is sufficient to probe the lower parts of the AGN luminosity range. 
    This is particularly important in the X-rays where extinction is large for obscured AGN. 
    For example, the final all-sky maps of \erositaa and \artxcc are expected to have depths of $\sim 1.6 \cdot 10^{-13}\fu$ and $\sim  3 \cdot 10^{-13}\fu$ at 2-10\,keV, respectively \citep{merloni_erosita_2012, pavlinsky_art-xc_2018}, which corresponds to a distance of $150-250$\,Mpc for an observed X-ray luminosity of $\lxo = 10^{42}\ergs$.
    However, CT AGN are suppressed by easily a factor of 10 to $\sim 100$ at these wavelengths.
    
    \item The MIR is much less affected by extinction, but sensitivity is the key restricting factor.
    I.e., the \wisee all-sky maps have an average depth capable of detecting an AGN with a 12$\um$ luminosity $\ltw = 10^{42}\ergs$ up to a distance of 220\,Mpc with $\ge 3\sigma$ in band\,3 (W3$\sim11.6$\,mag; \wisee documentation\footnote{\url{http://wise2.ipac.caltech.edu/docs/release/allsky/}}). 
    
    \item Another factor to take into account is that the completeness of our parent sample of galaxies directly restricts the completeness of our AGN search. 
    According to a recent estimate, our all-sky redshift completeness is only $78\%$ for galaxies with a redshift $z<0.03$ \citep{kulkarni_redshift_2018}, and the completeness is quickly dropping towards higher redshifts.

    \item Finally, once identified, we need to follow-up and characterise all the AGN in the volume. 
    We are especially interested in spatially resolving the in and outflows on sub-kiloparsec scales for as many objects as possible, which puts a feasibility-based upper limit on the volume.
    For example, at an object distance of $250\,$Mpc, one kiloparsec corresponds to one arcsec on sky, which is close to the effective resolution limit of most telescopes.
\end{itemize}
The above factors advocate to implement LASr as an all-sky survey with a spherical volume given by a radius between $\sim100\,$Mpc and $\sim 250\,$Mpc.
While, we plan to later use the larger value, 250\,Mpc, we start LASr first with the lower value, 100\,Mpc, to verify our approach.
Using the cosmological parameters of \cite{collaboration_planck_2016}, an object distance of 100\,Mpc corresponds to a redshift of $z=0.0222$.

\subsection{On the 2MRS galaxy sample}\label{sec:2MRS}
The current, commonly used reference sample for the local galaxy population is based on the 
the Two Micron All Sky Survey (2MASS; \citealt{skrutskie_two_2006}),  namely the 2MASS Redshift Survey (2MRS; \citealt{huchra_2mass_2012}).
It contains 45k galaxies which were selected from the 2MASS Extended Source Catalog (XSC) and the 2MASS Large Galaxy Atlas (LGA; \citealt{jarrett_2mass_2003}) according to the following criteria, a) detected in the $K$-band with $\mathrm{K}_\mathrm{S}  \le 11.75$\,mag; b) low foreground extinction with $E(B-V) < 1.0\,$mag; and c) sufficiently far from the Milky Way plane with $|b| > 5\degree$ for $30\degree < l < 330\degree$ and $|b| > 8\degree$ otherwise with $l$ and $b$ being the Galactic longitude and latitude, respectively. 
For $\sim 98\%$ of the objects, redshifts were collected from the literature or dedicated follow-up observations by \cite{huchra_2mass_2012}, so that the final galaxy sample with redshifts comprises 43.5k galaxies.
Out of those, 15k galaxies are within 100\,Mpc distance from us, to which we refer to simply as the ``2MRS sample'' in the following.

So far, it has been assumed that the 2MRS sample contains all major galaxies, at least outside of the Galactic plane.
However, $13\%$ of the host galaxies of the 191 B70 AGN within $D< 100$\,Mpc are not part of the 2MRS. 
This fraction increases to $26\%$ for $D<250$\,Mpc.
Since we aim at a final completeness of $>90\%$ for AGN-hosting galaxies, we have to complement the 2MRS sample (next section).
The necessity of this extension of the 2MRS is further discussed in Sect.~\ref{sec:gal_compa}.

\subsection{The LASr-GPS}\label{sec:uni}
In order to build our parent sample of galaxies for LASr, hereafter LASr-GPS, we combine the 2MRS with galaxies from the  major public astronomical databases.
Namely, the LASr-GPS is created by querying the December 2018 release of the NASA/IPAC Extragalactic Database (NED\footnote{\url{http://ned.ipac.caltech.edu/}}), the May 2018 release of the Centre de Donn\'ees astronomiques de Strasbourg (CDS) Set of Identifications, Measurements and Bibliography for Astronomical Data (SIMBAD\footnote{\url{http://simbad.u-strasbg.fr/simbad/}};  \citealt{wenger_simbad_2000}), and the most recent, i.e., 15th, data release of the Sloan Digital Sky Survey (SDSS\footnote{\url{https://www.sdss.org}}; \citealt{blanton_sloan_2017, aguado_fifteenth_2019}) for all objects within the redshift limit.

\begin{figure*}
   \includegraphics[angle=0,width=0.8\textwidth]{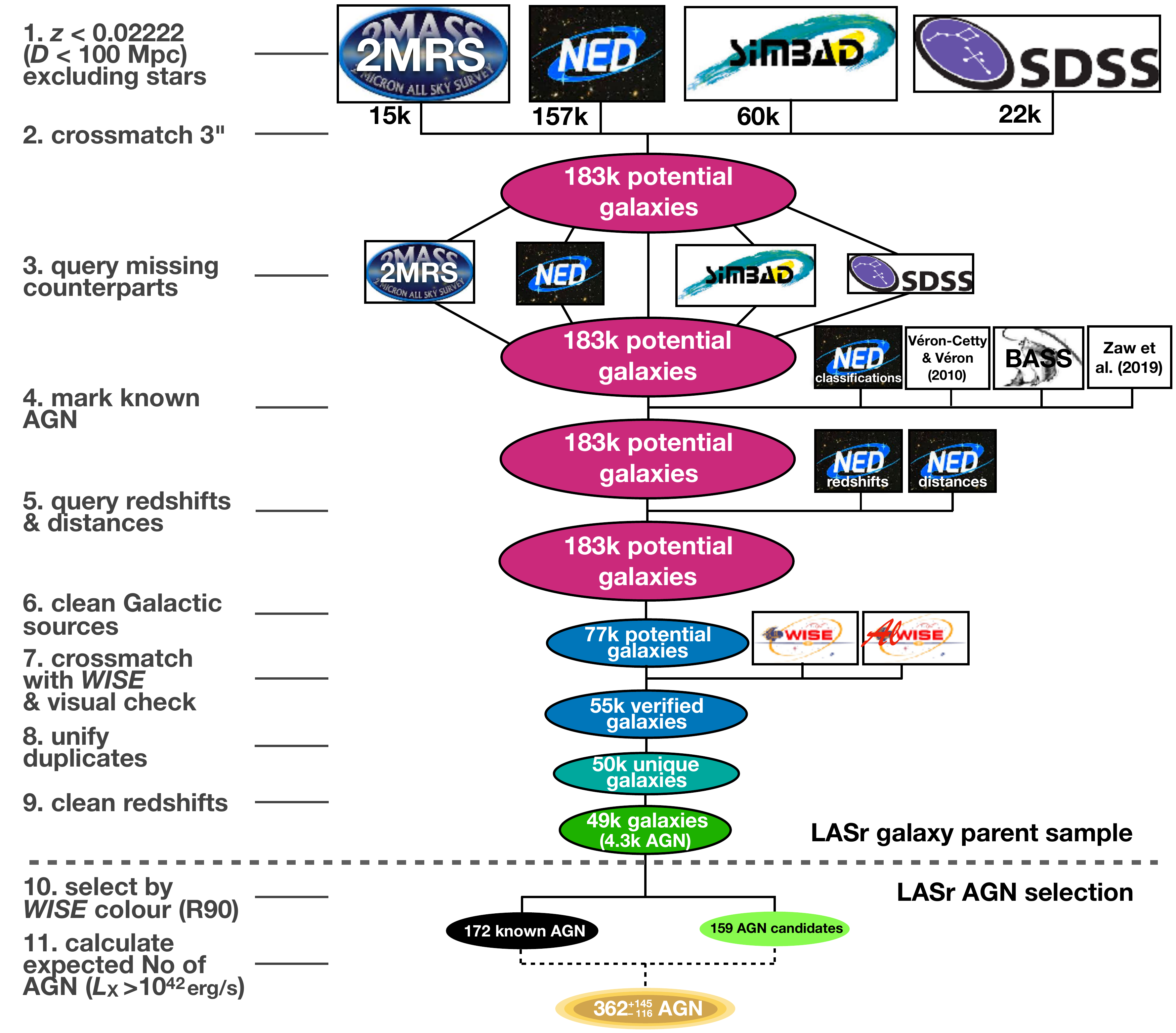}
    \caption{
  Schematic recipe for assembly of the LASr-GPS and following AGN selection. 
  It starts at the top with the  numbers of potential galaxies within the redshift limit found in the different databases 2MRS, NED, SIMBAD and SDSS.
  These are then merged into one sample of potential galaxies which is then cleaned and further information added in a number of steps, proceeding to the bottom, until the final parent sample of 49k verified galaxies, the LASr-GPS, is reached after step~9. 
  See Sect.~\ref{sec:uni} for a detailed description of each step in the LASr-GPS assembly, while the AGN selection that follows below the dashed line in steps~10 and 11 is described in Sect.~\ref{sec:R90} for the known AGN, Sect.~\ref{sec:candi} for the new candidates, and Sect.~\ref{sec:num} total AGN number estimate.
}
   \label{fig:sch}
\end{figure*}

A multi-stage cleaning process is necessary with iterations before and after merging of these different subsets to remove duplications, spurious redshifts and other contaminants in order to obtain a clean galaxy sample.
The full assembly process is illustrated in Fig.~\ref{fig:sch}.
Its order is partly dictated by practical aspects in the selection process.

In short, we first exclude all objects classified as stars if they have a redshift $z<0.001$ as well as objects with unreliable or photometric redshifts (step 1 in Fig.~\ref{fig:sch}),
yielding 157k, 60k, and 22k from NED, SIMBAD and SDSS respectively.
We crossmatch these subsets in step 2, unifying all matches within a cone of 3\,arcsec radius\footnote{
The radius is chosen to be well below the angular resolution of \wisee and prevent incorrect matches.
}.
Not for all objects a counterpart is found in every subset.
However, many of these objects actually have entries in NED, SIMBAD or SDSS but either without assigned redshift or are classified as stars in that database.
Thus, they were not selected in step 1.
In order to gather as much information as possible about each object, we therefore query for all still missing counterparts in the corresponding databases (step 3).
These steps yield 183k potential galaxies.

Next, we identify all known AGN within these potential galaxies by crossmatching with all major literature samples of AGN (step 4; see Sect.~\ref{sec:kAGN} for details)\footnote{
This is done at this early stage to ensure we are not losing any relevant objects in the following steps.
}.
In the next steps, 5 and 6, we first add the NED compiled redshifts and redshift-independent distances (NED-D; \citealt{steer_redshift-independent_2017}) and then use the added information from all the crossmatching to perform another cleaning of Galactic objects.
This is necessary because many objects are unclassified (or even erroneously classified as galaxies) in some of the databases but then identified as Galactic objects in others.
Most of these contaminants result from previous SDSS data releases included in NED and SIMBAD. 
Those contaminants we can now eliminate by using the most recent and improved classifications of SDSS DR15.
Specifically, we exclude all objects which have either (i) at least one classification as Galactic object but none as galaxy in NED, SIMBAD and SDSS (63k cases); (ii) at least two classifications as Galactic objects (4.8k cases); (iii) at least one classification as Galactic object and a redshift $< 0.0011$ (20k cases); or (iv) no classification as galaxy and a redshift $<0.0011$ (18k cases). 
This redshift threshold is determined from SDSS DR15 with the probability of being a genuine galaxy being $<1\%$ for all redshifts lower than that.
We make sure to keep all 2MRS galaxies during this step and check all doubtful cases individually to make sure that we do not erroneously exclude any genuine galaxy.
As a result of this cleaning, the sample is further reduced to 77k potential galaxies.
Then,  we perform crossmatching with \wisee (step 7; see Sect.~\ref{sec:WISE}).
During this step, we also check all objects visually and identify another 22k contaminants.
These are either entries from NED and SIMBAD where no optical counterpart is identifiable in the vicinity of the given coordinates, or cases where the coordinates point to a part of another galaxy in the sample. 
The reason for the latter can be inaccurate coordinates in NED and SIMBAD or multiple fibers placed on different parts of larger galaxies in SDSS.
This step is also used to correct coordinates of galaxies that are offset from its nucleus, or geometric centre (if the nucleus is unidentifiable).
The final two steps (8 and 9) clean the remaining duplicates, e.g., objects sharing the same \wisee counterpart, as well as objects with erroneous redshifts (see Sect.~\ref{sec:dist}).

The final galaxy sample contains 49k visually verified galaxies and includes all but 3 of the 15k 2MRS galaxies in the volume\footnote{
The excluded are: NGC\,6822 aka 2MASX\,J19445619-1447512 is a very nearby dwarf galaxy which is over-resolved in \wisee and 2MASS and thus can not be included.
2MASX\,J18324515-4131253 is actually part of ESO\,336-3. which is also in 2MRS, and, thus, it is excluded.
2MFGC\,02101 is most likely a foreground star in the outskirts of NGC\,1035 which is also in the 2MRS.
}.
Therefore, the LASr-GPS can indeed be seen as an extension of the 2MRS, and all the following steps performed with the LASr-GPS apply in equal measure to the 2MRS, unless mentioned otherwise.
The galaxies are listed in Table~\ref{tab_gal} which is available in its entirety online.
The LASr-GPS forms the parent sample for searching local AGN.


\subsection{Identification of known AGN}\label{sec:kAGN}
In our quest for a highly complete AGN sample, we benefit from the large amount of literature
that already identified many of the AGN in our volume.
NED and SIMBAD have collected a lot of these classifications which we obtained together with the object queries. 
In addition, for NED, we query the website of each individual object to extract the homogenized activity class as well as the basic description ("classifications") that also often contains information about any AGN in the system.
This results in 2617 AGN classifications from NED and 4398 from SIMBAD.
SDSS also provides AGN identifications based on an automatic assignment from the template fitting to the optical spectra, resulting in 271 automatic AGN classifications among the SDSS galaxies.

We complement these classifications with the two largest independent AGN collections, namely  \cite{veron-cetty_catalogue_2010} and  \cite{zaw_uniformly_2019}.
The former authors have collected 169k AGN from the literature of which 1135 are in our volume, while the latter have analysed all available optical spectra of the 2MRS galaxies, resulting in 8.5k AGN identifications of which 3078 are in our volume.
Finally, we add the new AGN identifications of the 191 B70 AGN within our volume.

In total, this leads to 4309 known AGN among the 49k galaxies of the LASr-GPS, of which 3887 are also in the 2MRS sample. 
Most of these have been identified using optical spectroscopic classifications.
We adopt optical AGN type classifications whenever they are available in the databases.
In addition, for the narrow-line AGN from \cite{zaw_uniformly_2019}, we perform the Seyfert, LINER (low-ionization nuclear emission-line region) and H\,II nucleus classification based on the emission line fluxes published in that work and the AGN diagnostics of \cite{kewley_host_2006}. 
This way, we could retrieve optical type classifications for $95\%$ (4101 of 4309) of the known AGN including 2409 Seyfert , 2053 LINER  and 1777 H\,II classifications\footnote{
The remaining 208 objects are simply classified as "AGN" in the databases without any optical type given.
}.
Here, we allow for multiple classifications of the same object\footnote{
There are 402 objects classified both as Seyfert and LINER, 993 as Seyfert and H\,II and 846 as LINER and H\,II.
} 
which is the case for $47\%$ of the AGN and to a large part the likely result of varying spectroscopic aperture, data quality and classification methods used.
In addition, some of the AGN identifications might be unreliable, in particular if the object has not been optically classified as Seyfert (1900 objects).
Most of the latter are optically classified as LINERs which is a controversial class with respect to its AGN nature because also stellar phenomena can produce similar emission line ratios (e.g., \citealt{stasinska_can_2008, cid_fernandes_alternative_2010, cid_fernandes_comprehensive_2011, yan_nature_2012, belfiore_sdss_2016, hsieh_sdss-iv_2017}).
These caveats have to be taken into account when using this compilation of classifications, and, in this work, we use them only for indicative purposes.
The same applies to the more detailed Seyfert obscuration type classifications, where we find 1012 objects classified as type~1 (Sy\,1.x) and 1545 as type~2 (Sy\,2) with $9\%$ (219) of the objects having both classifications or intermediate type (Sy\,1.8 or Sy\,1.9).
If we exclude all objects with multiple optical classifications, 490 type~1 and 475 type~2 AGN remain.

The known AGN are marked as such in Table~\ref{tab_gal} and their characteristics are further discussed in Sect.~\ref{sec:known_AGN_MIR}.

\subsection{Identification of known starbursts}\label{sec:kSB}
Not only AGN can produce significant MIR emission but also intense star formation does. 
Therefore, starbursts are the main source of contamination for AGN selection in the MIR (e.g., \citealt{hainline_mid-infrared_2016}).
In order to understand the effects of starbursts on the MIR appearance of galaxies, we also collect galaxies explicitly classified as starbursts in either of the databases, resulting in 4006 starbursts, mostly from SDSS (3762 objects).
Similar to the known AGN, the starburst sample is probably highly incomplete, but it shall serve us to understand the locus of starburst galaxies in the different parameter distributions in comparison to the AGN.
The corresponding objects are marked as well in Table~\ref{tab_gal}.

\subsection{Determination of redshifts \& distances}\label{sec:dist}
The most fundamental quantity that we require for each galaxy is its distance from us, not only to decide whether the galaxy is within our volume but also to determine its luminosity.
For most galaxies, the distance is estimated from the redshift for which we generally prefer the value provided by SDSS DR15, or NED if the former is not available. 
We consider a redshift robust if we either have a robust value in SDSS DR15 (their redshift confidence flag = 0), or we have at least two independent redshift measurements from all databases combined (including the redshift compilations in NED).
Otherwise, we consider the redshift somewhat uncertain and use a redshift confidence flag in Table~\ref{tab_gal} to mark these cases with a value of 1 ($0.5\%$ of the LASr-GPS), meaning that these values are not verified but there is no suspicion of a problem either.
In addition, there are several cases where the different redshift measurements are discrepant (standard deviation of measurements $>20\%$; $4.6\%$ of all galaxies).
In most of these cases, only one of the redshift measurements for the affected object is offset from the rest, often by a factor of two or more. 
In particular for the very nearby galaxies, we can thus often guess the ``right'' redshift from the visual size of the galaxy. 
For objects with discrepant redshifts, where we can not make a clear decision based on all available information, we assign a value of 2 to the redshift confidence flag ($0.3\%$ of the LASr-GPS), meaning that those redshifts are controversial and can not be trusted.
Therefore, we have robust redshifts for $99.2\%$ of the galaxies. 

With the redshifts, we compute the luminosity distance, $\dl$, for all galaxies.
However, in the nearby Universe $\dl$ can be inaccurate owing to the speed of the Hubble flow here being comparable to the peculiar motion of the galaxies.
Fortunately, a major effort of NED led to a large collection of 320k redshift-independent distance estimates for 182k galaxies dubbed NED-D \citep{steer_redshift-independent_2017}.

\begin{figure}
   \includegraphics[angle=0,width=\columnwidth]{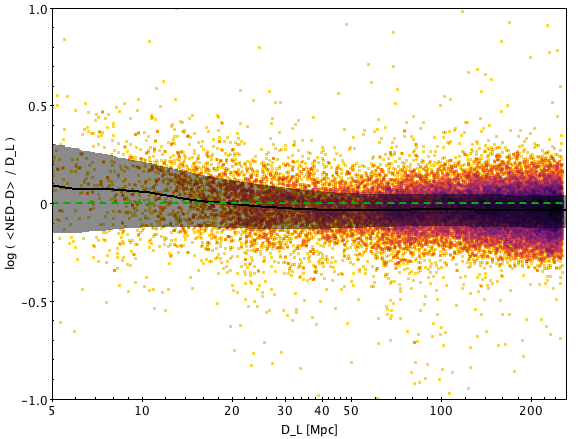}
    \caption{
    Distribution of the logarithmic ratio of the median NED-D redshift-independent distance and the luminosity distance, $\dl$, of each object as a function of $\dl$. 
    The colour scaling marks the density of the data points from yellow to black.
    The black line indicates the median value at a given $\dl$ with a width of 2\,Mpc, while the grey shaded area encompasses 2/3 of the population at each $\dl$. 
    The green dashed line marks the 1 to 1 correspondence. 
}
   \label{fig:NED-D}
\end{figure}

Of our 49k galaxies, NED-D values are available for 10.6k galaxies.
NED-D contains multiple measurements of very different methods for many galaxies, leading to a very heterogeneous data set. 
Unfortunately, it is not feasible here to perform a selection or weighting of different methods for each galaxy.
Instead, we simply compute the median of the different measurements.
Before adopting the NED-D values, we first compare them to our $\dl$ values in Fig.~\ref{fig:NED-D}.
As expected, we see that the deviation between NED-D and $\dl$ increases for small distances, while for larger distances the median ratio between the two converges to a constant value close to 1.
This happens roughly at $\dl = 50\,$Mpc.
Here, also the width of the scatter converges to 0.16\,dex (factor 1.44), indicating that above this value the scatter between the individual redshift-independent methods dominates over deviations from the Hubble Flow.
This motivates us to adopt the median NED-D value for the object distance if  $\dl < 50\,$Mpc (4.6k galaxies; $9.3\%$).
Otherwise we use  $\dl$. 
The resulting final redshifts and distances used are listed in Table~\ref{tab_gal}.

\subsection{Identification of WISE counterparts}\label{sec:WISE}
For the planned identification of AGN, we require the MIR properties of all the galaxies.
Therefore, we crossmatch our galaxy samples with the all-sky pointsource catalogues of \wise, specifically, the AllWISE catalogue (\citealt{cutri_vizier_2013}), and then visually verify the counterpart most likely corresponding to the nucleus of the galaxy.
In $93.3\%$ of the cases, this is the AllWISE source that is closest to the galaxy coordinates.
The median angular separation is 0.6\,arcsec and the 90th percentile is 2.7\,arcsec.
The large majority of the remaining $6.7\%$ are caused by inaccurate galaxy coordinates in the databases, which we clean manually.
Furthermore, for many small, late-type or disturbed galaxies, no nucleus can be robustly identified.
This is the case for $4\%$ of the LASr-GPS and $0.2\%$ of the 2MRS sample. 
We mark these galaxies with a corresponding warning flag in Table~\ref{tab_gal}. 
In these cases, we choose either the source closest to the approximate apparent geometric centre, or we take the brightest MIR emission knot overlapping with the optical counterpart (whichever seems more applicable).
Fortunately, these cases are predominantly small galaxies, which are the least relevant for our AGN search. 
Furthermore, in $0.6\%$ of the galaxies, the AllWISE catalogue failed to capture the nucleus for unknown reasons. 
For those, we fall back to the original data release catalogue (\citealt{cutri_vizier_2012}), which delivered a better counterpart in all cases.
This strategy allows us to allocate a \wisee counterpart to almost every object that is not rejected in any of the sample cleaning steps and iterations so that our final \wisee coverage is $99.94\%$.

However, we found that in 1.4\% of the galaxies, a nearby brighter source actually dominates the \wisee emission.
In those cases, the MIR emission of the latter is taken as upper limit for the fainter object.

Finally there are five cases\footnote{
2MASX\,J09181316+5452324,
AM\,1333-254,
IC\,1623,
IC\,2554,
VV\,662
} where the angular separation of two galactic nuclei was too small to be picked up as individual sources in the \wisee catalogues.
They are treated as one object, i.e., late-stage galaxy merger, in the following.

\subsection{Computation of MIR colours and luminosities} 
After having identified the most likely \wisee counterparts, we can now estimate the MIR emission of the galactic nuclei. 
The majority ($67\%$) of the galaxies are resolved in \wise, and, thus, their total MIR emission is not well captured in either of the \wisee catalogues (see, e.g., \citealt{cluver_galaxy_2014}).
However, since here we are mostly interested in the nuclear MIR emission we use the profile-fitting magnitudes in AllWISE, which roughly capture, and certainly not underestimate, the nuclear emission.
This was verified for nearby AGN by, e.g., \cite{ichikawa_complete_2017} through comparison with high angular resolution MIR data.
One might argue that the profile-fitting photometry is even superior for other purposes because it excludes most of the extended non-nuclear emission.

We calculate the observed central $3.4\um$ and $12\um$ luminosities, $\lwone$ and $\lwthree$, for each galaxy using the best estimate distance determined in Sect.~\ref{sec:dist} and the assigned  \wisee band 1 and 3 magnitudes,  \wonee ($\lambda = 3.4\um$)  and \wthreee ($\lambda = 11.56\um$), after first converting magnitudes to flux densities following the WISE documentation\footnote{\url{http://wise2.ipac.caltech.edu/docs/release/allsky/expsup/sec4_4h.html}}.
Owing to the low redshifts of our sources, no $K$ corrections are required.

\section{Characterisation of the parent sample of galaxies}\label{sec:gal_cha}
Before we study the AGN in our volume, we first compare the 2MRS and LASr-GPS and then address the completeness of the latter to better understand which limitations this might put on our subsequent AGN selection.

\subsection{Comparison of galaxy samples}\label{sec:gal_compa}
\begin{figure*}
   \includegraphics[angle=0,width=0.9\textwidth]{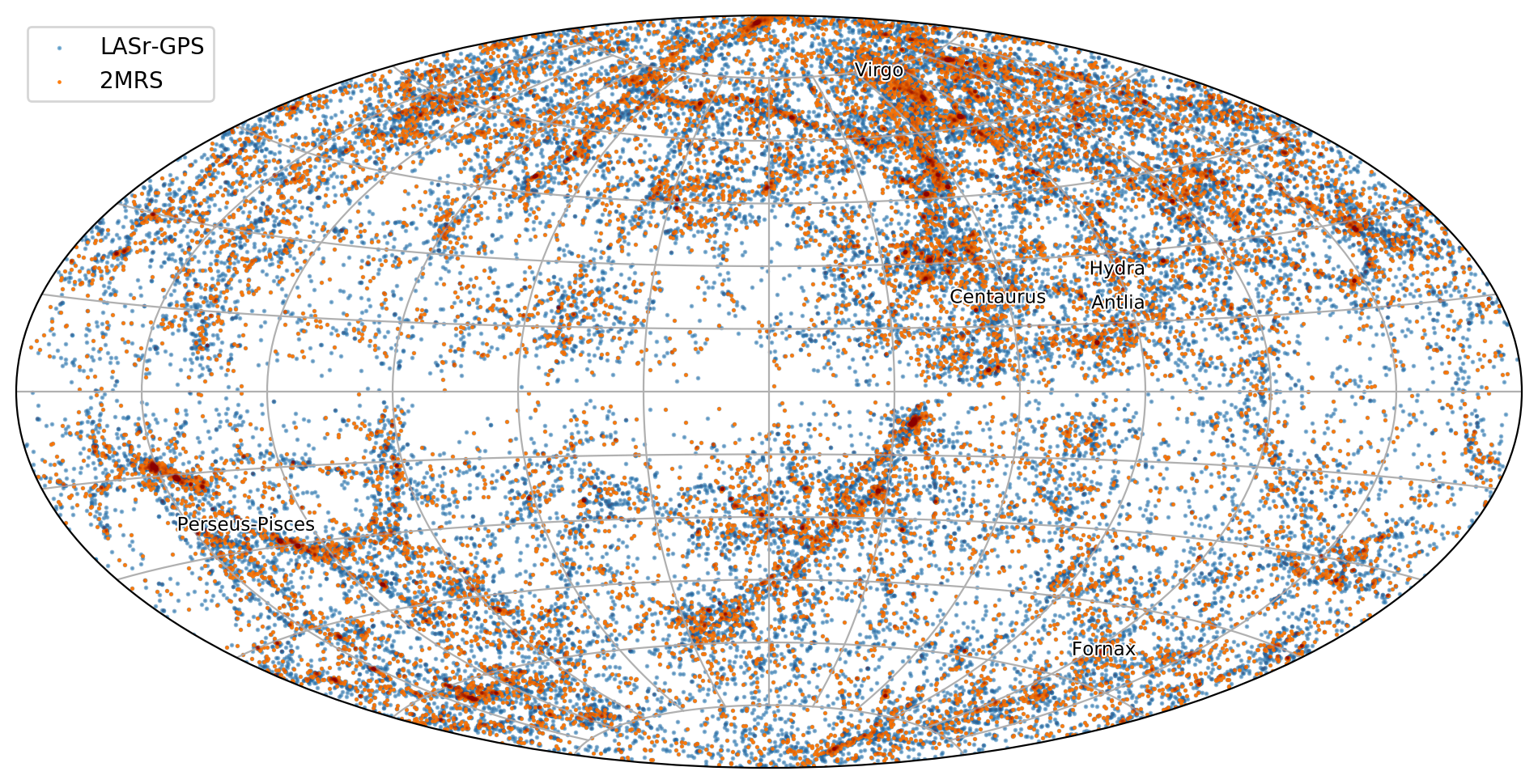}
    \caption{
  Aitoff projection of the Galactic coordinate distributions of all galaxies within the redshift limit from the 2MRS (orange) and LASr-GPS (blue).
  Darker colours mark areas of overdensity in linear scale, mostly marking the cosmic filaments within the volume.
  The center lines of the plot mark Galactic longitude $l=0$\,h and  Galactic latitude $b=0\degree$, respectively.
  Some nearby galaxy clusters are labelled.
}
   \label{fig:sky1}
\end{figure*}

\begin{figure*}
   \includegraphics[angle=0,width=0.7\textwidth]{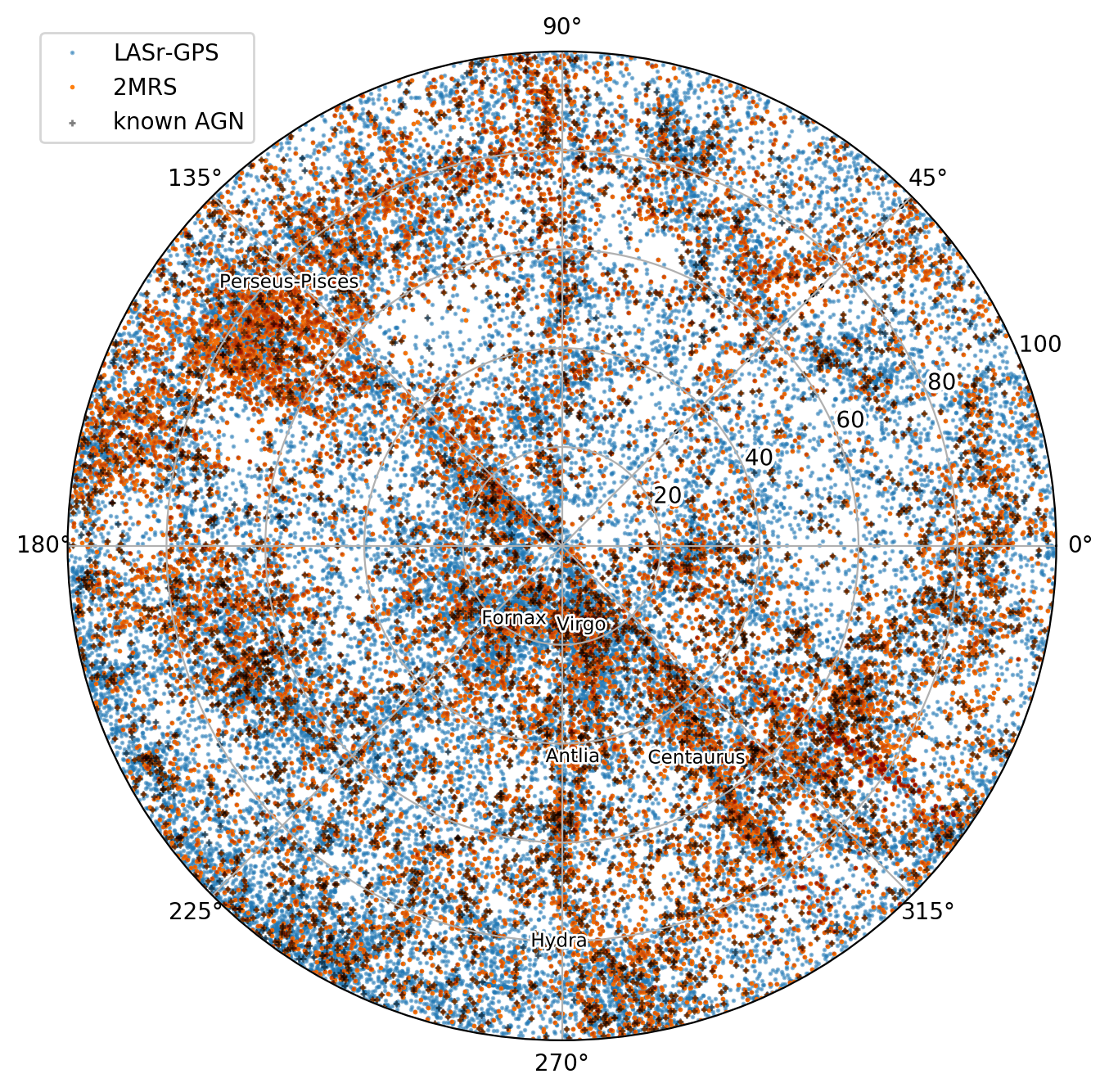}
    \caption{
  2D projection of the distributions of all galaxies into the Galactic longitude plane within the   redshift limit from the 2MRS (orange) and LASr-GPS (blue).
  Darker colours mark areas of overdensity in linear scale.
  Semi-transparent black crosses mark known AGN. 
  The radial axis states the radial object distance in Mpc. 
  Some nearby galaxy clusters are labelled.
}
   \label{fig:sky2}
\end{figure*}

\begin{figure*}
   \centering
   \includegraphics[angle=0,width=0.33\textwidth]{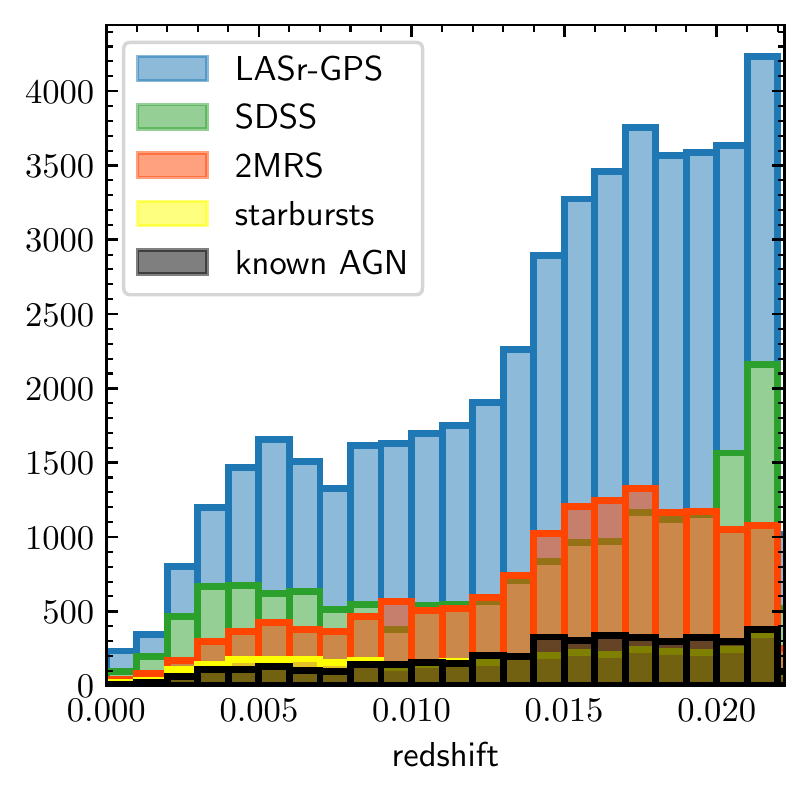}
    \includegraphics[angle=0,width=0.33\textwidth]{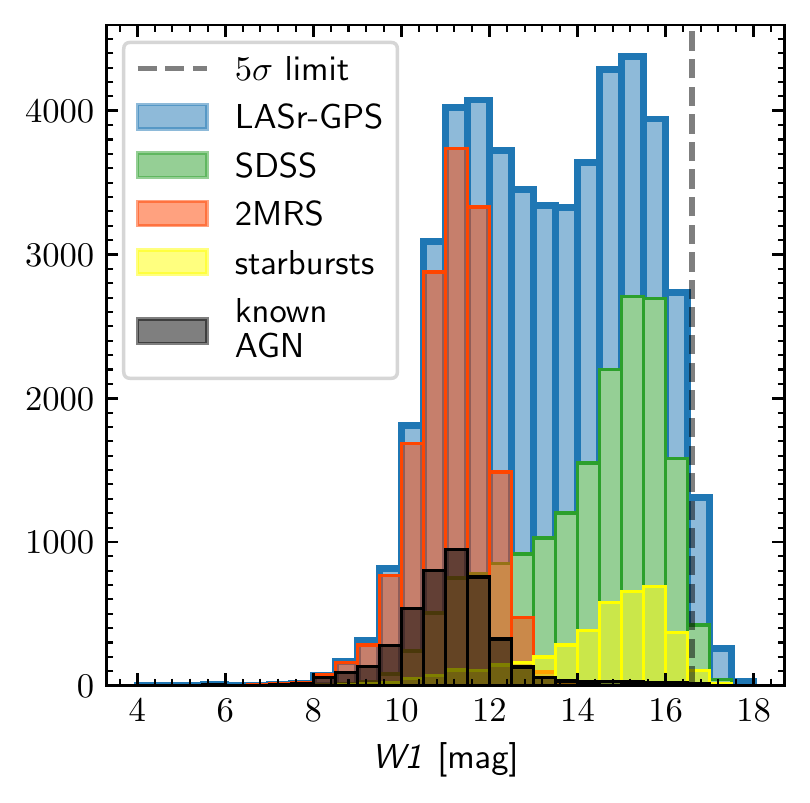}
    \includegraphics[angle=0,width=0.33\textwidth]{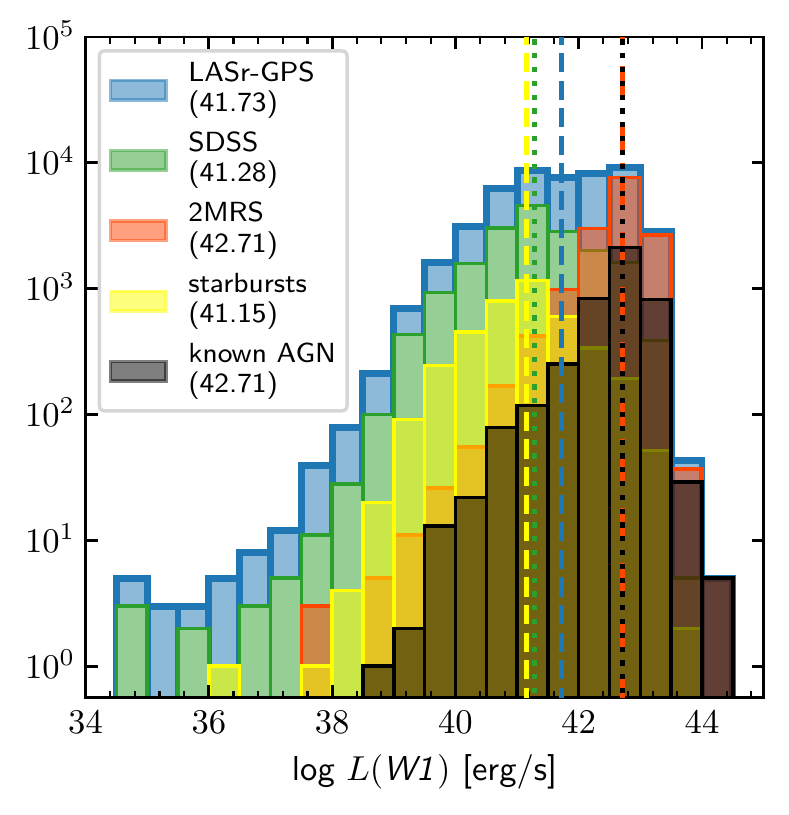}
    \caption{
    Redshift (left), \wisee central \wonee magnitude (middle) and luminosity (right) distributions of all galaxies  from the 2MRS (orange) and LASr-GPS (blue).
    For comparison, also the distribution of sources in SDSS DR15 is shown (green).
    The distribution of known starbursts is shown in yellow, while known AGN are shown in black.
    In addition, the middle plot shows the nominal $5\sigma$ depth of the AllWISE catalog as grey dashed line, while the right plot shows the median luminosities for each subsample as vertical dashed, dotted and dash-dotted lines of the corresponding colour.
}
   \label{fig:hist1}
\end{figure*}

First, we examine the spatial distribution of the galaxy parent samples in different projections, namely the all-sky map  (Fig.~\ref{fig:sky1}), the 2D projection onto the Galactic plane (Fig.~\ref{fig:sky2}) and the redshift distribution  (Fig.~\ref{fig:hist1}).
Most of the additional galaxies in the LASr-GPS compared to the 2MRS are in the Northern hemisphere (DEC$>0\degree$), which is mostly owing to SDSS. 
This is visible also in Galactic coordinates (Fig.~\ref{fig:sky1}), where the core area of SDSS is in the Galactic North ($b > 30\degree$).
In addition, both the LASr-GPS and 2MRS are clearly missing galaxies behind the Milky Way plane (we come back to that in Sect.~\ref{sec:gal_compl}).
Otherwise, the distribution of the 2MRS galaxies in particular follows the cosmological filaments and galaxy clusters contained in our volume (Fig.~\ref{fig:sky1}).
This is also visible in the Galactic plane projection (Fig.~\ref{fig:sky2}), although to a lesser degree probably owing to the collapse of the latitude dimension and the proper motions of the galaxies.
The latter can affect the redshift-based luminosity distances and, this way, artificially spread the filaments and clusters in radial direction (e.g., Centaurus, labelled in the figure).
Both sky projections indicate that our galaxy samples probe more or less well the cosmological structure of matter within the volume.

The redshift distribution (Fig.~\ref{fig:hist1}, left) illustrates that the number of galaxies in the LASr-GPS steeply rises with increasing distance up to the border of the volume. 
In addition, there is a dip in the redshift distribution around $z\sim0.01$ ($D\sim 45\,$Mpc) which is probably caused by the small scale anisotropy of the nearby Universe, namely voids to the Galactic East, North and West visible in Fig.~\ref{fig:sky2}.

\begin{figure}
   \centering
   \includegraphics[angle=0,width=1\columnwidth]{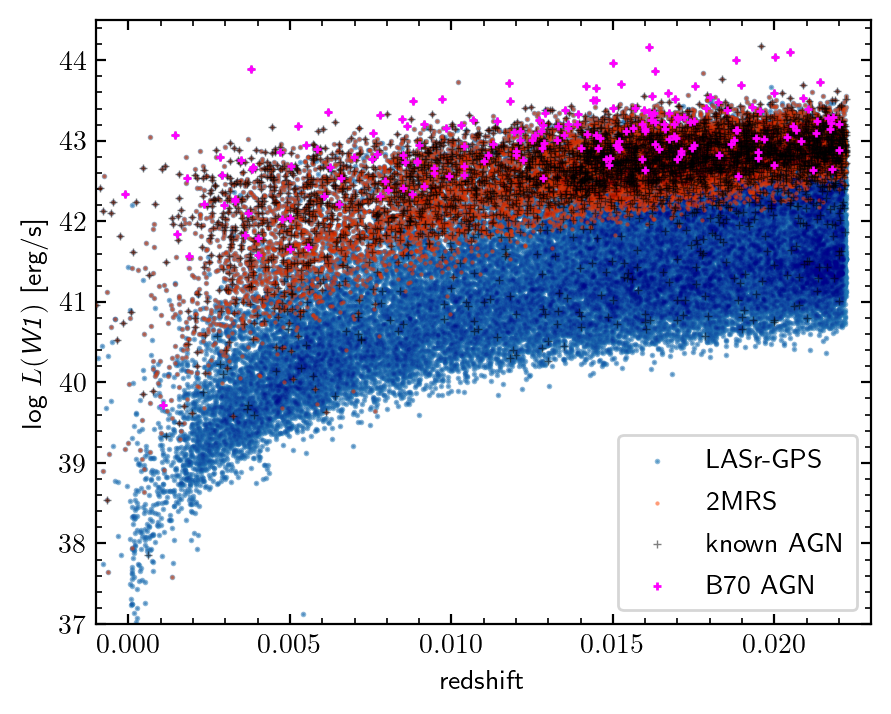}

    \caption{
    Central \wisee \wonee luminosity over redshift for all galaxies  from the 2MRS (orange) and LASr-GPS (blue).
    Semi-transparent black crosses mark known AGN, while magenta crosses mark B70 AGN.
}
   \label{fig:W_z}
\end{figure}

The redshift distribution of the 2MRS sample, on the other hand, levels off at $z\sim0.017$ and even decreases towards higher redshifts (Fig.~\ref{fig:hist1}, left).
This indicates that already at 100\,Mpc, the 2MRS starts missing galaxies owing to its $K$-band brightness cut.
The comparable shallowness of 2MRS with respect to the LASr-GPS is also visible in the \wisee central \wonee magnitude  and luminosity distributions (Fig.~\ref{fig:hist1}, middle and right),  as well as in the \wonee luminosity over redshift distribution  (Fig.~\ref{fig:W_z}).
The latter plot shows that the LASr-GPS probes the galaxy population down to central luminosities of $\lwone \sim 10^{41}\,$erg/s at a distance of 100\,Mpc, while the 2MRS has a depth of $\lwone\sim 10^{42.5}\,$erg/s.
The median central \wonee luminosity compared to the LASr-GPS and SDSS are also significantly higher for the 2MRS (by 1\,dex and 1.4\,dex, respectively). 
Similar trends apply as well to the other \wisee bands, just at higher magnitudes and lower luminosities (thus not shown here).
%

Interestingly, there are, however, also a significant number of galaxies well within the 2MRS brightness range but missing from 2MRS, as can be seen best in Fig.~\ref{fig:hist1}, middle and right.
Are all these galaxies situated in the Galactic plane?

\begin{figure}
   \centering
   \includegraphics[angle=0,width=1\columnwidth]{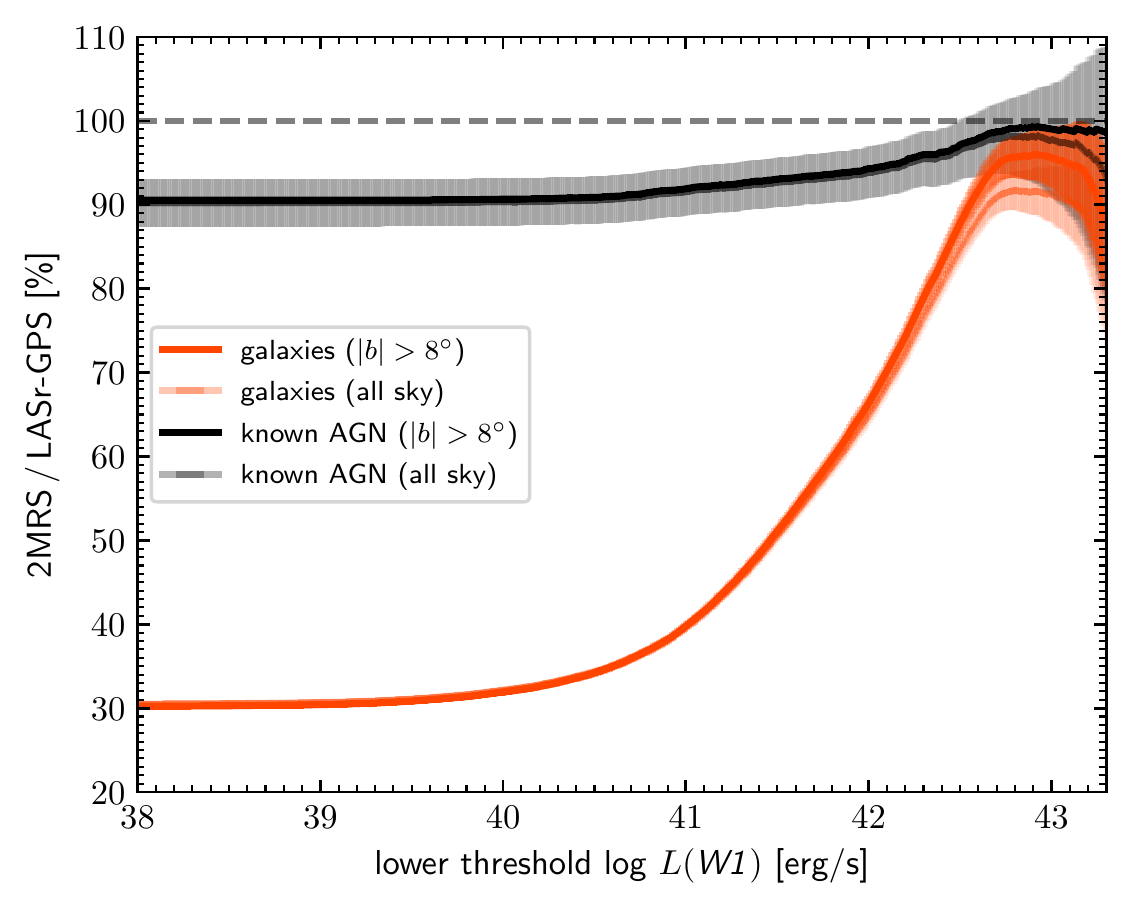}
    \caption{
    Galaxy and known AGN number ratios of 2MRS to LASr-GPS over a lower central \wonee luminosity limit (orange lines).
    The dark orange coloured line marks the galaxy ratio outside the Galactic plane, $|b|>8\degree$, while the light orange coloured line shows the galaxy ratio for the whole sky.
    Furthermore, the black (grey) line shows the known AGN ratio outside the Galactic plane, $|b|>8\degree$ (for the whole sky).
}
   \label{fig:2MRSfrac}
\end{figure}

To investigate this further, we examine how the galaxy number ratio of 2MRS over LASr-GPS evolves with luminosity in Fig.~\ref{fig:2MRSfrac}.
In the low luminosity regime, the galaxy ratio is $\sim 30\%$, while for $\lwone \gtrsim 10^{41}\ergs$, it starts to rise, surpassing $90\%$ at $\lwone > 10^{42.6}\ergs$ and finally reaching the maximum value of $96\%$ at $\lwone > 10^{42.9}\ergs$.
The latter numbers are for excluding the Galactic plane as defined for the 2MRS sample selection ($|b|>8\degree$).
If, we compare the 2MRS to LASr-GPS ratio over the whole sky, the maximum 2MRS fraction drops to $91.6\%$, reached at the same $\lwone$.
We can also look at the ratio of known AGN in 2MRS over LASr-GPS (also shown in Fig.~\ref{fig:2MRSfrac}).
Here, the minimum fraction is relatively high at $90\%$ already for low luminosity thresholds, i.e., $90\%$ of known AGN host galaxies are in the 2MRS. 
However, the ratio reaches its maximum of $99.1\%$ only at  $\lwone > 10^{42.9}\ergs$.

In conclusion, even outside the Galactic plane, the completeness of the 2MRS sample peaks only at $\lwone > 10^{42.9}\ergs$, which is well within the AGN regime, e.g., the B70 AGN host galaxies have a median of $\lwone = 10^{43}\ergs$.
This explains why the 2MRS is missing a significant fraction of B70 host galaxies and thus probably of the whole local AGN population, which justifies our extension to the LASr-GPS to maximise completeness.

\subsection{Completeness of LASr-GPS}\label{sec:gal_compl}
In the previous section, we have shown that the LASr-GPS provides a higher completeness in terms of potentially AGN hosting galaxies compared to the 2MRS sample.
However, how complete and deep is the LASr-GPS in absolute terms?

Optimally, one would want to express this depth in the physical galaxy property of total stellar mass.
However, here we simply use the unresolved \wisee emission which is missing significant stellar light depending on the galaxy size and distance. 
Furthermore, the mass-to-light ratio is not constant but depends on many galaxy parameters like galaxy type, metallicity and star formation rate and history (e.g., \citealt{wen_stellar_2013} and discussion therein).
Therefore, we refrain here from attempting stellar mass estimates but rather express the sample depth simply in the central \wonee luminosity.
For most galaxies, this quantity is probably dominated by the stellar bulge.

In Fig.~\ref{fig:W_z}, we already constrained the maximum depth of the LASr-GPS to be $\sim 10^{41}\ergs$ at a distance of 100\,Mpc.
The actual achieved completeness above this luminosity is  dictated by the redshift completeness in our case.
Owing to the heterogeneous nature of the public databases, their completeness is difficult to assess, and this can only be done empirically.
For example, \cite{kulkarni_redshift_2018} used a comparison of detected supernova events in galaxies with and without redshifts in NED to estimate the redshift completeness of the latter database to be $\sim78\%$ within a redshift of $z <0.03$.
This value provides a lower limit for our LASr-GPS, combining NED with other sources and being at lower redshift where completeness should be higher.
It particular, we will try to derive more accurate estimates here based on comparisons with  two highly complete galaxy surveys, one large-area survey (being representative of the volume), and one small-area survey (being very deep and highly complete).

\subsubsection{Comparison with SDSS}\label{sec:SDSS_comp}
\begin{figure}
   \centering
   \includegraphics[angle=0,width=0.33\textwidth]{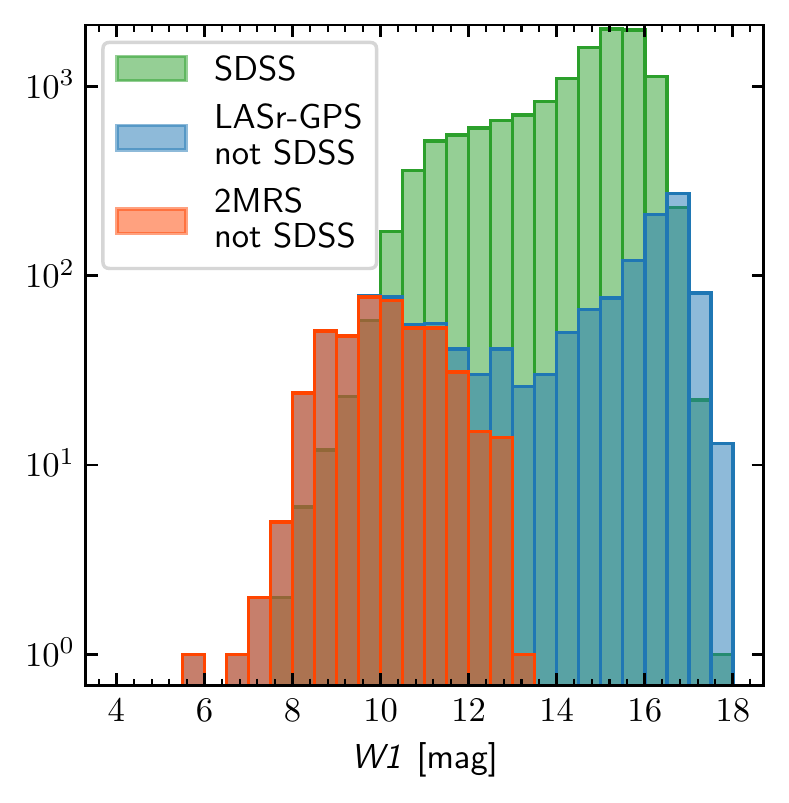}
    \caption{
    \wonee distribution of galaxies within the SDSS spectroscopic core area.
    Galaxies with SDSS spectra are shown in green, galaxies without but part of the LASr-GPS are shown in blue, and 2MRS galaxies without SDSS spectra in that area are shown in orange. 
}
   \label{fig:sdsscompl}
\end{figure}

The most powerful constraint for our redshift completeness is coming from the comparison to SDSS as reference for the highest available redshift completeness  at reasonable depth and representative sky coverage.
For simplicity, we here define the SDSS spectoscopic core area with simple cuts of $0\degree <$ DEC $< +60\degree$, 8:40\,h\,($130\degree$)\,$<$\,RA\,$<$\,16:00\,h\,($240\degree$).
This area comprises $13.2\%$ of the sky and contains 12.7k galaxies selected by LASr-GPS with SDSS spectroscopy in DR15 and a redshift placing them within our volume.   
The average redshift completeness of SDSS is $\sim 90\%$ but decreasing towards brighter galaxies for technical reasons \citep{montero-dorta_sdss_2009, reid_sdss-iii_2016}.
Indeed, we find that there are an additional 1503 galaxies of the LASr-GPS within this area but without SDSS redshifts, implying that the SDSS completeness is at most $88\%$.
As expected, these missing galaxies are bimodially distributed at the extremes of the galaxy brightness distribution (Fig.~\ref{fig:sdsscompl}), whereas the bright peak is almost completely made up by 2MRS galaxies that are not in SDSS.

\begin{figure}
   \centering
   \includegraphics[angle=0,width=1\columnwidth]{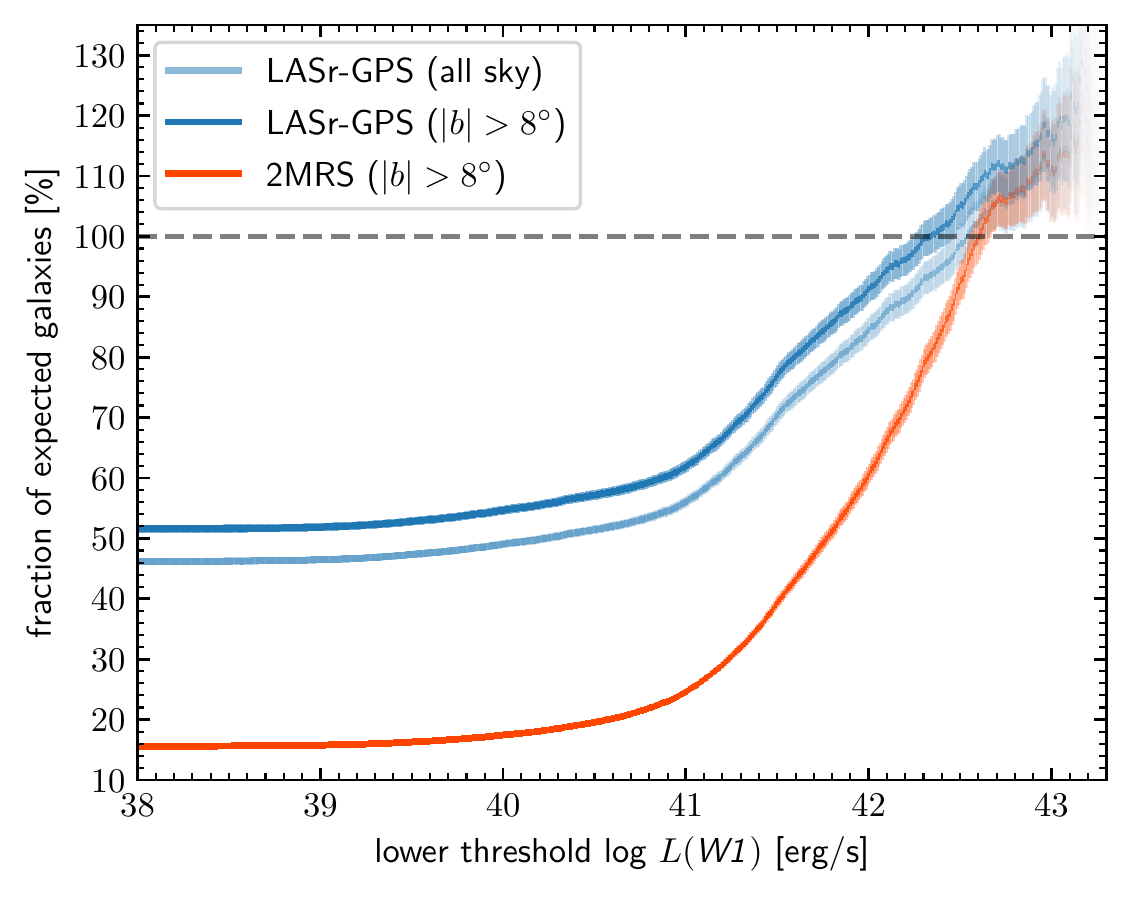}
    \caption{
    Fraction of expected galaxy included in the LASr-GPS, i.e., completeness of LASr-GPS, over a minimum central \wonee luminosity.
    The expected numbers are extrapolated from the SDSS core area (see main text for details).
    The solid coloured  lines mark the fraction of galaxies away from the Galactic plane, $|b|>25\degree$,  for the LASr-GPS (blue) and the 2MRS sample (orange), whereas the semitransparent surface gives the $1\sigma$ uncertainty.
    The the lighter coloured blue line shows the same without excluding the Galactic plane.
    The grey dashed line marks the 100$\%$ completeness level.
}
   \label{fig:galfrac}
\end{figure}

To mitigate the incompleteness of SDSS, we complement it with all galaxies from the 2MRS and LASr-GPS within the SDSS core area and assume that the result is $100\%$ complete within this area down to $\wone \lesssim 17$\,mag.
Further assuming that the SDSS core area define above is representative of the whole sky, we can use the above galaxy sample to estimate the galaxy \wonee luminosity distribution for the whole sky within our volume.
In Fig.~\ref{fig:galfrac}, we examine how the fraction of expected galaxies that are in the LASr-GPS above a lower \wonee luminosity limit, i.e., the completeness, depends on that lower luminosity threshold. 
For  $\lwone \lesssim 10^{41}\ergs$, the completeness is approximately constant between $50\%$ and $60\%$, if we  cut out the Milky Way plane ($|b|>8\degree$), and  $<50\%$ otherwise.
For higher $\lwone$, the completeness outside the Milky Way plane rises and reaches $90\%$ ($100\%$) at $\lwone = 10^{42}\ergs$ ($10^{42.3}\ergs$).

Maybe surprising, the observed to expected fraction continues rising above $100\%$ at higher luminosities. 
We interpret this behaviour as the result of a possible under-density of such luminous galaxies in the SDSS spectroscopic core area, which could lead to such an effect given the decreasing number statistics at high luminosities and the relatively small fraction of the sky contained in that area.
This would also explain why the  observed to expected fraction for the whole sky as well reaches $100\%$ despite the obvious under-sampling in the Milky Way plane.
Alternatively, this could imply that for galaxies with $\lwone > 10^{42.5}\ergs$, the under-sampled area does not contain a significant number of such luminous galaxies ($13.9\%$ of the sky for $|b|=8\degree$).
Finally, the 2MRS sample reaches $100\%$  completeness at only $\lwone > 10^{42.6}\ergs$, excluding the Milky Way plane (also shown in Fig.~\ref{fig:galfrac}).

\subsubsection{Comparison with GAMA}\label{sec:GAMA}
To further assess the completeness of the LASr-GPS, we also compare to a smaller area survey than SDSS with higher depth and completeness like the  Galaxy And Mass Assembly survey (GAMA; \citealt{liske_galaxy_2015}).
In particular, we use the two deep fields G12 and G15 from the latest release DR3\footnote{http://www.gama-survey.org/dr3/} \citep{baldry_galaxy_2018}.
The two fields combined cover a sky area of $\sim 120$\,deg$^2$ with a redshift completeness of $98.5\%$ for $r < 19.8\,$mag \citep{liske_galaxy_2015}.
Combined, they contain almost 100k galaxies, of which 811 are within $D<100\,$Mpc. 
The release versions of NED and SIMBAD used here do not include the GAMA DR3, allowing us to use them to test the completeness of LASr-GPS in an independent way.
For this, we cross-match the GAMA galaxies with the AllWISE catalogue, following the same method as throughout this work.
This yields counterparts for 720 of the 811 GAMA galaxies ($89\%$).
Out of those, 68 have $\lwone > 10^{42}\ergs$. 
Based on the SDSS-based result we would expect at least $90\%$  of them to be in the LASr-GPS.
Indeed, 64 out of 68 , i.e., 94$\%$,  are also in the LASr-GPS, verifying our high completeness above this lower luminosity threshold.

\subsubsection{Galactic plane shadow}\label{sec:MW}
\begin{figure}
   \includegraphics[angle=0,width=\columnwidth]{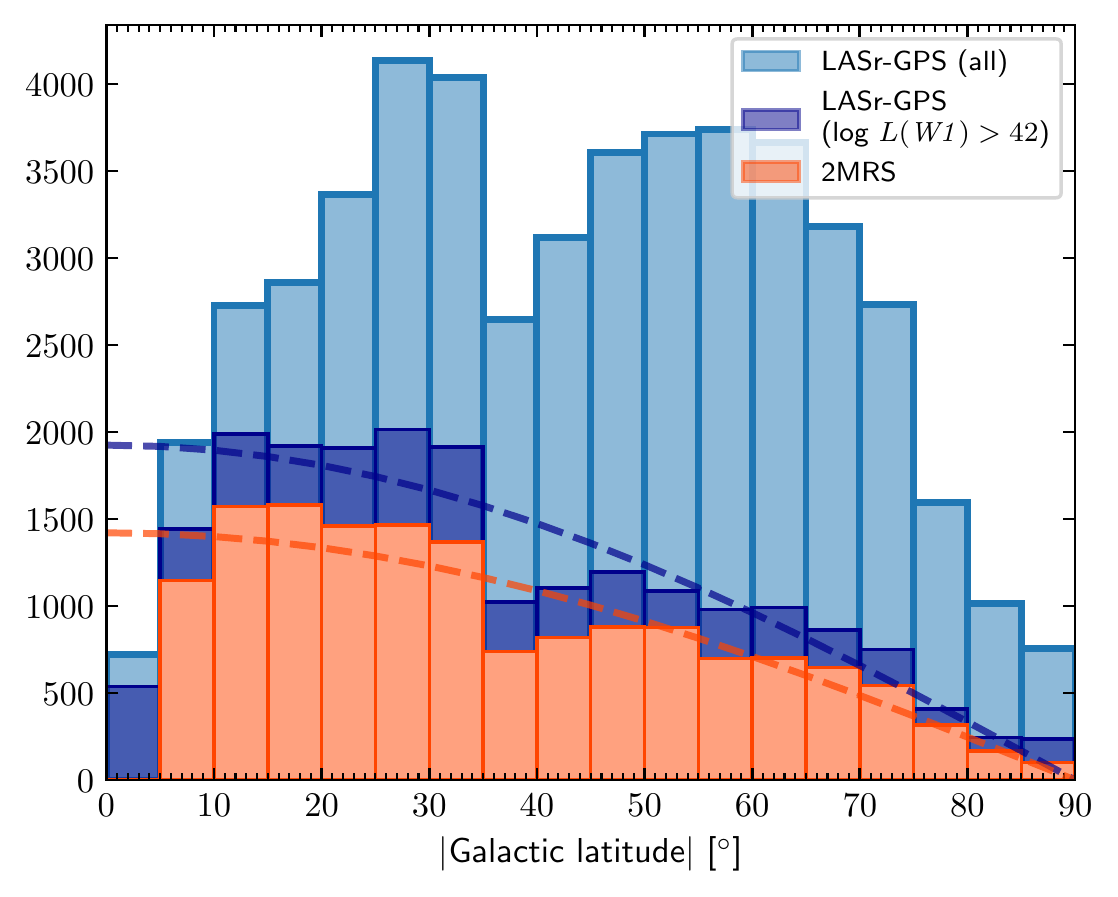}
    \caption{
   Absolute galactic latitude distribution of galaxies from the LASr-GPS (light blue) and the 2MRS sample (orange).
   The dark blue histogram marks galaxies from the LASr-GPS with $\lwone > 10^{42}\ergs$.
   The dashed lines mark fits of cosine shape to the corresponding distributions.
}
   \label{fig:MWO}
\end{figure}

The above results indicate that the LASr-GPS has a relatively high completeness for at least moderately luminous galaxies ($\lwone > 10^{42}\ergs$).
However, this statement excludes one big source of incompleteness of course, the shadow of the Milky Way, which through a combination of densely clustered foreground emission sources, and high values of extinction makes it very difficult to identify and characterize galaxies that have sky coordinates close to the Galactic plane.
To quantify this effect, we look at the absolute Galactic latitude distribution of galaxies (Fig.~\ref{fig:MWO}).
If the galaxies were distributed fully randomly in the sky, then the latitude distribution should describe a cosine, which is approximately the case, at least for the 2MRS and the LASr-GPS, if restricted to galaxies with  $\lwone > 10^{42}\ergs$.
At low latitudes ($|b| \lesssim 20\degree$), the observed distributions fall short of the expectations owing to the Galactic plane shadow.
In addition, the latitude distributions of all galaxy samples show a valley, i.e., an under-density between $35\degree \lesssim |b| \lesssim 45\degree$, caused by the voids in the local volume as already seen in the previous sky position and redshift distributions.

In order to quantify the Galactic plane shadowing, we fit a cosine functions to the bins with $|b| > 20\degree$ (shown as dashed lines in Fig.~\ref{fig:MWO}), and find the deficiency is $6.4 \pm 0.8\%$, whereas the uncertainty is estimated from using different binnings for the latitude distributions.
As expected, the 2MRS has a slightly higher missing fraction, owing to its latitude cut ($7.4 \pm 1.2\%$).
The Milky Way foreground will always be a problem for the study of galaxies behind it.
Therefore, is is probably easier to simply exclude this area from the volume when constructing samples for the AGN census in order to maximise completeness.  

\subsubsection{Concluding remarks on completeness}\label{sec:rem}
In the previous subsections, we addressed the completeness of the LASr-GPS empirically including the effect of the shadowing by the Galactic plane, leading to an all-sky completeness of $84\%$ for  $\lwone > 10^{42}\ergs$ and $96\%$ for $\lwone > 10^{42.3}\ergs$).
Outside the Galactic plane, we reach a completeness of at least $90\%$ for central luminosities of  $\lwone > 10^{42}\ergs$ and approximately $100\%$ for $\lwone > 10^{42.3}\ergs$.
These luminosities approximately correspond to stellar masses of $\log (M_*/M_\odot) \sim 9.4$ and $9.7$, respectively, using the simple relation provided by \cite{wen_stellar_2013}.
But again one has to keep in mind that these values are missing significant amounts of stellar light for most galaxies including only their bulges.

We plan to further increase the redshift completeness of the LASr-GPS in future work.
However, the above values mean that LASr can already now probe quite deep into the SMBH accretion regime, probing all galaxies where significant growth is occurring.
By going to smaller volumes, we could decrease lower luminosity limits further.
Although, at low luminosities, usually the stellar light by far dominates the total galaxy emission over the AGN, making it very difficult to isolate the AGN from its host.
We will address this as well in future follow-up works where we will try to use more sensitive (but complex) AGN identification techniques.
Here, we will utilize the simple but effective \wisee colour selection as a first probe of the AGN activity within the volume.

\section{Starting the AGN census}\label{sec:AGN}
With the depth and completeness of the LASr-GPS characterised, we can now move on to identify and characterise the AGN population within our volume.
We start with a brief summary of the already known AGN and then move on to the first AGN identificationn technique for LASr, namely \wisee MIR colour selection.
We estimate the efficiency of this technique and discuss possible limitations before applying it to the LASr-GPS to find new AGN candidates, in particular highly obscured and CT objects.
Next, we discuss possible host contamination and provide prospects for detecting these new AGN with the current X-ray all-sky surveys.
We conclude this section with an estimate of the total number AGN above a given luminosity limit within the volume, constraints on the CT fraction, and a comparison to luminosity functions from the literature.

\subsection{Luminosity estimates for the known AGN}\label{sec:known_AGN_MIR}
\begin{figure}
   \includegraphics[angle=0,width=0.95\columnwidth]{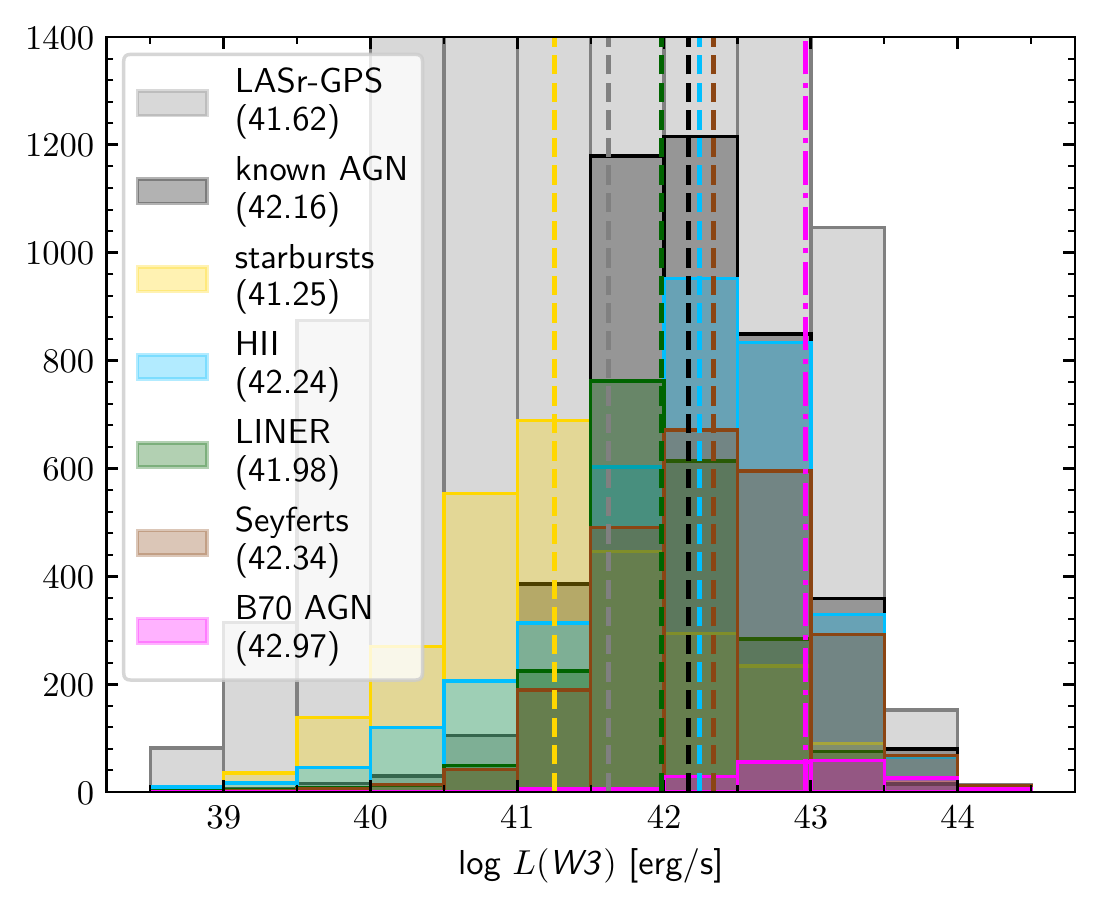}
      \includegraphics[angle=0,width=0.95\columnwidth]{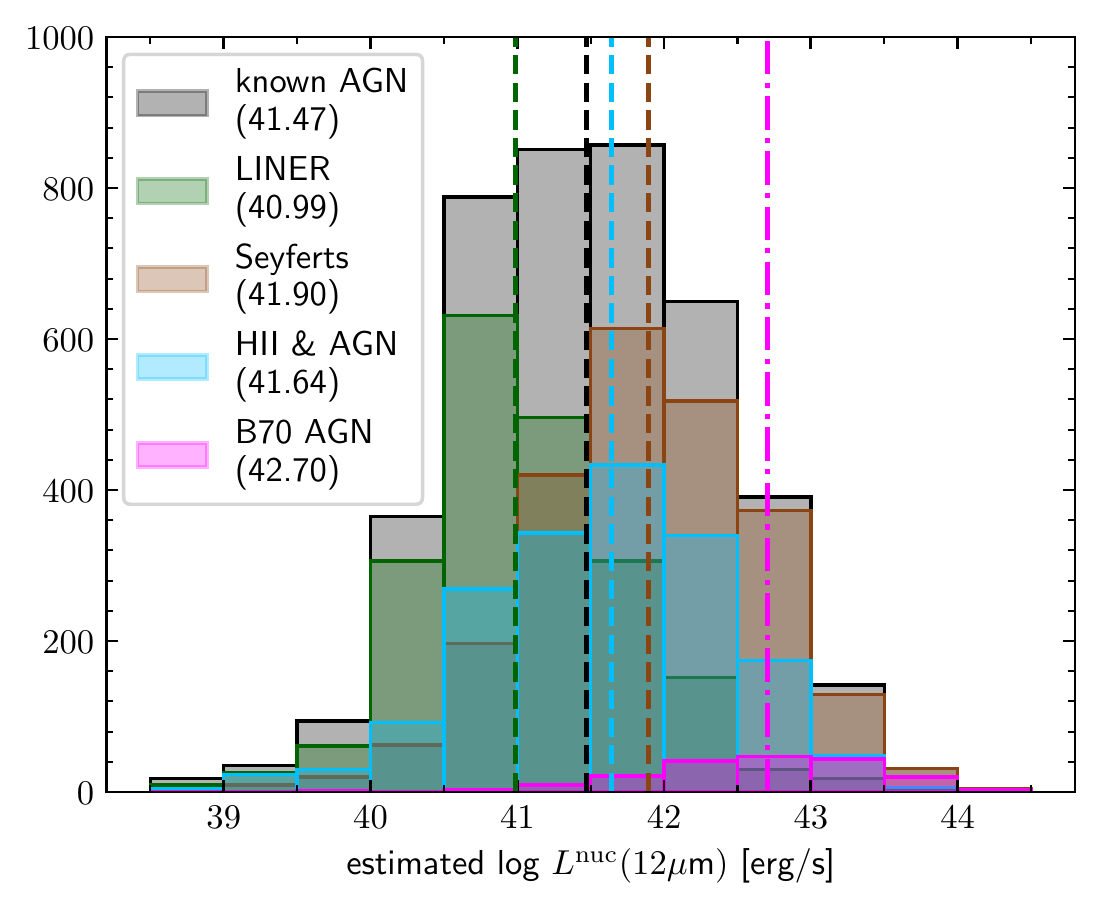}
    \caption{
      Top: \wthreee luminosity distribution for different AGN and star-formation hosting galaxy populations within LASr-GPS, namely known AGN (black), starbursts (yellow), H\,II nuclei (blue), LINERs (green), Seyferts (brown) and B70 AGN (magenta).
      The distribution of the whole LASr-GPS is shown in grey in the background.
      The dashed lines of the corresponding colour mark the median value which is also shown in the legend. 
      \newline Bottom: Estimated $\ltw$ distribution after decontamination of $\lwthree$ as described in Sect.~\ref{sec:known_AGN_MIR}.
}
   \label{fig:LW3hist}
\end{figure}

We know already from the collection of AGN identifications from the literature that there are at least 4.3k AGN within the volume (Sect.~\ref{sec:kAGN}).
The redshift and brightness distributions of their host galaxies are shown in Fig.~\ref{fig:hist1} and Fig.~\ref{fig:W_z}.
In order to further characterise the AGN population, we can use the \wthreee luminosities, tracing the $12\um$ continuum of the AGN, dominated by warm ($\sim 300\,$K) AGN heated dust.
Compared to the shorter bands, $\wthreee$ has the advantage of not being affected by stellar light. 
The $\wthreee$ luminosity distribution of known AGN is shown in Fig.~\ref{fig:LW3hist}, top panel.
As expected, the majority of known AGN seem to be relatively low-luminosity, e.g., compared to the B70 AGN. 
However, we know that star formation can also significantly contribute to $\wthreee$, in particular in large aperture measurements as used here\footnote{
The relatively high luminosities of the H\,II nuclei confirms this statement.
This does not apply to the systems classified as starbursts because many of them are compact dwarfs and, thus, do not reach such high luminosities.
}. 

\begin{figure}
   \includegraphics[angle=0,width=\columnwidth]{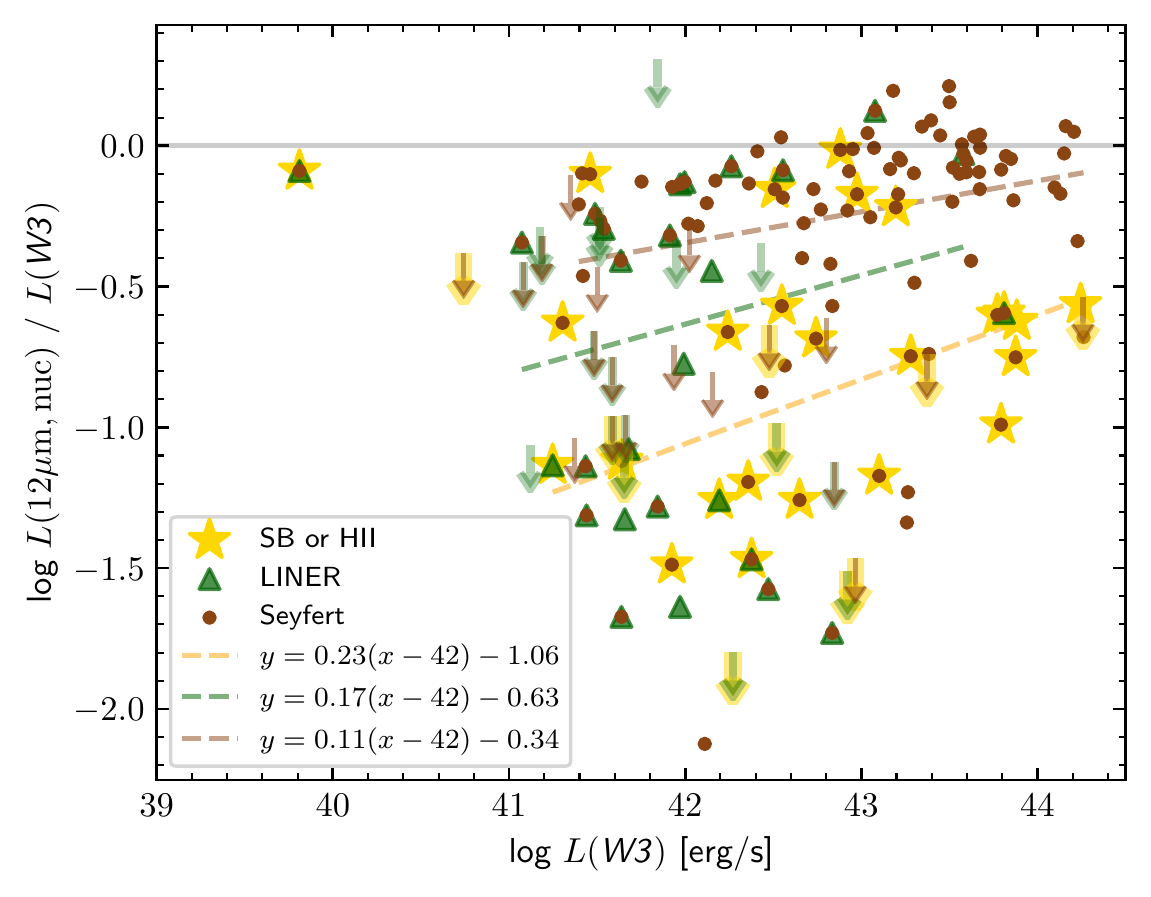}
    \caption{
Logarithmic ratio of nuclear $12\um$ luminosity, $\ltw$ from Asmus et al. (2014) to profile-fitting \wthreee luminosity, $\lwthree$, over $\lwthree$ for all 146 objects in common. 
Objects that are identified as star-forming in the literature (SB or HII) are marked with golden stars, while such with LINER classification have green triangle, and those with Seyfert classifications have brown dots. 
Objects can have several of these classifications in which cases the corresponding symbols are over-plotted. 
Nuclear $12\um$ non-detections are marked by arrows of the corresponding colours. 
The light grey line marks the zero line, while the dashed lines provide linear fits to various sub-populations of detections, namely the dark-gray line for all objects only classified as Seyferts, the green line for Seyferts that also have LINER classification but not star-forming and in orange for all objects without Seyfert classification.
}
   \label{fig:12rat}
\end{figure}

Decoupling AGN and star formation emission in \wthreee is a serious issue and requires detailed SED modelling, beyond the scope of this work.
However, we can attempt at least a rough decontamination of the \wthreee luminosities by computing statistical correction factors from the comparison of \wthreee to  high angular resolution measurements at the same wavelength.
In particular, \cite{asmus_subarcsecond_2014} presented a catalog of 253 nearby AGN with ground-based subarcsecond MIR photometry and estimated accurate estimates of the $12\um$ AGN luminosity, $\ltw$.
We crossmatch our AGN sample with this  catalog, finding 146 objects in common. 
In Fig.~\ref{fig:12rat} we show the ratios of $\ltw$ to the \wthreee profile-fitting luminosity, $\lwthree$.
While we already know from, e.g., \cite{asmus_subarcsecond_2014} that the nuclear to large aperture $12\um$ ratio is a strong function of the AGN luminosity, the same ratio shows only a weak increasing trend with increasing $\lwthree$ with a large scatter of $0.5$\,dex (Kendall's $\tauk=0.25$, null hypothesis probability $\log \pk = 4.1$).
On the other hand, we see that the ratio depends somewhat on the optical classification of the object with Seyferts having the highest and starbursts the lowest ratios.
Therefore, we determine a $\lwthree$ to $\ltw$ correction based on optical classification. 
Owing to the differing classifications in the literature, some of the objects are classified at the same time as Seyferts, LINERs, H\,II and/or starbursts (Sect.~\ref{sec:kAGN} and Sect.~\ref{sec:kSB}).
Therefore, we test different groupings and find a distinction in the following three subgroups leading to the best corrections: a) pure Seyferts (no other classification), b) Seyferts also classified as LINERs\footnote{
These probably correspond to objects situated in the Seyfert and LINER overlapping region in the BPT diagrams, i.e., Seyfert-LINER transition objects.
} (but not as H\,II or starburst classification), and c) non-Seyferts (no classification as Seyfert). 
Corresponding ordinary least-square linear regression in logarithmic space with treating $\lwthree$ as the independent variable leads to the following corrections:
\begin{equation*}
	\log \frac{\ltw}{\lwthree} =
	\begin{cases}
 		 0.11(\log \lwthree - 42) - 0.34 & \text{pure Seyfert}\\
 	 	 0.17(\log \lwthree - 42) - 0.63 & \text{Sy-LINER}\\ 
 	 	 0.23(\log \lwthree - 42) - 1.06 & \text{non-Seyfert}\\
 	 	 0.18(\log \lwthree - 42) - 0.57 & \text{no classif.}\\
	\end{cases}
\end{equation*}
The last case provides the general correction if no optical classification is available.
The $2/3$-of-the-population scatter around these best fit lines is 0.22\,dex, 0.57\,dex, 0.23\,dex and 0.43\,dex, respectively.
As said, this scatter is considerable and the above corrections should not be used for individual objects but only in a statistical sense.

Applying the above corrections to estimate $\ltw$ for all known AGN, we obtain the following distribution shown in the bottom panel of Fig.~\ref{fig:LW3hist}.
The estimated $\ltw$ distribution for all known AGN is on average 0.7\,dex lower than the one of $\lwthree$ with a median $\ltw$ of $10^{41.47}\ergs$. 
Only $18\%$ (781) of the known AGN have $\ltw > 10^{42.3}\ergs$, i.e., are at least moderately luminous.
For Seyferts, this number increases to $30\%$ (716 of 2385), while it is $68\%$ for the B70 AGN (130 of 190).
Again, these numbers should just provide a rough guidance for the luminosity ranges to be expected for the AGN in the volume.
More accurate numbers will become available in the future based on SED decomposition and MIR follow-up observations.


\subsection{MIR colours of AGN, galaxies and starbursts}\label{sec:galcol}
\begin{figure*}
   \includegraphics[angle=0,width=\textwidth]{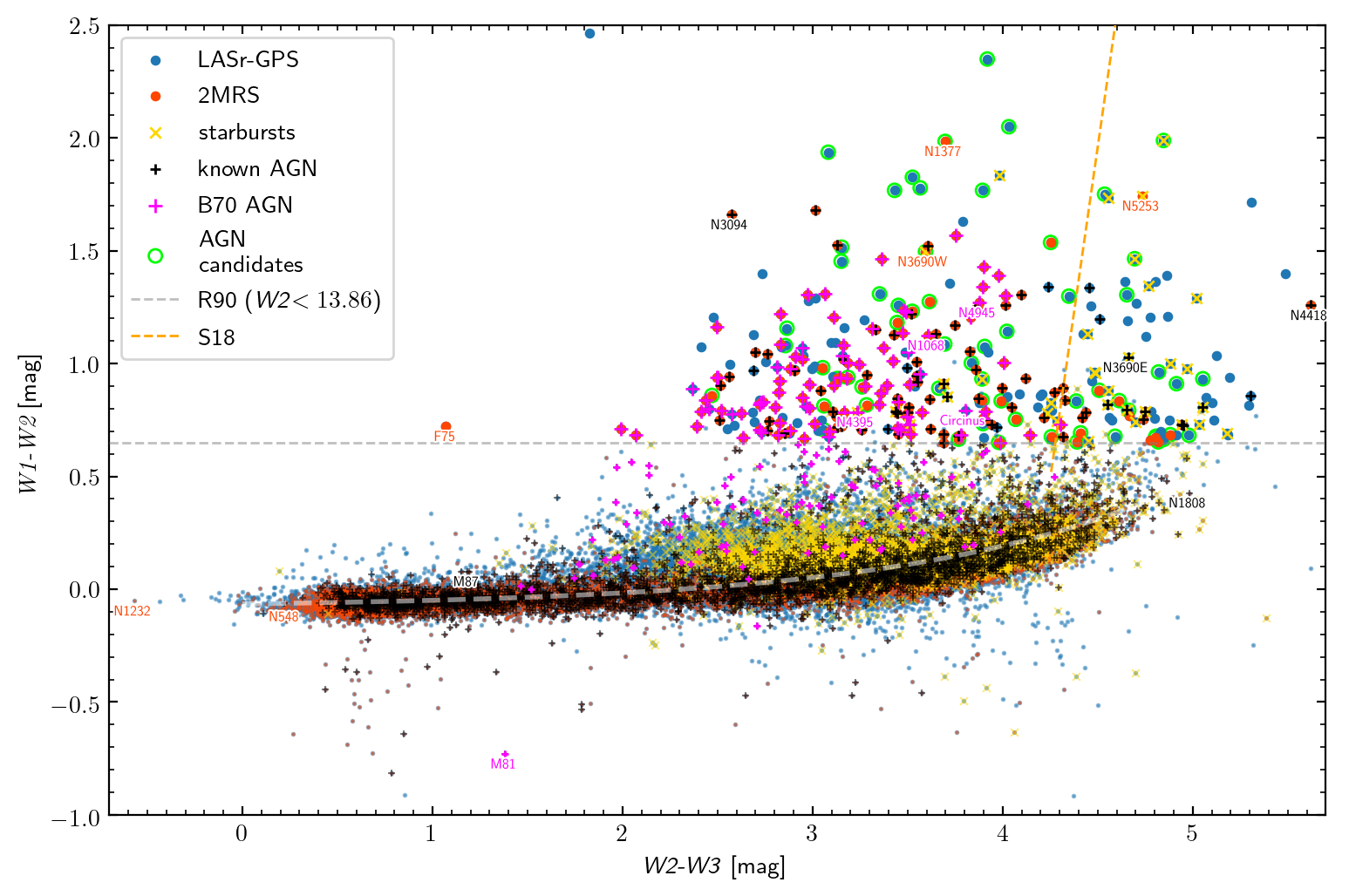}
    \caption{
    \wonetwoo versus \wtwothreee colour--colour distribution for all galaxies from the 2MRS (orange) and LASr-GPS (blue) detected in \wone, \wtwoo and \wthree.
    Yellow `X's mark starburst galaxies, while black crosses mark known AGN and magenta crosses mark B70 AGN.
    The R90 AGN colour selection criterion is shown as dashed, grey line (for $\wtwo < 13.86$\,mag), and galaxies that fulfill R90 are marked with large symbols.
    In addition, the R90 AGN candidates with $\lwthree > 10^{42.3}\ergs$ have green circles.
    The theoretical AGN/extreme-starburst discriminator line from Satyapal, Abel \& Secrest (2018) is shown as dashed orange line (AGN left, starbursts right).
    The star formation main sequence line from Jarret et al. (2019) is shown as white dashed line.
    Some notable galaxies are labelled with short names (``M'' stands for Messier, ``N'' for NGC, and ``F'' for Fairall. 
}
   \label{fig:col1}
\end{figure*}

\begin{figure*}
   \centering
   \includegraphics[angle=0,width=0.33\textwidth]{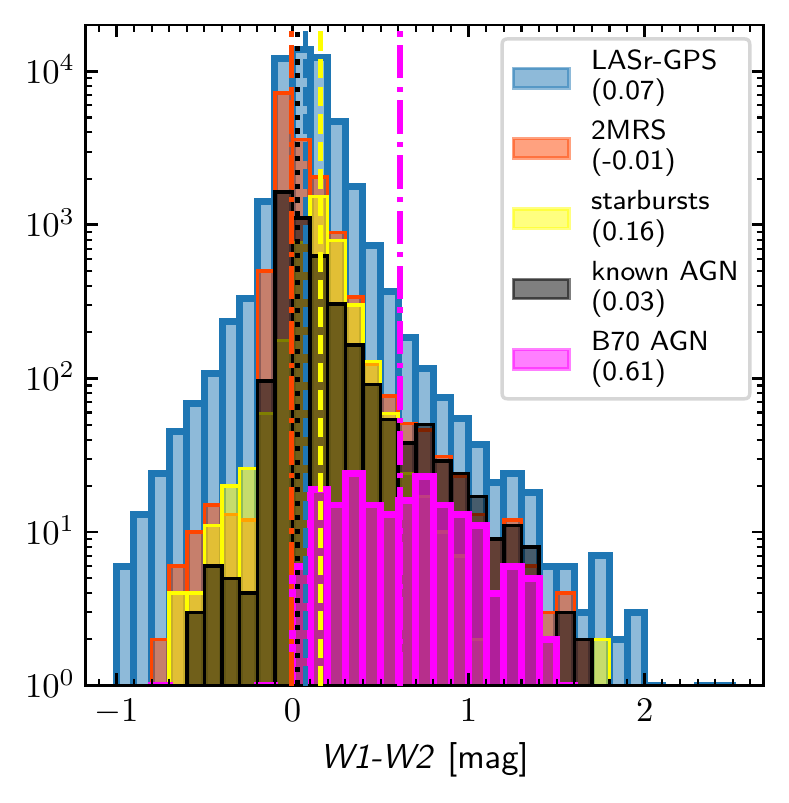}
    \includegraphics[angle=0,width=0.33\textwidth]{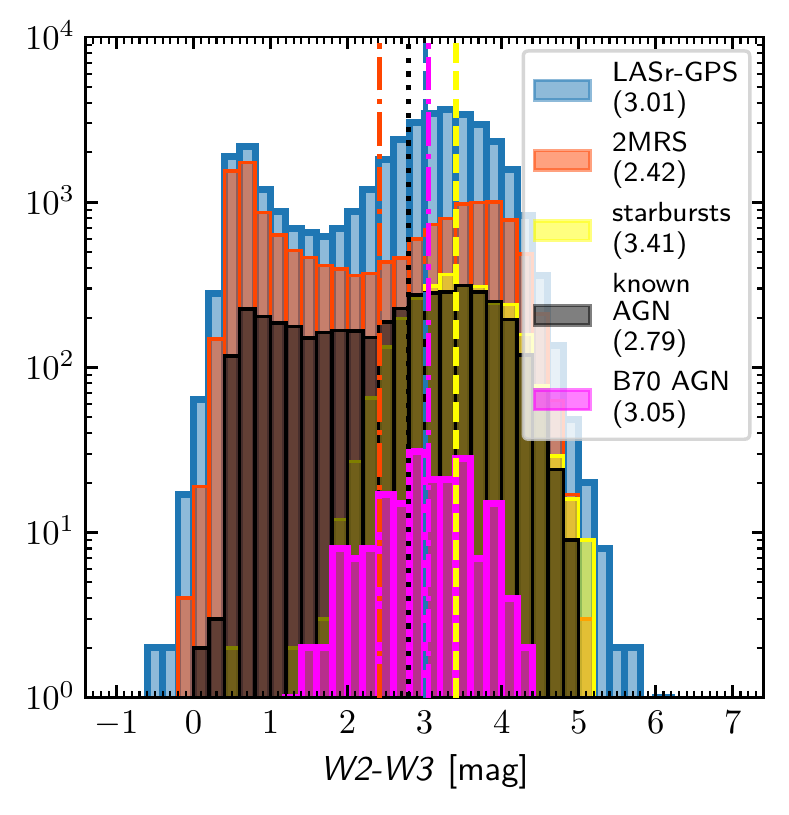}
    \includegraphics[angle=0,width=0.33\textwidth]{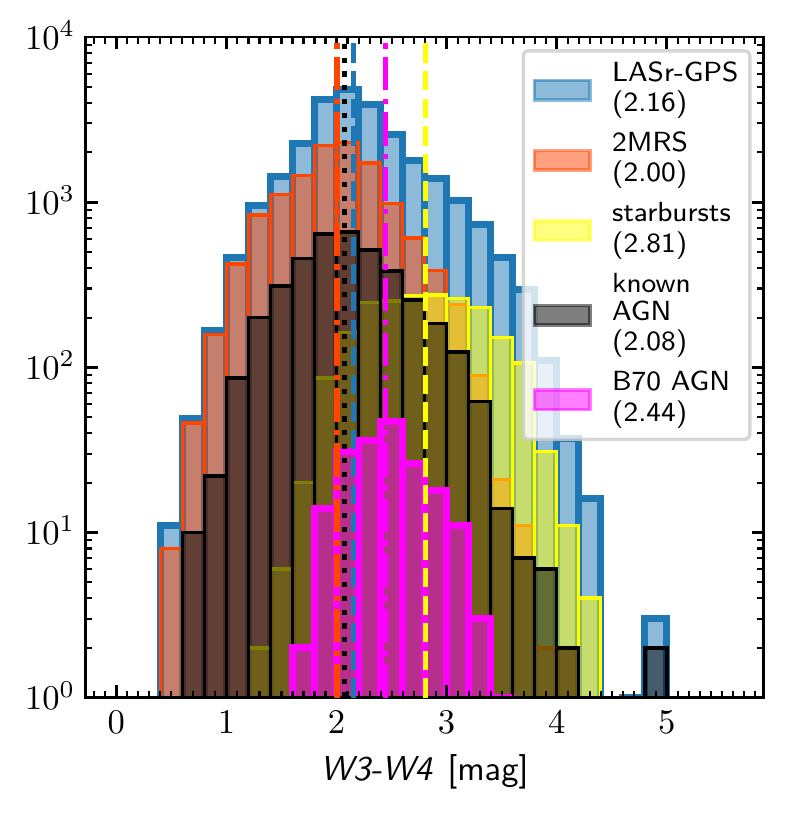}    
    \caption{
    \wisee colours \wonetwoo (left), \wtwothreee (middle) and \wthreefourr (right) distributions of all galaxies  from the 2MRS (orange) and LASr-GPS (blue) that have detections in the corresponding bands of each colour.
    The distribution of known starbursts is shown in yellow, while known AGN are shown in black and B70 AGN are shown in magenta.
    In addition, the median colours for each subsample are shown as vertical dashed, dotted and dash-dotted lines of the corresponding colour, and listed in the figure legends.
}
   \label{fig:colh}
\end{figure*}

Let us now examine the MIR colour distribution of the known AGN in the context of normal and starburst galaxies.
The MIR colour distribution in the \wonetwoo over \wtwothreee plane is shown in Fig.~\ref{fig:col1} for all galaxies that are detected in the three \wisee bands ($76\%$ of the LASr-GPS and $99.9\%$ of the 2MRS samples), while the distributions of the individual colours are shown in Fig.~\ref{fig:colh}.
The large majority of galaxies, and in particular the 2MRS galaxies, form a relatively narrow star formation main sequence from blue to red \wtwothreee colours at almost constant \wonetwoo colour (as already previously found in the literature, e.g., \citealt{jarrett_wise_2019}).
This sequence is caused by star formation which leads to an increasing amount of warm dust emission and, thus, redder \wtwothreee colours with increasing star formation intensity relative to the direct stellar emission of the galaxy.
For example, the bluest \wtwothreee objects are mostly passive, early-type galaxies like, e.g., NGC\,548  at $\wtwothreem=0.23$\,mag and $\wonetwom=-0.08$\,mag.
On the red side, the sequence is bending up to redder \wonetwoo colours of $\wonetwom\sim0.4$\,mag at its approximate reddest  end of $\wtwothreem\sim 5$\,mag.  
One of the most extreme objects here is the starbursting NGC\,1808 ($\wtwothreem=0.43$\,mag; $\wonetwom=4.98$\,mag).
In addition, the galaxy distribution of the LASr-GPS extends to redder \wonetwoo colours ($\wonetwom \sim 0.3$\,mag at intermediate \wtwothreee colours ($2 \lesssim \wtwothreem \lesssim 4$) filling up roughly the expected locus area of the spiral galaxies in Fig.~12 of \cite{wright_wide-field_2010}.
Most of the galaxies are dwarfs according to their \wonee luminosity and optical appearance.
The reason for the redder \wonetwoo colours is again star formation which can dominate \wtwoo if strong enough with respect to the stellar light of the host.
This  \wonetwoo reddening effect of star formation is the main source of contamination in AGN selections that are based on this colour, and will have to be taken into account (further discussed in Sect.~\ref{sec:SF}).

Most of the galaxies known to host an AGN follow the \wisee colour distributions of the 2MRS, i.e., rather massive galaxies.
Only in \wtwothreee colour, they are slightly redder on median (2.66\,mag vs. 2.42\,mag), i.e., they either prefer star-forming hosts, or contribute themselves the most to this colour.
%
Galaxies with hosting a luminous AGN, comparable MIR brightness at least in \wtwo, have a redder \wonetwoo colour. 
They leave the main sequence and move upward in the colour--colour plane of Fig.~\ref{fig:col1} with increasing AGN luminosity.
This trend motivates the colour selection based on \wonetwoo as discussed in the following.

\subsection{Identification of AGN by MIR colour}\label{sec:R90}
\begin{figure}
   \includegraphics[angle=0,width=0.91\columnwidth]{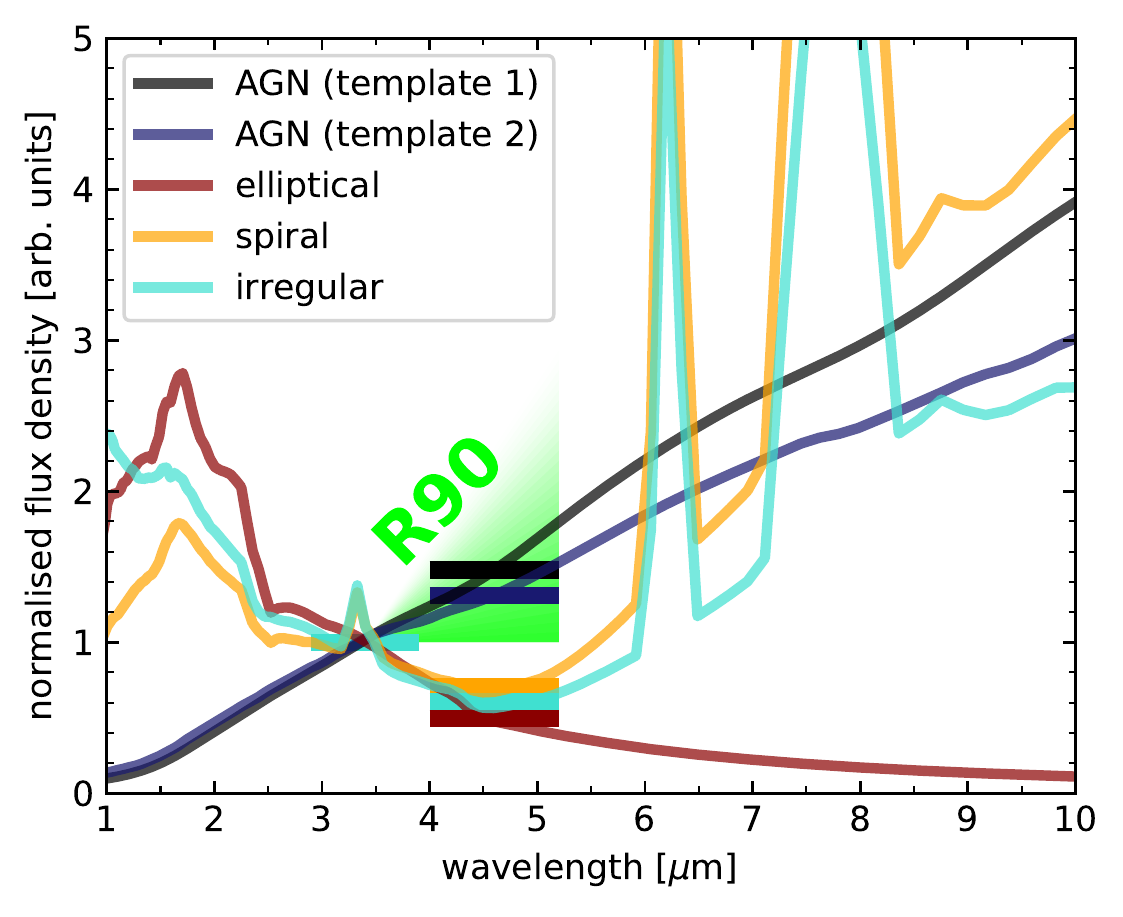}
    \caption{
      Illustration of the R90 \wisee colour selection criterion based in \wonetwo.
      The solid lines show the different SED templates for AGN and galaxy types taken from Assef et al. (2010), see that work for details.
      The thick horizontal bars indicate the synthetic photometry in \wonee and \wtwo, for each SED in the same color, respectively.
      The SEDs are normalised to the \wonee synthetic flux density.
      The green semitransparent triangle indicates the SED slopes that would be selected by the R90 criterion as AGN. 
      $\wonetwo = 0.65$\,mag corresponds to straight line in flux density space.
}
   \label{fig:R90illu}
\end{figure}

We now proceed to the MIR colour-based identification of AGN and quantify how its efficiency depends on the AGN luminosity.
Since the advent of the \wisee mission, many MIR colour selection methods have been put forward to find AGN (e.g., \citealt{stern_mid-infrared_2012,mateos_using_2012, assef_mid-infrared_2013}).
At the core, they are similar, building on the fact that the AGN-heated, warm dust emits significantly redder \wonetwoo colours than the light of the old stellar population in the host galaxy.
In addition, the \wonetwoo is little affected by extinction, in particular at low redshifts (e.g., \citealt{stern_mid-infrared_2012}), making \wonetwoo based AGN selection a formidable tool to select highly obscured and even CT AGN.
Here, we will use the most recent and refined selection criterion introduced in \cite{assef_wise_2018}, namely one that was designed to have $90\%$ reliability in selecting AGN (hereafter R90):
\begin{equation*}
	\wonetwom >
	\begin{cases}
 		 0.65 & \text{if}\,\,W2 < 13.86\,\mathrm{mag}, \\
 	 	 0.65 \exp[0.153(W2 - 13.86)^2] & \text{otherwise}, 
	\end{cases}
\end{equation*}
%
The R90 criterion is illustrated in Fig.~\ref{fig:R90illu} for typical galaxy and AGN SEDs from \cite{assef_low-resolution_2010}.
This criterion works best for \wtwoo detections with a signal-to-noise greater than 5 (otherwise biases can occur; see \citealt{assef_wise_2018} for details\footnote{
Normally, it is recommended to also remove objects for which the contamination and confusion flag in the AllWISE catalogue is set  (cc\_flags). 
However, here, to be inclusive, we keep such galaxies and examine them individually where necessary.
}).
All the 2MRS galaxies and $93\%$ of the LASr-GPS are above this limit ($99\%$ for $\lwthree> 10^{41}\ergs$).


The R90 criterion for $\wtwo < 13.86\,$mag is shown as grey dashed line in Fig.~\ref{fig:col1}.
Out of the 4.3k known AGN in the volume, 172 fulfil the R90 criterion as visualized in that figure with larger symbols\footnote{
Interestingly, only $83\%$ of the corresponding galaxies are in the 2MRS, once more confirming the incompleteness of 2MRS in terms of AGN.
}.
For $97\%$ (167) of the 172 R90 AGN, optical type classifications are available, $97\%$ (162) of which have a classification as Seyfert, while $9\%$ (15) are classified as LINERs and $18\%$ (29) as H\,II, i.e., $21\%$ (34) have multiple classifications in the literature.
The type~2 to type~1 ratio for the R90 AGN is 0.52 (similar as for the whole population of known AGN; Sect.~\ref{sec:kAGN}) with a significant population with intermediate (Sy~1.8 or Sy~1.9) or both type~1 and 2 classifications ($36\%$).
Depending on their treatment, the type~2 fraction among the known AGN\footnote{
The intrinsic type~2 fraction is probably higher because we expect the majority of AGN candidates to be obscured, i.e.,  type~2 (Sect.~\ref{sec:candi}).
} is between 38 and $62\%$.
The R90 AGN including their optical classifications are listed in Table~\ref{tab_R90AGN}.

\begin{figure*}
   \includegraphics[angle=0,width=\textwidth]{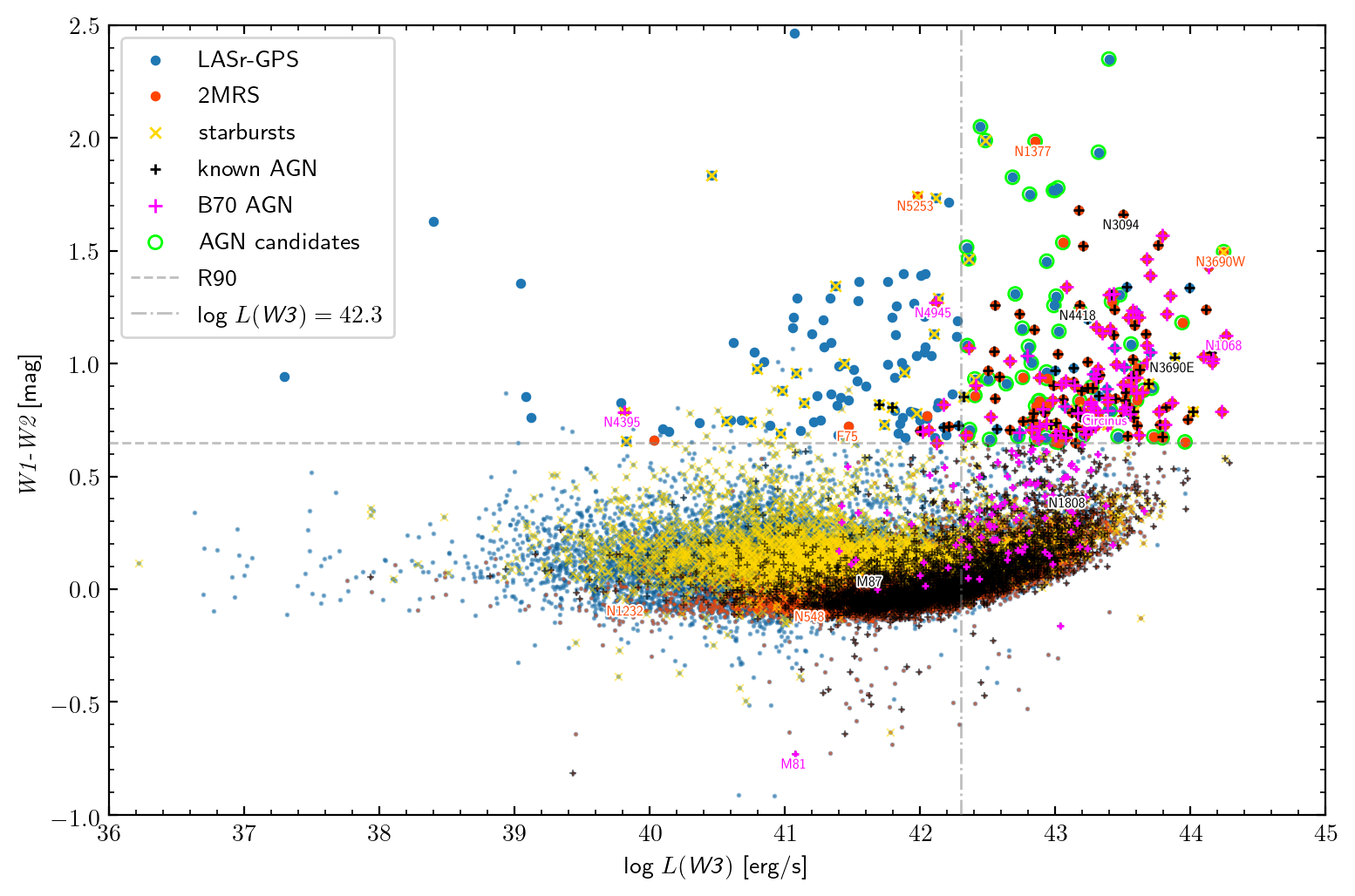}
    \caption{
     \wonetwoo  colour over \wthreee luminosity for all galaxies from the LASr-GPS detected in \wone, \wtwoo and \wthree.
     Description of the symbols is as in Fig.~\ref{fig:col1}.
     In addition, the vertical dot-dashed line marks $\lwthree = 10^{42.3}\ergs$.
}
   \label{fig:collum}
\end{figure*}

\begin{figure}
   \includegraphics[angle=0,width=0.95\columnwidth]{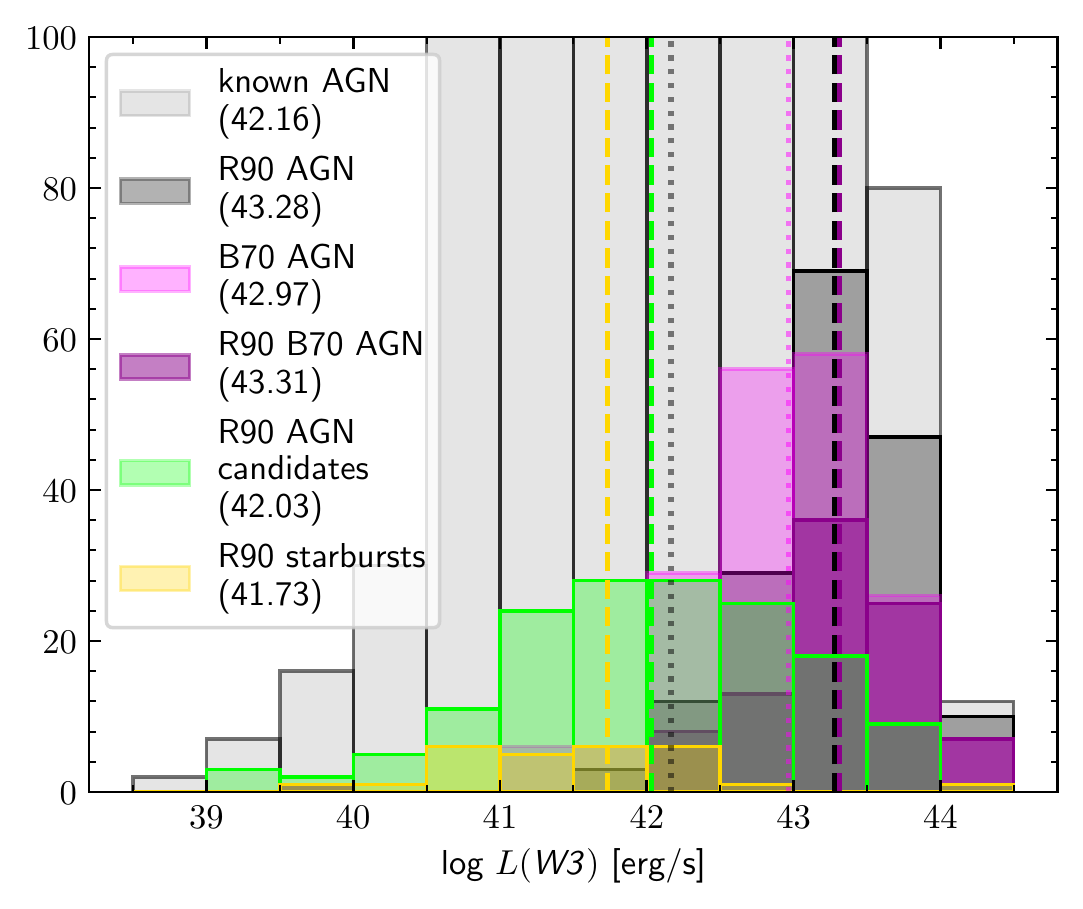}
      \includegraphics[angle=0,width=0.95\columnwidth]{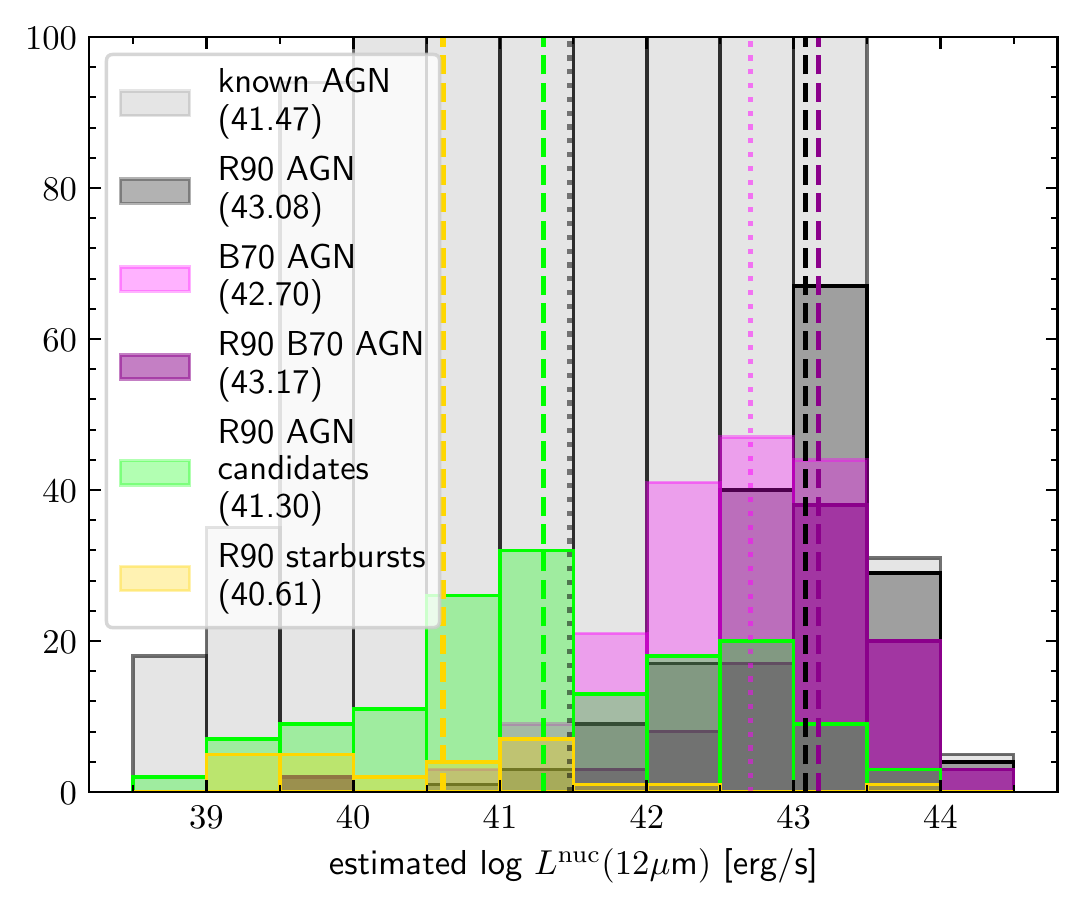}
    \caption{
      Top: \wthreee luminosity distribution for R90 selected AGN and candidates in comparison to all known AGN (grey) and B70 AGN (light magenta).
      R90 selected objects from all known AGN are shown in black, from B70 in dark magenta, from unidentified AGN in green and from known starbursts in gold.
      The dashed lines of the corresponding colour mark the median value which is also shown in the legend. 
      \newline Bottom: Corresponding estimated $\ltw$ distribution after decontamination of $\lwthree$ following Sect.~\ref{sec:known_AGN_MIR}.
}
   \label{fig:LW3hist2}
\end{figure}

As said, we expect the R90 criterion to preferentially select luminous AGN.
This effect is clearly visibile in Fig.~\ref{fig:collum}, where galaxies hosting known AGN, and in particular those from the B70 AGN sample, exhibit a trend of redder \wonetwoo colour for increasing \wthreee luminosity.
The $\lwthree$ and estimated $\ltw$ distributions of R90 selected objects are shown in Fig.~\ref{fig:LW3hist2}.
While the median $\lwthree$ of known AGN is $10^{42.2}\ergs$, the corresponding median luminosity for R90 selected AGN is more than 1\,dex higher, i.e., $10^{43.3}\ergs$.
For B70 AGN the trend is similar albeit smaller, i.e., 0.3\,dex.
If instead of the observed $\lwthree$, we use the estimated AGN luminosity, $\ltw$, the trend becomes even clearer, and the gap in luminosities larger (1.6\,dex for all known AGN and 0.5\,dex for the B70 AGN.

\begin{figure}
   \includegraphics[angle=0,width=\columnwidth]{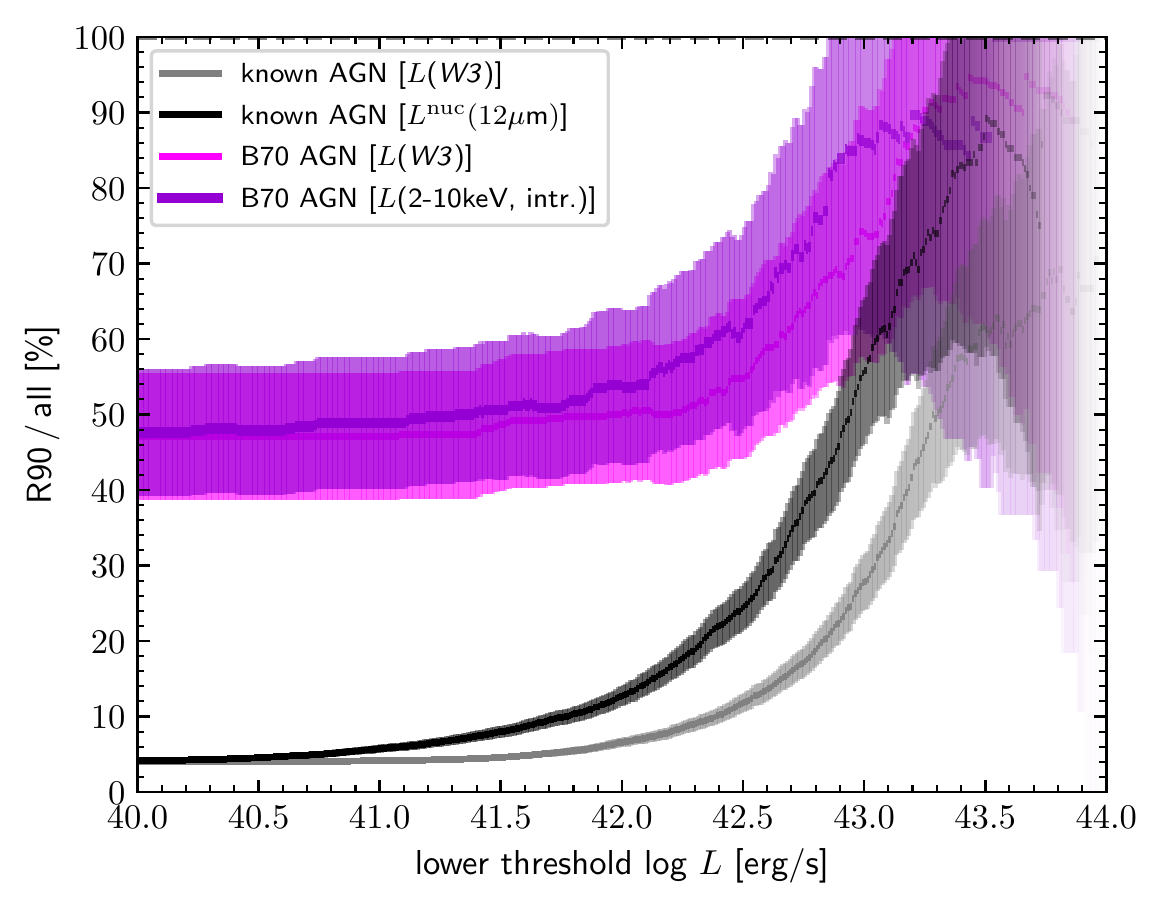}
    \caption{
Fraction of AGN selected by the R90 criterion depending on luminosity.
The grey line marks the fraction for all known AGN as a function of \wthreee luminosity, while the black marks the fraction as a function of the estimated $\ltw$.
The thin magenta line shows the fraction for the B70 AGN as a function of  \wthreee luminosity, while the thick, dark violet line  shows the same fraction but as a function of intrinsic 2-10\,keV X-ray luminosity. 
The shaded regions indicate the 1-$\sigma$ uncertainty on the number counts.
}
   \label{fig:r90}
\end{figure}

To quantify the fraction of AGN selected by the R90 criterion we look at its luminosity dependence in Fig.~\ref{fig:r90} for different subsamples of known AGN.
Independent of the AGN subsample and selected luminosity as AGN power tracer, for $L \lesssim 10^{42.5}\ergs$ the fraction of AGN selected by the R90 criterion is relatively low and constant, while for higher luminosities is rapidly increases.
For using $\wthree$ as AGN power tracer, the fraction of AGN selected by R90 levels off at a relatively low $60 - 70\%$ for $\lwthree > 10^{43.5}\ergs$ (grey line in Fig.~\ref{fig:r90}).
Most of the remaining $30 -40\%$ of AGN  not selected by R90 despite high \wthreee luminosity are situated in heavily star forming galaxies that dominate the MIR over the AGN. 
These are classified as HII in the optical indicating that the corresponding AGN are intrinsically much less luminous than $\lwthree$ values suggest.
Indeed, if we use the decontaminate $\ltw$ estimates from Sect.~\ref{sec:known_AGN_MIR}, the R90-selected fraction increases more rapidly, reaching $67\%$ at $\ltw > 10^{43.1}\ergs$ and peaking at $92\%$  (black line in Fig.~\ref{fig:r90}).
The completeness of the R90 selection even further increases if one only looks at the X-ray luminous B70 AGN (magenta line in Fig.~\ref{fig:r90}).
For this particular sample, we have the advantage of a better tracer of the AGN power than the \wthreee luminosity, namely the intrinsic 2-10\,keV X-ray luminosity (taken from \citealt{ricci_bat_2017}).
This allows us to assess the ``true'' efficiency of the R90 criterion (dark violet line in Fig.~\ref{fig:r90}).
Namely, R90 selects $54 \pm 9 \%$ of the AGN with $\lxi > 10^{42}\ergs$, while for $\lxi > 10^{43}\ergs$, $86 \pm 26\%$ are selected\footnote{
The relatively large uncertainties results from the small number statistics of the B70 within the volume at such high luminosities.
}.
Using our estimated $\ltw$ gives similar results to $\lxi$ which confirms the validity of our decontamination of the former in Sect.~\ref{sec:known_AGN_MIR}.
In addition, the comparison between the R90 fractions depending on $\lwthree$ and $\lxi$ for the B70 sample verifies that the $\lwthree$-based fractions are to be regarded as lower limit on the true efficiency of the R90 selection.

\subsection{New AGN candidates}\label{sec:candi}
Not all the galaxies selected by the R90 criterion are already known to host AGN.
There are 159 such galaxies, and thus new AGN candidates based on their \wonetwoo colour.
We double-check all galaxies individually to make sure that they are genuine galaxies with valid \wisee measurements and robust redshifts (as far as we can assess from the information at hand).
The resulting list of new AGN candidates and their properties can be found in Table~\ref{tab_R90candi}.
Only 31 ($19\%$) of the hosts of the new AGN candidates are in the 2MRS sample, indicating that they are relatively faint or compact galaxies.
Indeed, the median $\lwthree$ of the candidate systems is only $10^{42}\ergs$, so much lower than the median of the verified AGN systems that fulfil the R90 criterion ($10^{43.3}\ergs$; see Fig.~\ref{fig:LW3hist2}, top).
If we apply our \wthreee decontamination (Sect.~\ref{sec:known_AGN_MIR}), the resulting $\ltw$ distribution fractures into two peaks, one peaking at $\ltw\sim10^{41.3}\ergs$ and the other at $\ltw\sim10^{42.6}\ergs$ ( Fig.~\ref{fig:LW3hist2}, bottom).
This is caused by 51 of the AGN candidates having H\,II or starburst classifications and thus higher corrections to their $\lwthree$. 
It indicates that a significant fraction of objects with low $\lwthree$ luminosities might be contaminants, i.e., not AGN but star-formation dominated systems. 

\subsubsection{On contamination by starbursts}\label{sec:SF}
The R90 criterion was designed for selecting distant, luminous and point-like AGN. 
Its $90\%$ reliability in selecting AGN might not hold for local, extended galaxies.
We saw in Fig.~\ref{fig:col1} that the large majority of star-forming galaxies lie on the red \wtwothreee tail of the main sequence but significantly below typical AGN \wonetwoo colours .
However, it was argued by \cite{hainline_mid-infrared_2016} that strong star formation, in particular in dwarf galaxies, can also lead to red, AGN-like \wonetwoo colours.
These systems would then have as well very red \wtwothreee colours ($\gtrsim 4$\,mag) which would motivate to add a \wtwothreee colour cut to improve the reliability of a \wonetwoo-based AGN selection.
\cite{satyapal_star-forming_2018} further investigated this with theoretical colour tracks of extreme starburst systems and determined a theoretical  \wtwothreee colour criterion (hereafter S18):
\begin{equation*}
\wtwothreem< 0.17\,(\wonetwom + 24.5),
\end{equation*} 
to separate AGN and starbursts.
This criterion is plotted in Fig.~\ref{fig:col1} as orange dot-dashed line and marked for individual  known R90 AGN and R90 AGN candidates in Table~\ref{tab_R90AGN} and  Table~\ref{tab_R90candi}, respectively.
Of the 159 R90 AGN candidates,  100 ($63\%$) fulfil the S18 criterion and, thus, are expected to not be starburst dominated. 
Among the R90 AGN candidates not fulfilling S18, there are indeed some of those compact star-forming galaxies that \cite{hainline_mid-infrared_2016} identified as ``AGN imposters'' (e.g., II\,Zw\,40, Mrk\,193, SBS\,0335-052, and UGC\,5189).
In total, 28 ($18\%$)  of the AGN candidates are classified as starbursts or blue compact dwarfs in our literature collection.
However, only 14 ($50\%$) of them would be excluded by the S18 criterion. 

Among the known AGN, $89\%$ (154 of 172) fulfil the S18 criterion\footnote{
One object, Mrk\,3 aka UGC\,3426 has no valid $\wthreee$ measurement and thus S18 can not be computed.
}.
For the B70 AGN, fulfilment is even $98\%$ (89 of 91).
Of the 17 R90 AGN not fulfilling S18, nine show signs of strong star formation in the literature and are in fact controversial concerning the existence of luminous AGN in these galaxies\footnote{
These are Arp\,220, CGCG\,032-017, Mrk\,93, NGC\,253, NGC\,3256, NGC\,3690E, NGC\,7130, NGC\,7552, and TOLOLO\,1220+051.
}.
On the other hand, two of the remaining galaxies, NGC\,4418 (aka NGC\,4355) and 2MASX\,J04282604-0433496, show no signs of strong star formation, judging from their \spitzer/IRS spectra \citep{asmus_subarcsecond_2014}.
Instead, NGC\,4418 hosts a highly obscured nucleus with the obscuration probably causing  the red \wtwothreee colour (see e.g., \citealt{roche_silicate_2015}). 
In fact, both objects are among the reddest in terms of \wtwothreee colour ($>5$) of all galaxies in the LASr-GPS.
While, the nature of the dominating MIR emitter in NGC\,4418 is still somewhat controversial (e.g., \citealt{sakamoto_submillimeter_2013, varenius_radio_2014}), the case of object makes clear that also heavy obscuration can lead to very red \wtwothreee colours\footnote{
See also the similarly mysterious Arp\,220; e.g., \citealt{martin_unbearable_2016, paggi_x-ray_2017, sakamoto_resolved_2017, yoast-hull_-ray_2017}).
}.
Thus, the application of the S18 criterion might exclude the most obscured AGN, which are the ones we are hunting for!

In addition, for a complete, unbiased  sample of AGN, one wants to include even star-formation dominated galaxies, as long as the intrinsic luminosity of the AGN is above the selected  lower threshold\footnote{ 
Finding such objects is difficult with \wisee colour selection alone but might require high angular resolution data over a wide wavelength range, something we plan for the future with dedicated follow-up of these red objects.
}.
We conclude from this discussion that applying a \wtwothree-based criterion like the S18 in addition to the R90 criterion indeed increases pureness of AGN selection.
However, a significant fraction of starbursts still remains while many AGN that are either heavily obscured or live in hosts with dominating star formation are excluded.

Instead, we notice that in the \wthreee luminosity distribution in Fig.~\ref{fig:LW3hist2} that most of the starbursts have relatively low luminosities.
For example, $90\%$ (25 of 28) of the starbursts and BCDs selected by R90 have $\lwthree < 10^{42.3}\ergs$.
This suggests that a lower luminosity cut could be more successful at removing contaminating non-AGN galaxies with dominating starbursts.
Using $\lwthree > 10^{42.3}\ergs$ as threshold, leaves 61 of the R90 AGN candidates which according to the above number should be genuine AGN with $90\%$ probability. 
They are marked with green circles in Fig.~\ref{fig:col1} and Fig.~\ref{fig:collum}.

\subsubsection{Prospects for detection in X-rays}\label{sec:eros}
There is a close correlation between the observed MIR and intrinsic X-ray luminosities for local AGN (e.g. \citealt{lutz_relation_2004, gandhi_resolving_2009}), allowing us to estimate the intrinsic X-ray AGN luminosities of our new AGN candidates and infer the chances to detect them with the X-ray all-sky missions, \swift/BAT, \srg/\artxc/\erosita.
In the following, we detail our Monte Carlo simulation per source to estimate the detection rates for the 61 R90 AGN candidates with $\lwthree > 10^{42.3}\ergs$ (corresponding to intrinsic X-ray luminosities above the nominal sensitivity of \erositaa after eight passes; $\fxo \gtrsim 1.6 \cdot 10^{-13}\fu$ \citep{merloni_erosita_2012}).
\begin{figure*}
   \includegraphics[angle=0,width=0.29\textwidth]{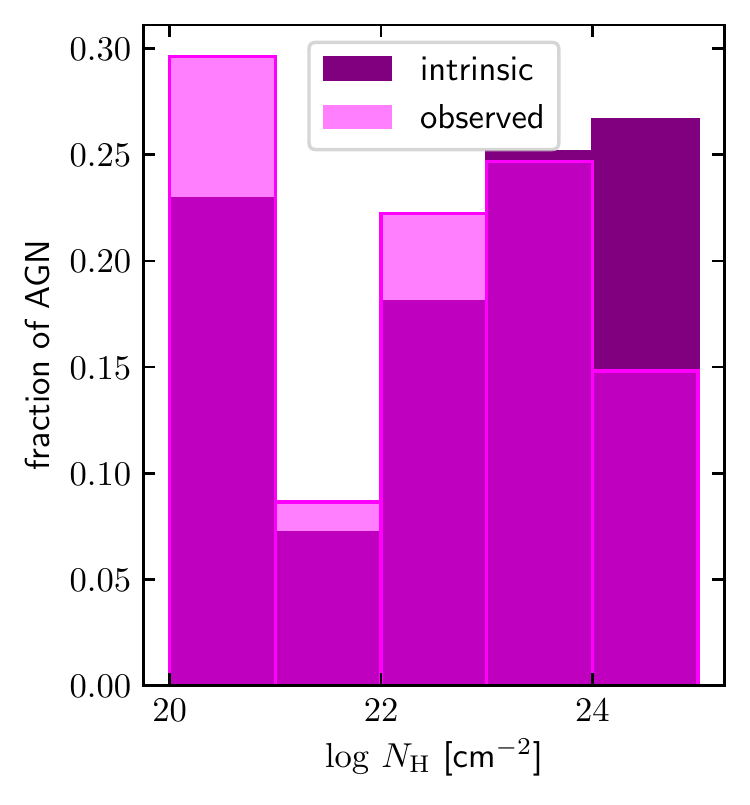}
   \includegraphics[angle=0,width=0.33\textwidth]{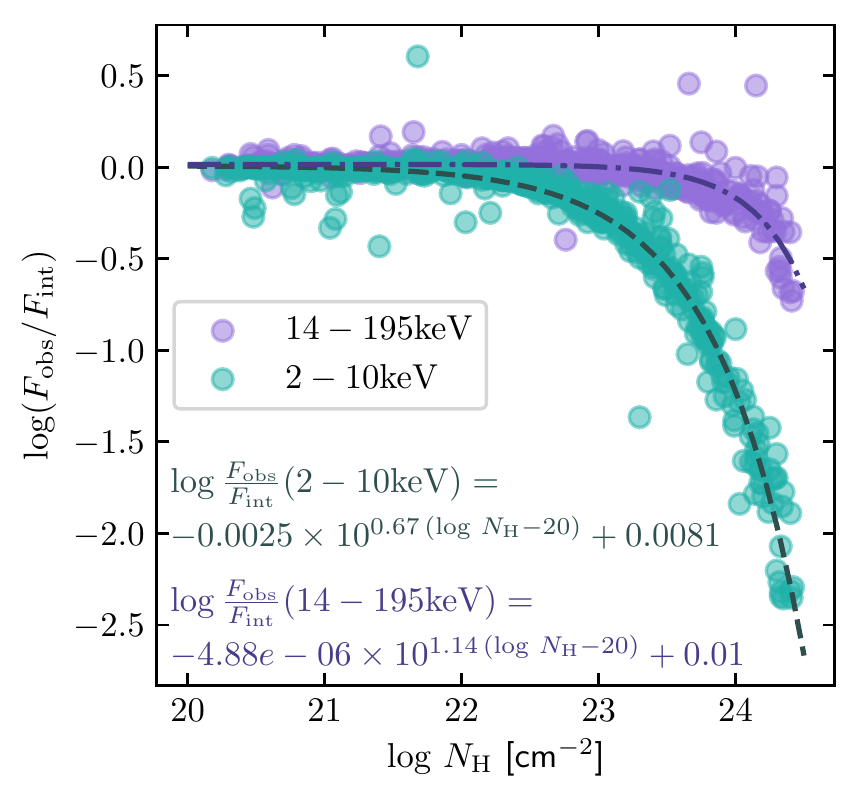}  
   \includegraphics[angle=0,width=0.35\textwidth]{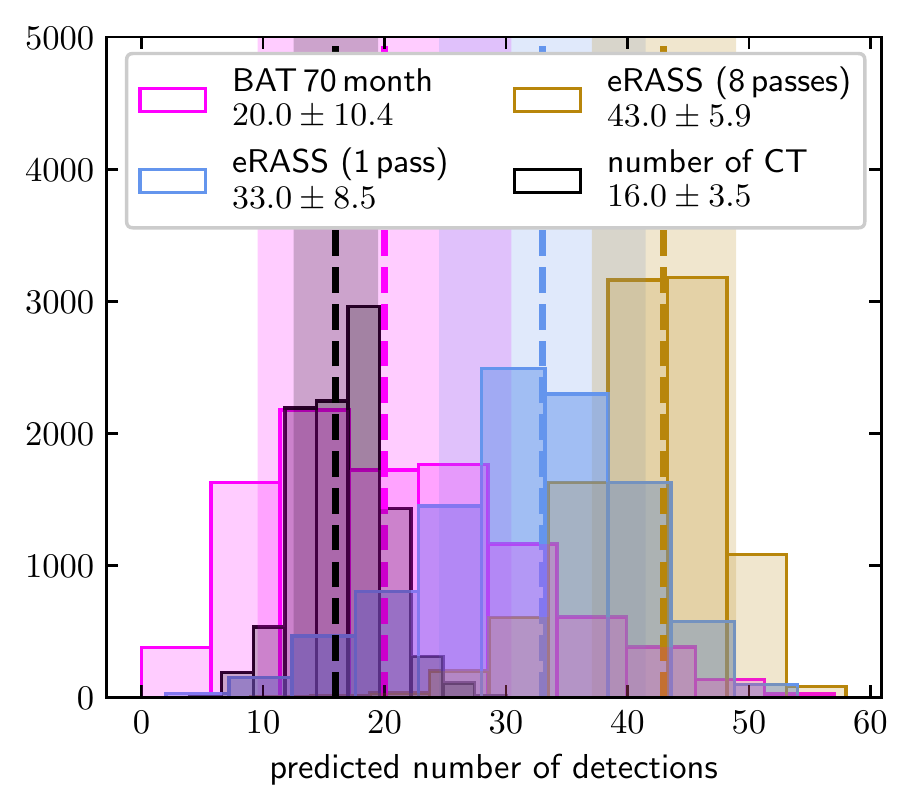}
    \caption{
{\it Left:} normalised distributions of obscuring column density, $\nh$, for the BAT 70\,month AGN sample.
In light magenta is shown the observed distribution for the B70 AGN in LASr fulfilling R90, while in dark purple is shown the inferred intrinsic distribution from Ricci et al. (2015).\newline
{\it Middle:} empirical X-ray extinction curves for the BAT 70\,month AGN sample with $10^{20} \le \log (\nh \cm{^2}) \le 24.5$ for observed and intrinsic fluxes as well as $\nh$ values taken from Ricci et al. (2017). 
In purple is the shown the observed to intrinsic flux ratio for the 14-195\,keV energy range, while in green the same is shown for the 2-10\,keV range.
The darker colored dot-dashed and dashed lines give exponential fits to the data, respectively, with the corresponding best parameters shown as well in the same color.\newline
{\it Right:} simulated detection rates for \swift/BAT after 70\,months (magenta) and \srg/\erositaa after one (blue) and eight all-sky passes (gold). 
The distributions show the results of the iterations of the Monte Carlo simulation using the intrinsic $\nh$ distribution, while the dashed lines give the median and the shaded areas the standard deviation. 
These values are given in the legend as well.
In addition, the expected number of CT AGN is shown in black.
}
   \label{fig:sim}
\end{figure*}
In particular, these steps are performed:

\begin{enumerate}

\item \textbf{$\ltw$ prediction:} we use the $\lwthree$ decontamination method from Sect.~\ref{sec:known_AGN_MIR} to estimate $\ltw$.

\item \textbf{Intrinsic $L_{\rm X}$ prediction:} we use the most accurate determination of the MIR--X-ray luminosity relation by \cite{asmus_subarcsecond_2015} to convert $\ltw$ into $\lxi$: 
\begin{equation*}
 \log \left( \frac{\lxi}{10^{43}\textrm{erg}\,\mathrm{s}^{-1}}\right) =  -0.32  + 0.95 \log \left( \frac{\lnn}{10^{43}\textrm{erg}\,\mathrm{s}^{-1}}\right)
\end{equation*}
with an observed scatter of 0.4\,dex.

\item \textbf{$N_{\rm H}$ assignment:} to estimate the observed X-ray fluxes from $\lxi$, we have to assign an obscuring column density, $\nh$. 
Here, we use the bias-corrected intrinsic $\nh$ distribution from the BAT 70\,month AGN \citep{ricci_compton-thick_2015} as reference probability function to draw a random $\nh$ (shown in Fig.~\ref{fig:sim}, left). 

\item \textbf{Application of extinction:} In Fig.~\ref{fig:sim}, middle, we show the B70 AGN observed to intrinsic X-ray flux ratios vs. $\nh$ from \citet{ricci_bat_2017}. This was fit with an exponential function, which was found to give a good description of the data yielding a theoretical extinction curve.
\end{enumerate}
We then performed a Monte Carlo resampling of the above steps. 
We assumed the probability distributions of each $\ltw$ and $\lxi$ value to be Gaussian-distributed with width equal to the observed scatter in both conversions (much larger than the individual source X-ray fit uncertainties).
For 10$^4$ iterations, the resulting observed X-ray flux distributions are stable (Fig.~\ref{fig:sim}, right) and can be compared to the flux limits provided for the all-sky surveys of \swift/BAT and \srg/\erosita\footnote{
We omit \artxc\ here because its different energy band would require futher conversion with additional uncertainties but given the flux limit of its all-sky survey \citep{pavlinsky_art-xc_2018}, we expect detection rates to be a factor two to three lower than with \erosita.
}. 

According to this simulation, we would expect to detect $33\pm9$ of the 61 R90 AGN candidates already in the first pass of the \erositaa all-sky survey, and $43\pm6$ in the full survey. 
The remaining objects would then expected to be highly obscured, with $16\pm3.5$ objects expected to be CT obscured.
However, if we convert the intrinsic 2-10\,keV fluxes into 14-195\,keV fluxes using the median ratio $0.42 \pm 0.25$\,dex as determined from the BAT 70\,month AGN, then we would expect that $20\pm10$ of the candidates\footnote{
The large uncertainty on this expected number of detections is caused by the scatter of the flux ratio in the X-ray bands.
} 
would have been detected already in the 70\,month \swift/BAT all-sky map with the nominal detection limit is $\fxh = 1.34 \cdot 10^{-11}\fu$ \citep{baumgartner_70_2013}.
This might indicate that a larger fraction of the R90 AGN candidates are highly obscured than assumed.
On the other hand, the fact that none are detected in the 70\,month BAT map is in fact consistent with the design-based expectation that only $90\%$ of the 221 galaxies that fulfill the R90 criterion indeed host an AGN.
In other words, we have to expect that $\sim 22$ of the 221 R90 objects are contaminants, and all of them would be among the R90 AGN candidates.

Alternatively, one could argue that possibly many of the CT obscured AGN that are missing according to the difference of the intrinsic to observed $\nh$ distribution (Fig.~\ref{fig:sim}) are among the R90 AGN candidates.
If we assume that the R90 selection is independent of X-ray obscuration, we expect 54 CT objects according to the intrinsic $\nh$ distribution from \cite{ricci_compton-thick_2015}, while only 18 AGN are currently known to be CT obscured, as we further discuss in Sect.~\ref{sec:CT}. 
Therefore, easily twice as many CT AGN might be present among the candidates as assumed in the above simulation which would then lower the expected detection rates correspondingly, and, in particular, remove any expected detections in the BAT 70\,month map.

\subsection{On the CT AGN fraction and CT candidates}\label{sec:CT}
As discussed in Sect.~\ref{sec:intro}, one of the main caveats of current AGN samples is the bias against the most obscured, i.e. CT, objects\footnote{
Note that CT AGN are likely not a special class of AGN but just the high end of a continuous obscuration distribution in the AGN population which is hard to detect because obscuration becomes opaque even at the highest photon energies. 
}. 
The real fraction of the CT AGN is still highly uncertain with estimates ranging from $10\%$ to $50\%$ of all AGN (e.g., \citealt{burlon_three-year_2011, ricci_compton-thick_2015, akylas_compton-thick_2016, lansbury_nustar_2017, georgantopoulos_nustar_2019,gandhi_x-ray_2003,gilli_synthesis_2007,ueda_toward_2014,ananna_accretion_2019}; Boorman et al., in prep.). 
The effort of building a complete AGN sample, starting with this work, will hopefully help to narrow down the uncertainty on this fraction.
In the meantime, we can derive lower limits on the CT fraction by simply adding up the number of known CT AGN in the volume.
The first lower limit comes from the B70 AGN sample.
It has 20 out of 153 AGN with $\lxi>10^{42}\ergs$ within the volume determined to be CT, i.e., a fraction of $13\%$.
Among the R90 galaxies with $\lwthree > 10^{42.3}\ergs$, 10 of the 84 B70 AGN are CT obscured, i.e., $12\%$.
In addition, there are eight more known AGN that are not in the B70 but are CT and fulfill the R90 and luminosity cuts\footnote{
These are IC\,3639, Mrk\,573 aka UGC\,1214, NGC\,660, NGC\,1320, NGC\,1386, NGC\,4418, NGC\,5135, and NGC\,5347
(in order of the object list: \citealt{boorman_ic_2016, guainazzi_x-ray_2005}; Annuar et al., in prep.; \citealt{balokovic_nustar_2014, levenson_penetrating_2006, maiolino_elusive_2003, singh_suzaku_2012, levenson_penetrating_2006}).
}.
Together, this means at least 18 of the 160 R90 AGN with $\lwthree > 10^{42.3}\ergs$ are CT obscured, i.e., $11\%$. 

However, it is likely that the true CT fraction is significantly higher as was indicated in Sect.~\ref{sec:eros} already, since none of the (predicted) intrinsically X-ray-bright R90 AGN candidates have been detected by BAT.
In particular, if we assume the bias-corrected $\nh$ distribution of \cite{ricci_compton-thick_2015}, i.e., a CT fraction of $27\%$, to apply for all 221 R90 objects with $\log \lwthree > 10^{42.3}\ergs$, then we would expect 60 CT AGN in total.
Since in the BAT detected subset, there are only 10, there should be 50 CT AGN among the 137 R90 objects not in B70.
To test whether this is consistent with the observations, we repeat the Monte Carlo simulation of Sect.~\ref{sec:eros} for these 137 objects assuming 50 CT AGN among them.
As a result we would  still expect $60\pm 25$ objects to have been detected in the 70\,month BAT map.
Even if we assume again 22 contaminants as a result of the R90 selection, this leaves $38\pm 25$.
In fact, it would take an intrinsic CT fraction of $40\%$ to become consistent with no BAT detection within $1\sigma$ uncertainty.
On the other hand, we do not expect more than $\sim 100$ CT AGN among the 137 because of at least 14 of the known AGN being optically classified as type~1 AGN and thus unlikely CT. 
This would translate into an intrinsic CT fraction of $55\%$ which we regard as an upper limit.
These findings suggest that the intrinsic CT fraction is between $40-55\%$ in the here probed luminosity regime.
However, these numbers should be regarded as indicative only owing the large number of very simplified assumptions made here.

Let us examine some of the objects in more detail.
The most promising CT candidates are those with the highest MIR-to-X-ray ratio, for example 
sources that are not detected by \swift/BAT after 70\,months but are 1\,dex brighter than the \wthreee magnitude corresponding to the nominal detection limit of $\fxh = 1.34 \cdot 10^{-11}\fu$, namely $\wthree< 4.7$\,mag.
Indeed, we find that six out of the eight known CT AGN that remained undetected in the BAT 70\,month map fulfil this criterion, so a $75\%$ success rate.
If we apply this magnitude limit to the whole R90 AGN sample excluding B70, we identify a further nine CT candidates among the known AGN.
Six of them do not fulfil the S18 criterion and are in fact known to host starbursts (Arp\,220, IC\,1623B, NGC\,253, NGC\,3256, NGC\,3690E, and NGC\,7552)\footnote{
However, \cite{teng_nustar_2015} find that the X-ray data of Arp\,220 is consistent with a CT AGN being present in this source.
}.
So their $\wthree$ emission could be star-formation dominated.
This leaves three more robust CT AGN candidates (ESO\,420-13, NGC\,1377, and NGC\,3094).

We can also apply this diagnostic to the R90 AGN candidates which yields six galaxies, of which three are known to host starbursts (MCG\,+12-02-001, NGC\,520 and NGC\,3690W), leaving another three candidates for CT AGN (ESO\,127-11, ESO\,173-15, and ESO\,495-5).
We plan to investigate these candidates further in the future.


\subsection{Total number of AGN estimate}\label{sec:num}
Even without having confirmed all the R90 objects as AGN, we can make a rough estimate of the total number of AGN above a given luminosity limit within the volume based on the characterisation of the criterion and found numbers from the previous sections.
The main assumption is that the defining feature of the R90 criterion is valid also in our volume, namely that $90\%$ of galaxies with such a red \wonetwoo colour indeed host an AGN, at least for objects with  $\lwthree > 10^{42.3}\ergs$ as concluded in Sect.~\ref{sec:SF}.
This lower luminosity limit matches well with our completeness limit for the LASr-GPS (Sect.~\ref{sec:gal_compl}).
Furthermore, R90 selects the majority of AGN with luminosities greater than this threshold (Sect.~\ref{sec:R90}).
Therefore, we use $\lwthree > 10^{42.3}\ergs$ here in the absence of a more accurate AGN power tracer.

There are 221 R90 galaxies with  $\lwthree > 10^{42.3}\ergs$, of which 160 are known to host an AGN, 84 of which are in the B70 sample.
According to the R90 definition, we expect that 199 ($90\%$) of them host genuine AGN. 
This number is consistent with applying the S18 cut to the R90 sample instead, which would return 186 objects, i.e., $84\%$, which is slightly lower but we know that S18 also removes some AGN.
Owing the complications of S18 discussed in Sect.~\ref{sec:SF}, we stick with the simple R90-based estimate in the following, i.e., our initial estimate for the total number of AGN in the volume is $N_\mathrm{ini} = 199$.
For the final best estimate, this number has to be corrected by the various factors of incompleteness as discussed in the following.

%

\subsubsection{Colour selection incompleteness}
The main source of incompleteness is the colour selection.
The R90 criterion from \cite{assef_wise_2018} was designed for high reliability. This reliability comes at the price of a significant level of incompleteness, which we have seen already in Sect.~\ref{sec:R90}.
Namely, for a lower luminosity of $\lwthree > 10^{42.3}\ergs$, only $51 \pm 10\%$ of AGN are selected.
Thus, we require $N_\mathrm{ini}$ to be multiplied by a colour selection incompleteness correction factor, $c_\mathrm{CSI} = 1.95^{+0.48}_{-0.32}$.
This factor does not yet account for contamination of the $\lwthree$ flux, which is addressed next.

\subsubsection{Host contribution to \wthree}
Host contribution, mostly through star formation, to \wthreee leads us to overestimate the intrinsic AGN luminosity and, thus, to the inclusion of AGN with intrinsic luminosities below our completeness limit.
We have already estimated this effect statistically in Sect.~\ref{sec:known_AGN_MIR} and applied a corresponding correctop in the Monte Carlo simulations of Sect.~\ref{sec:eros}.
Thus, we here just repeat the first part of the Monte-Carlo simulation of that section to estimate the $\ltw$ distribution for the AGN in our R90 galaxies, where no direct measurement is available.
This way, we find that $187\pm 22$ out of the 221 ($85\pm10\%$) of the R90 galaxies have expected $\ltw > 10^{42.3}\ergs$ .
Therefore, the corresponding host contamination correction factor is $c_\mathrm{HC} = 0.85 \pm 0.1$.

\subsubsection{Parent sample incompleteness}
Another source of incompleteness is of course the galaxy parent sample used for the AGN selection.
The level of incompleteness of the LASr-GPS was estimated in Sect.~\ref{sec:gal_compl}.
There, we used \wonee as rough tracer of the stellar mass of the galaxies, while here we want to know the completeness with respect to the  $\lwthree$ luminosity threshold.
Thus, we repeat the completeness analysis of Sect.~\ref{sec:gal_compl} but using \wthree, and find that for $\lwthree > 10^{42.3}\ergs$ and  $|b| > 8\degree$, the galaxy parent sample is $96.1 \pm 4.2\%$ complete\footnote{
The crossmatching with \wisee is normally another source of incompleteness but we found \wisee counterparts for all galaxies in the LASr-GPS.
On the other hand, the \wisee counterparts for $1.4\%$ of the galaxies were drowned by brighter nearby objects (Sect.~\ref{sec:WISE}).
However, we do not consider this effect in the total number of AGN estimate because it is much smaller than the uncertainties of the other corrections.
}.
As discussed in Sect.~\ref{sec:MW}, the shadow of the Milky Way further increases the incompleteness of the parent sample by $6.4\pm 0.8\%$.
Therefore, we adopt a total galaxy parent sample incompleteness correction factor, $c_\mathrm{PSI} = 1.11 \pm 0.04$.

\subsubsection{Other corrections, not accounted for}
We did not attempt to correct for the fact that redshift-independent distances are not available for all of the galaxies. 
This is the case for $71\%$ of the R90 AGN and candidates. 
We found that the redshift independent distances are on average $10\%$ smaller than the redshift-based distances. 
This would mean that the luminosities of these galaxies would decrease by 0.04\,dex, leading to the loss of 3 candidate AGN but none of the known AGN. 
At the same time, 31 additional known AGN and 20 candidates would fall into the volume. 
However, since we did not consider redshift-independent distances for galaxies with $\dl >50$\,Mpc, a correction is not straightforward and thus not applied here.

In addition to the above incompleteness effects, there are also object intrinsic effects like obscuration in the MIR.
The latter, however, has little effect on the \wonetwoo colour at low redshifts as shown in \cite{stern_mid-infrared_2012}, because extinction at the wavelengths of \wonee and \wtwoo is low and approximately constant in typical extinction laws (e.g., \citealt{fritz_line_2011}).
Thus, no correction for that is applied here. 

Finally, one might ask, what about beamed MIR emission, i.e., blazars? 
Out of the full sample of 838 AGN in the full BAT 70\,month catalogue, 105 are classified as beamed sources according to BZCAT \citep{massaro_roma-bzcat:_2009}, and 5 are in our $D<100\,$Mpc volume, implying a beamed fraction of $\sim3\%$! 
On the other hand, all of these 5 objects are known to have SEDs that are not dominated by beamed emission (Cen\,A, Mrk\,348, NGC\,1052, NGC 1275, NGC 7213). 
This suggests that the true beamed fraction is $\ll 1\%$, and, thus we ignore this effect here.

\subsubsection{Best estimate}
We applied all the above correction factors to our initial average estimate, $N_\mathrm{ini} = 199$, to arrive at our best estimate: 
\begin{equation*}
N_\mathrm{best} = N_\mathrm{ini} \cdot c_\mathrm{CSI} \cdot c_\mathrm{HC} \cdot c_\mathrm{PSI} = 1.82 N_\mathrm{ini} = 362^{+145}_{-116}
\end{equation*}
AGN with $\ltw>10^{42.3}\ergs$ (equivalent to $\lxi>10^{42}\ergs$) in our $D<100\,$Mpc volume.
This corresponds to a number density of $8.6^{+3.5}_{-2.8}\,\times$\,10$^{-5}$\,Mpc$^{-3}$.

We also repeat the above estimation for  $\lxi>10^{43}\ergs$  and  $\lxi>10^{44}\ergs$, resulting in $101^{+55}_{-25}$ and $4^{+2}_{-1}$ AGN above these luminosity thresholds, respectively.
These compare to 53 and 2 AGN known with $\lxi>10^{43}\ergs$  and  $\lxi>10^{44}\ergs$, respectively, within the volume.

\subsection{Comparison to estimates from luminosity functions}\label{sec:lfunc}
Finally, with these purely observational estimates for the number of AGN within 100\,Mpc, one might want to compare to the predictions from currently used AGN luminosity functions.
First, we compare to an optical luminosity function, namely the one derived by \cite{palanque-delabrouille_luminosity_2013} for luminous AGN in the redshift range $0.7 < z < 4$, whereas its shape was assumed to be a standard double power law following \cite{boyle_2df_2000}.
For a redshift of 0.01, they found a break magnitude of -22.1 and the power-law indices $\alpha=3.5$ and $\beta = 1.43$, while the break value for the bolometric luminosity is $\sim 10^{45}\ergs$.
Here, we used the lower cut-off of $10^{43}\ergs$ for the bolometric luminosity which with the simple assumption of   $\lbol = 10 \lxi$ (e.g., \citealt{vasudevan_piecing_2007}) corresponds to the same lower luminosity cut used for our total AGN number estimate in the previous section, i.e., $\lxi = 10^{42}\ergs$.
We then integrated the luminosity function over the whole sky up to a redshift of 0.0222 (corresponding to our distance limit of 100\,Mpc).
This results in an estimated number of optical AGN of 82. 
The latter number corresponds only to the unobscured AGN, so we need to correct for the obscuration fraction which is somewhere between $\sim 50\%$ to $80\%$ (e.g., \citealt{schmitt_testing_2001, hao_active_2005-1}), resulting in 164 to 410 objects. 

Instead of an optical luminosity function, using an X-ray luminosity function has the advantage of also including obscured AGN (e.g., \citealt{ueda_cosmological_2003}).
There is a large variety of such functions available in the literature.
For simplicity, we here choose only one of the recent works that attempted to incorporate the CT fraction as well, namely \cite{aird_evolution_2015}.
This work compares several different approaches for determining an X-ray luminosity function, and we refer the reader to that work for more details.
We try several of those functions, for example the luminosity-dependent density evolution model which returns an estimate of 125 AGN including obscured objects, while the flexible double power law (FDPL) yields a total number estimate of 175 AGN above our luminosity limit. 
Finally, \cite{aird_evolution_2015} put forward a model that includes a description of the absorption distribution function (XLAF), allowing to compute the number of unobscured and obscured AGN separately.
It results in an estimate of 97 unobscured and 264 obscured AGN, i.e., 361 AGN in total.
This number is indeed very close to our best estimate of 362 AGN in our volume and also agrees well with a corresponding estimate using the luminosity function from \cite{ueda_toward_2014}.
Interestingly, the best fitting CT/Compton-thin obscured fraction found in \cite{aird_evolution_2015} of $34\%$ predicts that 90 out of the 361 AGN are CT, i.e., a total CT fraction of $25\%$.
Once, the R90 AGN sample has been better characterised and the candidates verified, more constraining tests will be possible.


\section{Summary \& Conclusions}\label{sec:concl}
The recent and ongoing sensitive all-sky surveys including \wise, \erositaa and \artxcc, in combination with the collected knowledge of large astronomical databases, now allow us to obtain a complete census of significantly accreting SMBHs manifesting as AGN in the local Universe.
This is the goal of LASr, and this work has presented the first steps in this project.
In particular, we first created a LASr galaxy parent sample, LASr-GPS, of $\sim49$k galaxies by combining NED, SIMBAD, SDSS and 2MRS for a volume of $D<100\,$Mpc.
We then crossmatched the sample with \wisee to obtain the MIR properties of the host galaxy bulges.
The analysis based on this sample leads to the following main results:
\begin{itemize}

\item First, we estimated the resulting LASr-GPS is $\sim 90\%$ complete for galaxies with central (bulge) luminosities of $\lwone > 10^{42}\ergs$ (Sect.~\ref{sec:gal_compl}), a factor $\sim 4$ deeper than the 2MRS galaxy sample (Sect.~\ref{sec:gal_compa}).

\item The 20.6k galaxies above this luminosity harbour 4.3k known AGN collected from identifications in the literature (Sect.~\ref{sec:kAGN}).
However, we caution the reader that not all of these AGN identifications might be reliable which is particularly true for the controversial class of the LINERs.
Of these $56\%$ have an optical classification as Seyfert with the apparent type~2 to type~1 ratio between 49 to $60\%$.

\item We compute optical classification-based corrections to estimate the nuclear $12\um$ luminosities of the AGN from the $\wthreee$ profile fitting magnitudes, and find that the majority of the known AGN have low luminosities, i.e., only $18\%$ are estimated to have $\ltw > 10^{42.3}\,$erg\,s$^{-1}$ (Sect.~\ref{sec:known_AGN_MIR}).

\item We then proceed to use \wise-based AGN identification by MIR colour to find new AGN candidates.
For this purpose we employ the R90 criterion from \cite{assef_wise_2018}, which is based on the \wonetwoo and selects AGN with a $90\%$ pureness.
We estimate that this criterion has an average efficiency of $51\pm10\%$ to select AGN with $\lxi > 10^{42}\ergs$ (Sect.~\ref{sec:R90}).  

\item The R90 criterion selects 172 galaxies known to host AGN (Sect.~\ref{sec:R90}), and 159 AGN {\em candidates} (Sect.~\ref{sec:candi}).  
Of the R90 selected AGN, $97\%$ are classified optically as Seyferts with an apparent type~2 fraction between 38 and $62\%$, depending on how objects with multiple or intermediate classifications are treated.
The intrinsic optical type~2 fraction is likely higher than $50\%$ because we expect most of the R90 candidates to be type~2.
It could be up to $71\%$, depending how many of the R90 candidates are genuine AGN and obscured.

\item We find that the \wtwothree-based criterion presented by \cite{satyapal_star-forming_2018} to exclude strong starbursts  indeed further increases the pureness of R90 selected AGN but also excludes some highly obscured AGN and AGN hosted in star-forming galaxies (Sect.~\ref{sec:SF}).

\item A lower luminosity cut of $\lwthree > 10^{42.3}\ergs$  is $90\%$ efficient at removing compact star-forming galaxies, so that remaining contamination in our R90 sample should be low (Sect.~\ref{sec:SF}).
This luminosity cut leaves 61 robust AGN candidates. 

\item We predict detection rates for the \erositaa all-sky survey, and find that the majority of the AGN candidates are expected to be highly obscured, in order to explain their non-detection by \swift/BAT and reach the expected intrinsic CT fraction for the whole sample (Sect.~\ref{sec:eros}). 

\item The discussion of constraints on the CT fraction based on the R90 selected AGN sample indicates the intrinsic CT fraction is likely higher than the $27\%$ estimated from the BAT 70\,month sample, and could be up to $55\%$ (Sect.~\ref{sec:CT}). 

\item Finally, we use the R90 selection to estimate the total number of AGN with $\lxi > 10^{42}\ergs$ within 100\,Mpc to be $362^{+145}_{-116}$, corresponding to a number density of $8.6^{+3.5}_{-2.8}\,\times$\,10$^{-5}$\,Mpc$^{-3}$ (Sect.~\ref{sec:num}).
This estimate is consistent with estimates from recent X-ray luminosity functions for AGN in the literature (Sect.~\ref{sec:lfunc}).

\end{itemize}

In future LASr work, we plan to follow up the new AGN candidates, e.g., with optical spectroscopy and present a full characterisation of the R90 AGN sample, before adding additional AGN identification techniques, e.g., based on MIR variability to increase the fraction of identified AGN within 100\,Mpc.
In the long term, data from the X-ray missions will complement the MIR-based identification of AGN and provide intrinsic AGN power estimates, allowing us to combine MIR and X-ray diagnostics to identify and characterise the majority of CT AGN.
The final volume-limited sample of LASr AGN should provide a robust redshift zero anchor for AGN population models.

\section*{Acknowledgements}
The authors would like to especially thank the user support teams of NED and SIMBAD, without whom this work would not have been possible.
DA would also like to give special thanks to the development teams of the two fantastic and ever-improving tools TOPCAT and Aladin, both of which played a fundamental part in this research.
Furthermore, we cordially thank the referee for valuable comments that helped to improve the manuscript.
DA acknowledges funding through the European Union’s Horizon 2020 and Innovation programme under the Marie Sklodowska-Curie grant agreement no. 793499 (DUSTDEVILS).
PG acknowledges support from STFC (ST/R000506/1) and a UGC-UKIERI Thematic Partnership. PG also acknowledges discussions with the {\em eROSITA} AGN team, especially A.\,Merloni and A.\,Georgakakis. 
P. B. acknowledges financial support from the STFC and the Czech Science Foundation project No. 19-05599Y.
RJA was supported by FONDECYT grant number 1191124.
SF acknowledges support from the Horizon 2020 ERC Starting Grant DUST-IN-THE-WIND (ERC-2015-StG-677117).
FS acknowledges partial support from a Leverhulme Trust Research Fellowship. 
JA acknowledges support from an STFC Ernest Rutherford Fellowship, grant code: ST/P004172/1.  
This publication makes use of data products from the {\it Wide-field Infrared Survey Explorer}, which is a joint project of the University of California, Los Angeles, and the Jet Propulsion Laboratory/California Institute of Technology, funded by the National Aeronautics and Space Administration.
We acknowledge the usage of the HyperLeda database (http://leda.univ-lyon1.fr; \citealt{makarov_hyperleda._2014}).
This research made use of the NASA/IPAC Extragalactic Database
(NED), which is operated by the Jet Propulsion Laboratory, California Institute of Technology, under contract with the National Aeronautics and Space Administration. 
This research has made use of the SIMBAD database,
operated at CDS, Strasbourg, France  \citep{wenger_simbad_2000}.
This research has made use of ``Aladin sky atlas" developed at CDS, Strasbourg Observatory, France \citep{bonnarel_aladin_2000, boch_aladin_2014}.
This research made use of Astropy, a community-developed core Python package for Astronomy \citep{astropy_collaboration_astropy:_2013, astropy_collaboration_astropy_2018}.
This research has made use of NASA's Astrophysics Data System.
This research has made use of TOPCAT \citep{taylor_topcat_2005}. 
%
%
\bibliographystyle{mnras} 
\bibliography{my_lib_ref.bib} 

\appendix

{\onecolumn
\begin{landscape}
\input{tab_gal.tex}		                              
\begin{minipage}{1.25\textwidth}
\normalsize
{\it -- Notes:} 
(1) object name, mostly following NED nomenclature;
(2) origin of object with preference to NED if available;
(3) flag whether the galaxy is in 2MRS;
(4) and (5) flag whether galaxy is known to host an AGN or starburst, respectively;
(6) and (7) equatorial coordinates of the object centre in J2000 in degrees;
(8) redshift;
(9) redshift confidence flag: 0 means  that the value is robust, 1 means that the value is not robust but there is no reason to doubt, and 2 means that the redshift is controversial;
(10) object distance in Mpc;
(11) No nucleus flag: If true, the source does not show any clear centre or nucleus in the optical/infrared images;
 (12), (13), (14), and (15) \wisee profile-fitting photometric magnitudes;
(16) and (17) \wisee \wonetwoo and \wtwothreee colours;
(18) and (19)  observed \wonee and \wthreee continuum luminosities , calculated from the selected distance and the profile-fitting magnitudes.
\end{minipage}
\end{table*}
\end{landscape}

\begin{landscape}
\scriptsize
\input{tab_R90AGN.tex}		                              
\begin{minipage}{1.25\textwidth}
%
\normalsize
{\it -- Notes:} 
(1) object name, mostly following NED nomenclature;
(2) flag whether the galaxy is in 2MRS;
(3), (4), (5), (6), (7) (8) flags whether galaxy has Seyfert, Sy\,1, Sy\,2, LINER, H\,II or starburst classification, respectively;
(9) and (10) equatorial coordinates of the object centre in J2000 in degrees;
(11) redshift;
(12) object distance in Mpc;
(13) and (14) \wisee \wonetwoo and \wtwothreee colours;
(15) and (16)  observed \wonee and \wthreee continuum luminosities , calculated from the selected distance and the profile-fitting magnitudes.
(17) nuclear $12\um$ luminosity of the AGN either taken from \cite{asmus_subarcsecond_2014}, if uncertainty $\le 0.1$\,dex, or estimated from $\lwthree$ and optical type (see Sect.~\ref{sec:known_AGN_MIR} for details);
(18) intrinsic 2-10\,keV X-ray luminosity for AGN from the BAT70 sample \citep{ricci_bat_2017}, and from \cite{asmus_subarcsecond_2015} otherwise, where available;
(19) flag whether the source fulfils the S18 criterion or not (see Sect.~\ref{sec:SF} for details);
\end{minipage}

\scriptsize
\input{tab_R90candi.tex}		                              
\begin{minipage}{1.25\textwidth}
%
\normalsize
{\it -- Notes:} 
(1) object name, mostly following NED nomenclature;
(2) flag whether the galaxy is in 2MRS;
(3) and (4) flags whether galaxy has a H\,II or starburst classification, respectively;
(5) and (6) equatorial coordinates of the object centre in J2000 in degrees;
(7) redshift;
(8) object distance in Mpc;
(9) and (10) \wisee \wonetwoo and \wtwothreee colours;
(11) and (12)  observed \wonee and \wthreee continuum luminosities, calculated from the selected distance and the profile-fitting magnitudes.
(13) nuclear $12\um$ luminosity of an assumed AGN estimated from $\lwthree$ and optical type (see Sect.~\ref{sec:known_AGN_MIR} for details);
(14) flag whether the source fulfils the S18 criterion or not (see Sect.~\ref{sec:SF} for details);
\end{minipage}
\label{lastpage}
\end{landscape}


\end{document}

%% file: macros.tex
\newcommand{\degree}{^\circ}
\newcommand{\um}{\,\mu\mathrm{m}}
\newcommand{\cm}{\,\mathrm{cm}}
\newcommand{\erg}{\,\mathrm{erg}}
\newcommand{\ergs}{\erg\,\mathrm{s}^{-1}}

\newcommand{\fu}{\erg\,\cm^{-2}\mathrm{s}^{-1}}

\newcommand{\lbol}{L_\mathrm{bol}}

\newcommand{\dl}{D_L}

\newcommand{\nh}{N_\mathrm{H}}

\newcommand{\lnn}{L^\mathrm{nuc}(12\,\mu\mathrm{m})}

\newcommand{\lxi}{L^\mathrm{int}(\textrm{2-10\,keV})}

\newcommand{\lxo}{L^\mathrm{obs}(\textrm{2-10\,keV})}
\newcommand{\fxo}{F^\mathrm{obs}(\textrm{2-10\,keV})}

\newcommand{\fxh}{F^\mathrm{obs}(\textrm{14-195\,keV})}

\newcommand{\wonee}{\textit{W1}\ }
\newcommand{\wtwoo}{\textit{W2}\ }
\newcommand{\wthreee}{\textit{W3}\ }

\newcommand{\wone}{\textit{W1}}
\newcommand{\wtwo}{\textit{W2}}
\newcommand{\wthree}{\textit{W3}}
\newcommand{\wfour}{\textit{W4}}
\newcommand{\wonetwoo}{\textit{W1-W2}\ }
\newcommand{\wtwothreee}{\textit{W2-W3}\ }
\newcommand{\wthreefourr}{\textit{W3-W4}\ }
\newcommand{\wonetwo}{\textit{W1-W2}}
\newcommand{\wtwothree}{\textit{W2-W3}}

\newcommand{\wonetwom}{\mathit{W1-W2}}
\newcommand{\wtwothreem}{\mathit{W2-W3}}

\newcommand{\lwthree}{L(\mathit{W3})}
\newcommand{\lwone}{L(\mathit{W1})}
\newcommand{\ltw}{L^\mathrm{nuc}(12\um)}

\newcommand{\pk}{p_\mathrm{K}}
\newcommand{\tauk}{\tau_\mathrm{K}}

\newcommand{\spitzer}{{\it Spitzer}}                    

\newcommand{\swiftt}{\textit{Swift}\ }
\newcommand{\swift}{\textit{Swift}}
\newcommand{\wise}{\textit{WISE}}
\newcommand{\wisee}{\textit{WISE}\ }

\newcommand{\srg}{\textit{SRG}}

\newcommand{\erosita}{eROSITA}
\newcommand{\erositaa}{eROSITA\ }
\newcommand{\artxc}{ART-XC}
\newcommand{\artxcc}{ART-XC\ }

%% file: tab_gal.tex
\begin{table*}
\caption{LASr-GPS}
\label{tab_gal}
\scriptsize
\centering
\begin{tabular}{l c c c c c c c c c c c c c c c c c c}
\hline\hline
 &  & in & known &  &  &  &  & z &  & No &  &  &  &  &  &  & $\log L$ & $\log L$\\
Name & Origin & 2MRS & AGN & starburst & RA & DEC & z & Flag & D & Nucleus & \wone & \wtwo & \wthree & \wfour & \wonetwo & \wtwothree & (\wone) & (\wthree)\\
 &  &  &  &  & [$\degree$] & [$\degree$] &  &  & [Mpc] &  & [mag] & [mag] & [mag] & [mag] & [mag] & [mag] & [erg/s] & [erg/s]\\
(1) & (2) & (3) & (4) & (5) & (6) & (7) & (8) & (9) & (10) & (11) & (12) & (13) & (14) & (15) & (16) & (17) & (18) & (19)\\
\hline
2dFGRS S805Z417 & NED & False & False & False & 0.00162 & -56.14106 & 0.01010 & 0 & 45.0 & False & 15.98 & 15.92 & $\ge$12.56 & $\ge$9.00 & 0.07 & 3.36 & 40.4 & $\le$40.3\\
UGC 12889 & NED & True & False & False & 0.00696 & 47.27479 & 0.01673 & 0 & 75.0 & False & 11.07 & 11.13 & 9.04 & 7.03 & -0.05 & 2.09 & 42.8 & 42.1\\
KUG 2357+156 & NED & False & False & False & 0.00905 & 15.88188 & 0.02002 & 0 & 89.9 & False & 12.18 & 12.09 & 8.11 & 6.02 & 0.09 & 3.98 & 42.6 & 42.7\\
SDSS J000003.22-010646.9 & NED & False & False & False & 0.01342 & -1.11303 & 0.02178 & 0 & 98.0 & True & 15.41 & 15.11 & $\ge$11.98 & $\ge$8.88 & 0.30 & 3.13 & 41.3 & $\le$41.2\\
KUG 2357+228 & NED & False & False & False & 0.01464 & 23.08753 & 0.01488 & 0 & 66.6 & False & 14.77 & 14.69 & $\ge$11.81 & $\ge$8.62 & 0.08 & 2.88 & 41.3 & $\le$40.9\\
MCG -01-01-016 & NED & True & False & False & 0.03600 & -6.37400 & 0.02179 & 0 & 98.0 & False & 11.95 & 11.90 & 8.53 & 6.66 & 0.04 & 3.37 & 42.7 & 42.6\\
MCG -01-01-017 & NED & True & True & False & 0.04708 & -5.15875 & 0.01898 & 0 & 85.2 & False & 12.43 & 12.45 & 9.40 & 7.46 & -0.02 & 3.05 & 42.4 & 42.1\\
2MASX J00001215+0205503 & NED & False & False & False & 0.05067 & 2.09742 & 0.02170 & 0 & 97.6 & False & 12.01 & 11.75 & 7.67 & 4.57 & 0.26 & 4.09 & 42.7 & 42.9\\
CGCG 548-023 & NED & True & False & True & 0.05404 & 46.96514 & 0.01790 & 0 & 80.3 & False & 11.29 & 11.13 & 8.60 & 6.33 & 0.16 & 2.54 & 42.8 & 42.4\\
CGCG 498-057 & NED & False & False & False & 0.05542 & 33.13417 & 0.01684 & 0 & 75.5 & False & 12.12 & 12.00 & 9.05 & 7.05 & 0.12 & 2.95 & 42.4 & 42.1\\
2MFGC 00003 & NED & False & False & False & 0.05975 & 70.03300 & 0.01530 & 0 & 68.5 & False & 11.80 & 11.57 & 7.50 & 5.59 & 0.23 & 4.07 & 42.5 & 42.7\\
GALEXASC J000017.22+272403.0 & NED & False & False & False & 0.07208 & 27.40083 & 0.01552 & 0 & 69.5 & False & 15.18 & 14.95 & $\ge$12.02 & 8.97 & 0.22 & 2.93 & 41.1 & $\le$40.9\\
GALEXASC J000019.31-315611.3 & NED & False & False & False & 0.08050 & -31.93667 & 0.01230 & 0 & 54.9 & False & 15.42 & 15.06 & 12.30 & $\ge$9.07 & 0.36 & 2.76 & 40.8 & 40.6\\
FAIRALL 1061 & NED & True & False & False & 0.09838 & -47.01881 & 0.01998 & 0 & 89.8 & False & 11.04 & 11.12 & 10.54 & $\ge$8.51 & -0.08 & 0.58 & 43.0 & 41.7\\
2MASX J00002482-0451473 & NED & False & False & False & 0.10351 & -4.86313 & 0.01892 & 0 & 85.0 & False & 13.10 & 13.17 & $\ge$11.48 & $\ge$7.97 & -0.06 & 1.69 & 42.1 & $\le$41.3\\
UGC 12893 & NED & False & False & False & 0.11638 & 17.21869 & 0.00367 & 0 & 16.3 & False & 14.95 & 15.00 & $\ge$12.58 & $\ge$8.22 & -0.05 & 2.42 & 40.0 & $\le$39.4\\
LEDA 089491 & NED & False & False & False & 0.12167 & -60.68076 & 0.02210 & 0 & 99.5 & False & 14.64 & 14.44 & 10.91 & 7.58 & 0.20 & 3.53 & 41.7 & 41.6\\
ESO 293- G 027 & NED & False & False & False & 0.12283 & -40.48447 & 0.01061 & 0 & 47.3 & False & 12.60 & 12.58 & 9.42 & 7.22 & 0.02 & 3.15 & 41.8 & 41.6\\
KUG 2357+225 & NED & False & False & False & 0.13946 & 22.77844 & 0.02020 & 0 & 90.8 & False & 13.20 & 13.03 & 9.43 & 6.73 & 0.18 & 3.59 & 42.1 & 42.1\\
UGC 12898 & NED & False & False & False & 0.15600 & 33.60127 & 0.01594 & 0 & 71.4 & False & 15.08 & 14.88 & 12.39 & $\ge$8.22 & 0.20 & 2.48 & 41.2 & 40.7\\
2dFGRS S357Z026 & NED & False & False & False & 0.19546 & -30.64639 & 0.01428 & 0 & 63.9 & False & 16.00 & 16.07 & $\ge$12.63 & $\ge$8.86 & -0.08 & 3.45 & 40.7 & $\le$40.6\\
ESO 193- G 009 & NED & True & False & False & 0.22192 & -47.35681 & 0.01972 & 0 & 88.6 & False & 11.28 & 11.31 & 9.28 & 7.85 & -0.03 & 2.03 & 42.9 & 42.2\\
APMUKS(BJ) B235824.83-412603.8 & NED & False & False & False & 0.24577 & -41.15485 & 0.00050 & 0 & 2.2 & True & 17.39 & 17.05 & $\ge$12.66 & $\ge$9.24 & 0.33 & 4.39 & 37.3 & $\le$37.6\\
NGC 7802 & NED & True & False & False & 0.25175 & 6.24206 & 0.01776 & 0 & 79.7 & False & 10.46 & 10.48 & 9.54 & 7.66 & -0.02 & 0.94 & 43.1 & 42.0\\
SDSS J000103.59+143448.6 & NED & False & False & True & 0.26500 & 14.58018 & 0.00573 & 0 & 25.5 & False & 15.58 & 15.40 & $\ge$11.98 & $\ge$8.54 & 0.18 & 3.42 & 40.1 & $\le$40.0\\
GALEXASC J000109.10-162721.7 & NED & False & False & False & 0.28813 & -16.45619 & 0.01574 & 0 & 70.5 & False & 15.34 & 15.27 & $\ge$11.99 & $\ge$8.68 & 0.07 & 3.28 & 41.1 & $\le$40.9\\
KUG 2358+128A & NED & False & False & False & 0.30575 & 13.14406 & 0.01830 & 0 & 82.1 & False & 13.13 & 13.05 & 9.65 & 7.12 & 0.08 & 3.40 & 42.1 & 42.0\\
MCG +02-01-010 & NED & False & False & False & 0.31238 & 13.11256 & 0.01873 & 0 & 84.1 & False & 14.20 & 14.06 & 11.02 & $\ge$8.36 & 0.14 & 3.04 & 41.7 & 41.4\\
IC 5376 & NED & True & True & False & 0.33237 & 34.52572 & 0.01678 & 0 & 75.2 & False & 10.95 & 11.00 & 8.95 & 7.28 & -0.05 & 2.05 & 42.9 & 42.2\\
NGC 7803 & NED & True & False & False & 0.33321 & 13.11125 & 0.01790 & 0 & 80.3 & False & 10.27 & 10.17 & 6.92 & 4.48 & 0.11 & 3.25 & 43.2 & 43.0\\
2MASX J00012334+4733537 & NED & True & True & False & 0.34764 & 47.56505 & 0.01747 & 0 & 78.3 & False & 11.77 & 11.18 & 7.76 & 5.19 & 0.59 & 3.42 & 42.6 & 42.7\\
MRK 0934 & NED & False & False & False & 0.35850 & 13.11300 & 0.01753 & 0 & 78.6 & False & 12.59 & 12.44 & 8.65 & 6.25 & 0.15 & 3.80 & 42.3 & 42.3\\
NGC 7805 & NED & True & False & False & 0.36154 & 31.43375 & 0.01605 & 0 & 71.9 & False & 10.55 & 10.61 & 10.13 & 8.89 & -0.06 & 0.49 & 43.0 & 41.7\\
UGC 12910 & NED & False & False & False & 0.36833 & 5.38944 & 0.01317 & 0 & 58.9 & False & 15.48 & 15.08 & 12.02 & $\ge$8.38 & 0.40 & 3.06 & 40.9 & 40.7\\
NGC 7806 & NED & True & False & False & 0.37521 & 31.44186 & 0.01590 & 0 & 71.2 & False & 10.82 & 10.80 & 8.62 & 6.89 & 0.03 & 2.17 & 42.9 & 42.2\\
CGCG 433-016 & NED & False & False & True & 0.39146 & 15.08156 & 0.02119 & 0 & 95.3 & False & 14.25 & 14.03 & 10.20 & 7.52 & 0.22 & 3.82 & 41.8 & 41.9\\
UGC 12913 & NED & False & False & False & 0.40292 & 3.50558 & 0.02115 & 0 & 95.1 & False & 14.21 & 14.05 & 11.32 & $\ge$8.47 & 0.16 & 2.73 & 41.8 & 41.4\\
GALEXASC J000137.80+172918.9 & NED & False & False & False & 0.40708 & 17.48861 & 0.02151 & 0 & 96.7 & False & 16.50 & 16.43 & $\ge$12.50 & $\ge$9.01 & 0.06 & 3.94 & 40.9 & $\le$41.0\\
UGC 12914 & NED & True & True & False & 0.40967 & 23.48364 & 0.01458 & 0 & 65.2 & False & 10.28 & 10.25 & 7.78 & 5.34 & 0.03 & 2.47 & 43.0 & 42.5\\
AGC 748776 & SIMBAD & False & False & False & 0.42229 & 13.84256 & 0.02112 & 0 & 95.0 & False & 15.27 & 14.95 & $\ge$12.17 & $\ge$8.80 & 0.33 & 2.78 & 41.4 & $\le$41.1\\
\multicolumn{19}{c}{\textbf{Abridged. Full table available online in digitial format.}}\\
\hline	
\end{tabular}

%% file: tab_R90AGN.tex
\scriptsize
\begin{longtable}{l c c c c c c c c c c c c c c c c c c}
\caption{known AGN selected by R90\label{tab_R90AGN}}\\
\hline\hline
 & in &  &  &  &  &  &  &  &  &  &  &  &  & $\log L$ & $\log L$ & $\log L^\mathrm{nuc}$ & $\log L^\mathrm{int}$ & \\
Name & 2MRS & Seyfert & Sy\,1 & Sy\,2 & LINER & H\,II & starburst & RA & DEC & z & D & \wonetwo & \wtwothree & (\wone) & (\wthree) & ($12\um$) & (2-10\,keV) & S18\\
 &  &  &  &  &  &  &  & [$\degree$] & [$\degree$] &  & [Mpc] & [mag] & [mag] & [erg/s] & [erg/s] & [erg/s] & [erg/s] & \\
(1) & (2) & (3) & (4) & (5) & (6) & (7) & (8) & (9) & (10) & (11) & (12) & (13) & (14) & (15) & (16) & (17) & (18) & (19)\\
\hline\endfirsthead\caption{continued.}\\\hline\hline
 & in &  &  &  &  &  &  &  &  &  &  &  &  & $\log L$ & $\log L$ & $\log L^\mathrm{nuc}$ & $\log L^\mathrm{int}$ & \\
Name & 2MRS & Seyfert & Sy\,1 & Sy\,2 & LINER & H\,II & starburst & RA & DEC & z & D & \wonetwo & \wtwothree & (\wone) & (\wthree) & ($12\um$) & (2-10\,keV) & S18\\
 &  &  &  &  &  &  &  & [$\degree$] & [$\degree$] &  & [Mpc] & [mag] & [mag] & [erg/s] & [erg/s] & [erg/s] & [erg/s] & \\
(1) & (2) & (3) & (4) & (5) & (6) & (7) & (8) & (9) & (10) & (11) & (12) & (13) & (14) & (15) & (16) & (17) & (18) & (19)\\
\hline\endhead\hline\endfoot
UGC 00006 & True & True & True & True & False & True & False & 0.79009 & 21.96016 & 0.02195 & 98.7 & 0.81 & 3.51 & 43.6 & 43.8 & 43.58$\pm$0.57 & \dots & True\\
MCG -01-01-069 & True & True & True & False & False & True & False & 4.36917 & -3.23050 & 0.02129 & 95.8 & 1.68 & 3.02 & 42.8 & 43.2 & 42.81$\pm$0.57 & \dots & True\\
2MASX J00193596-0440105 & True & True & False & True & True & False & False & 4.89999 & -4.66957 & 0.02052 & 92.2 & 1.02 & 3.16 & 43.4 & 43.6 & 43.21$\pm$0.57 & \dots & True\\
2MASX J00253292+6821442 & False & True & False & True & False & False & False & 6.38696 & 68.36228 & 0.01200 & 53.6 & 0.68 & 3.09 & 42.9 & 42.9 & 42.62$\pm$0.22 & 43.2 & True\\
ESO 350-IG 038 & False & False & False & False & False & True & False & 9.21958 & -33.55472 & 0.02060 & 92.6 & 1.34 & 4.46 & 43.2 & 44.0 & 43.39$\pm$0.23 & \dots & False\\
NGC 0253 & True & True & False & True & False & True & True & 11.88800 & -25.28822 & 0.00081 & 3.2 & 0.73 & 4.94 & 42.2 & 43.0 & $\le$41.50 & 39.7 & False\\
NGC 0262 & True & True & True & True & False & False & False & 12.19642 & 31.95697 & 0.01502 & 67.2 & 1.21 & 3.11 & 43.4 & 43.6 & 43.45$\pm$0.22 & 43.5 & True\\
IC 1623B & True & True & True & False & False & True & False & 16.94817 & -17.50697 & 0.02025 & 91.0 & 1.24 & 3.85 & 43.6 & 44.1 & 43.92$\pm$0.57 & \dots & True\\
\end{longtable}

%% file: tab_R90candi.tex
\scriptsize
\begin{longtable}{l c c c c c c c c c c c c c}
\caption{R90 AGN candidates\label{tab_R90candi}}\\
\hline\hline
 & in &  &  &  &  &  &  &  &  & $\log L$ & $\log L$ & $\log L^\mathrm{nuc}$ & \\
Name & 2MRS & H\,II & starburst & RA & DEC & z & D & \wonetwo & \wtwothree & (\wone) & (\wthree) & ($12\um$) & S18\\
 &  &  &  & [$\degree$] & [$\degree$] &  & [Mpc] & [mag] & [mag] & [erg/s] & [erg/s] & [erg/s] & \\
(1) & (2) & (3) & (4) & (5) & (6) & (7) & (8) & (9) & (10) & (11) & (12) & (13) & (14)\\
\hline\endfirsthead\caption{continued.}\\\hline\hline
 & in &  &  &  &  &  &  &  &  & $\log L$ & $\log L$ & $\log L^\mathrm{nuc}$ & \\
Name & 2MRS & H\,II & starburst & RA & DEC & z & D & \wonetwo & \wtwothree & (\wone) & (\wthree) & ($12\um$) & S18\\
 &  &  &  & [$\degree$] & [$\degree$] &  & [Mpc] & [mag] & [mag] & [erg/s] & [erg/s] & [erg/s] & \\
(1) & (2) & (3) & (4) & (5) & (6) & (7) & (8) & (9) & (10) & (11) & (12) & (13) & (14)\\
\hline\endhead\hline\endfoot
2MASX J00042025+3120313 & False & False & False & 1.08440 & 31.34204 & 0.01692 & 75.8 & 1.14 & 4.02 & 42.5 & 43.0 & 42.64$\pm$0.57 & True\\
ESO 409-IG 015 & False & False & False & 1.38273 & -28.09818 & 0.00244 & 8.2 & 0.83 & 3.88 & 39.4 & 39.8 & 38.81$\pm$0.57 & True\\
HS 0017+1055 & False & False & False & 5.08917 & 11.20583 & 0.01885 & 84.6 & 1.27 & 4.70 & 41.1 & 41.9 & 41.37$\pm$0.57 & False\\
2MASS J00203545-2751440 & False & False & False & 5.14779 & -27.86225 & 0.02079 & 93.5 & 0.69 & 2.88 & 41.9 & 41.8 & 41.23$\pm$0.57 & True\\
UGC 00521 & False & False & False & 12.80075 & 12.02525 & 0.00226 & 33.0 & 0.66 & 1.97 & 40.8 & $\le$40.3 & 39.44$\pm$0.57 & True\\
MCG +12-02-001 & True & False & False & 13.51664 & 73.08485 & 0.01576 & 70.6 & 0.65 & 4.82 & 43.3 & 44.0 & 43.74$\pm$0.57 & False\\
UM 296 & False & True & True & 14.76710 & 1.00119 & 0.01782 & 79.9 & 0.93 & 3.89 & 42.0 & 42.4 & 41.44$\pm$0.23 & True\\
MCG -01-03-072 & False & False & False & 15.59546 & -4.50859 & 0.00584 & 25.9 & 1.54 & 3.52 & 39.4 & $\le$39.9 & 38.96$\pm$0.57 & True\\
\end{longtable}